\documentclass[a4paper,11pt]{article}
\pdfoutput=1 

\usepackage{jcappub} 

\usepackage[T1]{fontenc} 

\usepackage{ae,aecompl}
\usepackage{enumitem}
\usepackage{mathtools,slashed}
\usepackage{physics}
\usepackage{longtable}
\usepackage{float}
\usepackage{subfig}
\hypersetup{colorlinks,linkcolor={blue},citecolor={blue},urlcolor={blue}}  
\usepackage{epsf,palatino,url,wrapfig,array,setspace,amsmath,amssymb,fancyhdr,multirow,lscape,appendix,rotating,etoolbox}
\usepackage{graphicx, graphics,epsfig,txfonts, xcolor}
\usepackage{threeparttable}
\usepackage{extarrows}

\newcommand{\code}{ATHE$\nu$A}
\renewcommand{\d}[1]{\ensuremath{\operatorname{d}\!{#1}}}

\title{A closer look at the electromagnetic signatures of Bethe-Heitler pair production process in blazars}

\author[a]{D. Karavola}
\author[a,b]{M. Petropoulou}

\affiliation[a]{Department of Physics, National and Kapodistrian University of Athens, \\ University Campus Zografos, GR 15783, Greece}
\affiliation[b]{ Institute of Accelerating Systems \& Applications, \\ University Campus Zografos, GR 15783, Greece}

\emailAdd{dkaravola@phys.uoa.gr}
\emailAdd{mpetropo@phys.uoa.gr}

\abstract{The ``twin birth'' of a positron and an electron by a photon in the presence of a nucleus, known as Bethe-Heitler pair production, is a key process in astroparticle physics. The Bethe-Heitler process offers a way of channeling energy stored in a population of relativistic protons (or nuclei) to relativistic pairs with extended distributions. Contrary to accelerated leptons, whose maximum energy is limited by radiative losses, the maximal energy of pairs is determined by the kinematics of the process and can be as high as the parent proton energy. We take a closer look at the features of the injected pair distribution, and provide a novel empirical function that describes the spectrum of pairs produced by interactions of single-energy protons with single-energy photons. The function is the kernel of the Bethe-Heitler pair production spectrum that replaces a double numerical integration involving the complex differential cross section of the process, and can be easily implemented in numerical codes. We further examine under which conditions Bethe-Heitler pairs produced in blazar jets can emit $\gamma$-ray photons via synchrotron radiation, thus providing an alternative to the inverse Compton scattering process for high-energy emission in jetted active galactic nuclei. For this purpose, we create 36 numerical models using the code \code \, optimized so that the Bethe-Heitler synchrotron emission dominates their $\gamma$-ray emission.
After taking into consideration the broadband spectral characteristics of the source, the jet energetics, and the properties of radiation fields present in the blazar environment, we conclude that $\gamma$-rays  in high-synchrotron-peaked blazars are unlikely to be produced by Bethe-Heitler pairs, because the emitting region is found to be opaque in photon-photon pair production at photon energies $\gtrsim 10$~GeV. On the contrary, $\gamma$-ray spectra of low-synchrotron-peaked blazars may arise from Bethe-Heitler pairs in regions of the jet with typical transverse size $\sim 10^{15}-10^{16}$ cm and co-moving magnetic field $50-500$~G. For such cases, an external thermal target photon field with temperatures $T\sim 4 \cdot 10^2 - 6 \cdot 10^3$K is needed. The latter values could point to the dusty torus of the AGN. Interestingly, a Bethe-Heitler-dominated high-energy component is mostly found in models of intermediate-synchrotron peaked blazars, for a wide range of magnetic fields and source radii.}

\begin{document}

   
\maketitle
\flushbottom

\section{Introduction}
The ``twin birth'' of a positron and an electron by a photon in the presence of a nucleus,  known as Bethe-Heitler pair production, is a key process in astroparticle physics. While the cross section of the interaction has been provided by Hans Bethe and Walter Heitler in 1934 \cite{Bethe_Heitler}, it was not until the discovery of the cosmic-wave background radiation in the mid-sixties \cite{penzias_wilson_1965} and the detection of ultra-high-energy cosmic rays \citep{PhysRevLett.10.146}, that Bethe-Heitler pair production received attention by the astrophysics community, and was discussed as an energy loss mechanism for cosmic rays during their propagation to Earth \cite[e.g.][]{Greisen_1966, Zatsepin_Kuzmin_1966, blumenthal_1970}. 

With the discovery of X-ray non-thermal compact sources, such as active galactic nuclei (AGN), there was a focus shift to pair cascades \cite[e.g.][]{1987MNRAS.227..403S, 1990ApJ...363L...1Z}. Because acceleration of electrons to relativistic energies can be radiation-limited, other processes that can efficiently channel energy to relativistic pairs have been discussed, including Bethe-Heitler pair production~\citep[e.g.][]{1987ApJ...320L..81S, 1990ApJ...362...38B, Kirk_Mastichiadis_1992}. Because this is a subdominant energy loss channel for relativistic protons that satisfy the energy threshold for pion production, Bethe-Heitler pair injection was often neglected in AGN radiation models. Nevertheless, Bethe-Heitler pairs are injected with a different energy distribution than the pairs from photopion production interactions \citep[e.g.][]{kelner_energy_2008}. Therefore, Bethe-Heitler pairs may still leave a radiative signature on the AGN spectrum as illustrated in more recent numerical works \cite{petropoulou_bethe-heitler_2015, 2018ApJ...865..124M, 2019ApJ...874L..29R, 2020ApJ...889..118Z, 2023MNRAS.524...76Z}.

To the best of our knowledge no analytical description of the pair production spectrum arising from the interaction of a single proton with a single photon exists in the literature. Numerical codes for multi-messenger emission that take into account the injection of pairs from Bethe-Heitler interactions \citep[e.g.][]{refId0, cerruti_hadronic_2015, zacharias_exhale-jet_2022, lehamoc, am3} employ numerical integration of at least two integrals\footnote{Four integrals are computed when dealing with extended distributions of interacting particles.}, one involving the energy- and angle-dependent cross section of the process \citep{blumenthal_1970}. In works aiming at the study of non-linear pair cascades (where the target photon field evolves with time and depends on the parent proton distribution) or in cases where high-accuracy is needed, the calculation of the Bethe-Heitler spectrum can be a computational bottleneck (for ways to speed up the computation, see Refs.~\cite{lehamoc} and \cite{am3}). 

In this work we construct an empirical analytical function for the Bethe-Heitler pair injection spectrum that can be easily implemented in numerical codes (section \ref{sec:BH-spec}). Our function is benchmarked with numerical results from the code \code \ that are based on Monte-Carlo simulations of proton-photon interactions performed by \cite{PJ96}. The function is not a unique mathematical representation of the pair production spectrum. Instead, it was conceptualized by studying the shape of the pair production spectrum, and identifying scaling relations with the proton Lorentz factor and the target photon energy as seen in the proton rest frame (i.e. the interaction energy). Using the empirical function we take a closer look at key features of the pair injection spectrum produced in interactions between protons and photons with power-law distributions, which are typically expected in astrophysical sources. By considering these key features of the injection spectrum, we examine under which conditions Bethe-Heitler pairs can produce $\gamma$-ray spectra as those observed from AGN jets (section \ref{sec:application}), and conclude with a discussion of our findings (section \ref{sec:discussion}).

\section{The Bethe-Heitler pair production spectrum}\label{sec:BH-spec}
The distribution of electrons or positrons 
produced by a single proton with Lorentz factor $\gamma_p$ interacting with a population of photons with distribution $f_{ph}(\epsilon)$ is given by \cite{kelner_energy_2008}:
    \begin{equation}
    \frac{\d N}{\d \gamma_e}=\frac{1}{2 \gamma_p^3} \int_{\frac{(\gamma_{p}+\gamma_e)^2}{4\gamma_{p}^2\gamma_e}} ^{\frac{m_p}{\gamma_{p}m_e}} \d \epsilon \frac{f_{ph}(\epsilon)}{\epsilon^2}\int_{\frac{(\gamma_{p}+\gamma_e)^2}{2\gamma_{p}\gamma_e}}^{2\gamma_{p}\epsilon}\d\omega\ \omega\int_{\frac{(\gamma_{p}^2+\gamma_e^2)}{2\gamma_{p}\gamma_e}} ^{\omega-1} \d E_{-} \frac{W(\omega ,E_{-} ,\xi )}{p_{-}} \label{dNdE_full}
    \end{equation}
where $W(\omega ,E_{-} ,\xi )$ is the cross section of the interaction  \citep{blumenthal_1970}, $E_{-}$ is the electron energy, $\omega$ is the photon energy in the proton rest frame, and $\xi\equiv \frac{\gamma_{p} E_{-}-\gamma_{e}}{\gamma_{p} p_{-}}$ where $\gamma_p$ and $\gamma_e$ are the proton and pair Lorentz factors, respectively. In the expression above\footnote{The limits of integration are obtained in the regime $1 \ll  \gamma_p \epsilon \ll m_p/m_e$.}  all energies are normalized to $m_e c^2$. 

In codes that model the time evolution of particle distributions the pair distribution at injection has to be computed at every time step \citep{mastichiadis1995synchrotron, refId0, cerruti_hadronic_2015, gao_direct_2017}. Numerical integration of the last integral involving the cross section can slow down the computations, especially if these aim at high accuracy. The last two integrals can be instead pre-calculated for various combinations of $\gamma_e$, $\gamma_p$, and $\omega$ values \citep{cerruti_hadronic_2015, gao_direct_2017}. By using interpolation in a three-dimensional space, the desired integral values can then be obtained by reducing the computational time by a factor of $\sim 20-30$ \citep{lehamoc}. Nevertheless, the computation would be even faster, if an empirical analytical function describing the pair distribution was available. A similar approach was adopted by \cite{kelner_energy_2008}, who derived analytical parametrizations for the 
energy distributions of secondary particles produced in proton-photon interactions that lead to the production of pions.

Our goal is to provide an analytical function for the Bethe-Heitler pair production spectrum (section \ref{sec:function}) that could be easily implemented in numerical codes or used in semi-analytical calculations. Moreover, by using this empirical function we will study key features of the pair production spectrum, such as characteristic energies and power-law slopes (section \ref{sec:features}) for various energy distributions of interacting particles.

\subsection{An empirical function for the pair production spectrum}\label{sec:function}

We present an analytical function that describes the total injection spectrum of particles (i.e., electrons and positrons, later referred to as pairs), $q_{\rm BH}(\gamma_e)$, produced by the Bethe-Heitler interaction process of a single proton of Lorentz factor $\gamma_p$ and a single target photon of energy $\epsilon$ (in units of $m_e c^2$). Our function is benchmarked with numerical results from \code~\cite{mastichiadis_spectral_2005,refId0} that are based on Monte-Carlo simulations of proton-photon interactions performed by \cite{PJ96}. Details about the function design and its performance against \code \ results for different cases are presented in Appendix~\ref{App:BH_Analyt}. We also compare the results of our novel function, $q_{BH}$, against those expected by the numerical implementation of the Bethe-Heitler production spectrum presented in \cite{kelner_energy_2008} (see Appendix~\ref{app:KA-comparison}).

\begin{figure}[h!]
    \centering
        \includegraphics[width=0.48\textwidth]{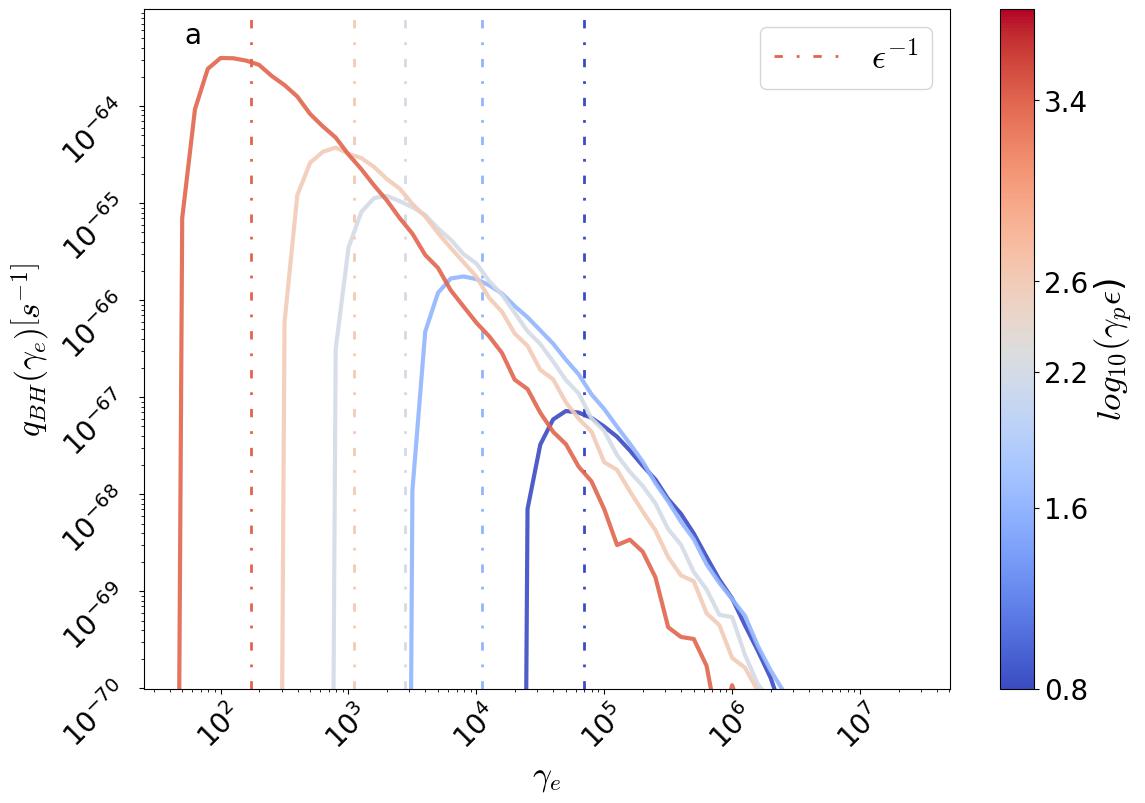}
        \includegraphics[width=0.48\textwidth]{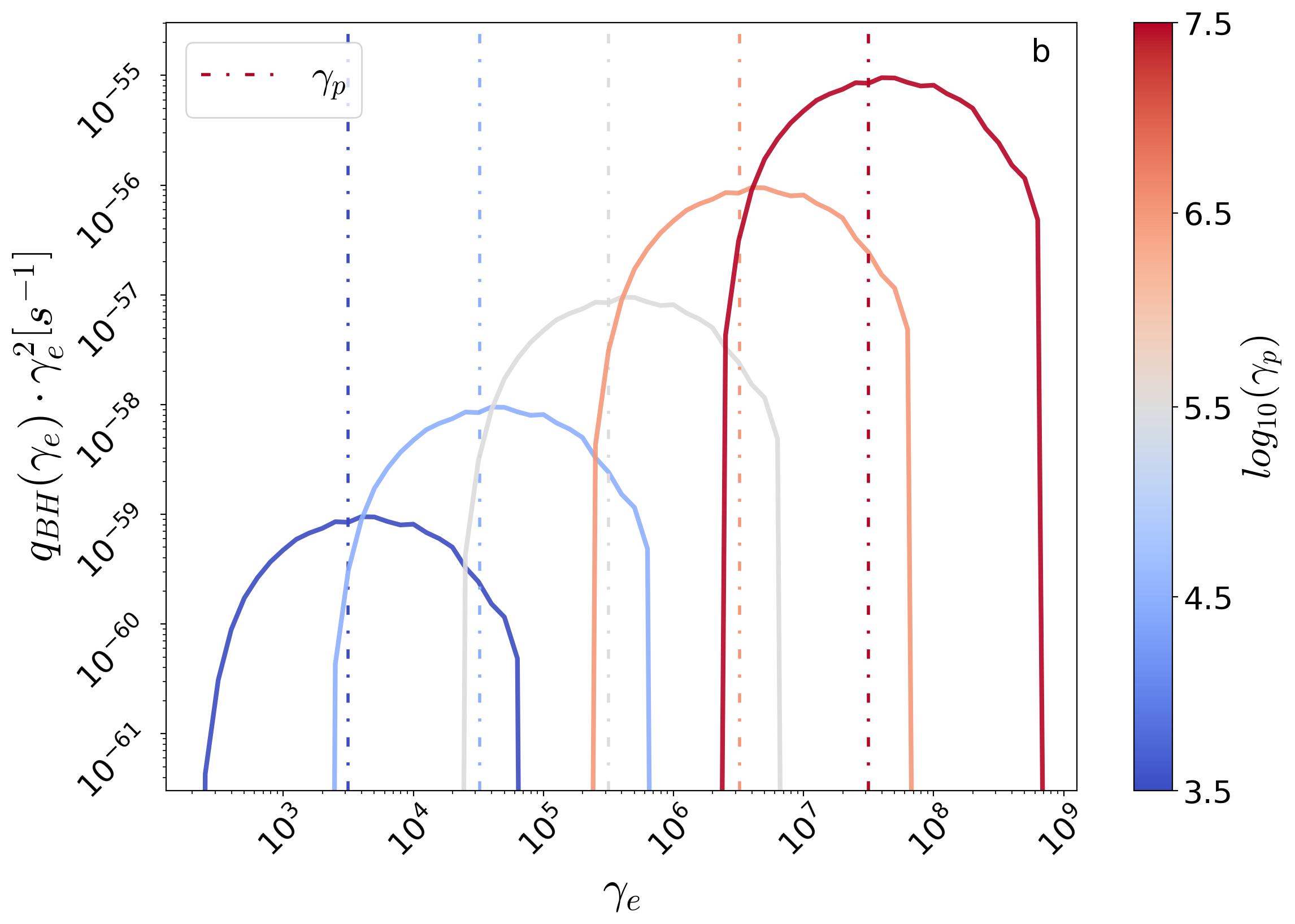}
        \includegraphics[width=0.48\textwidth]{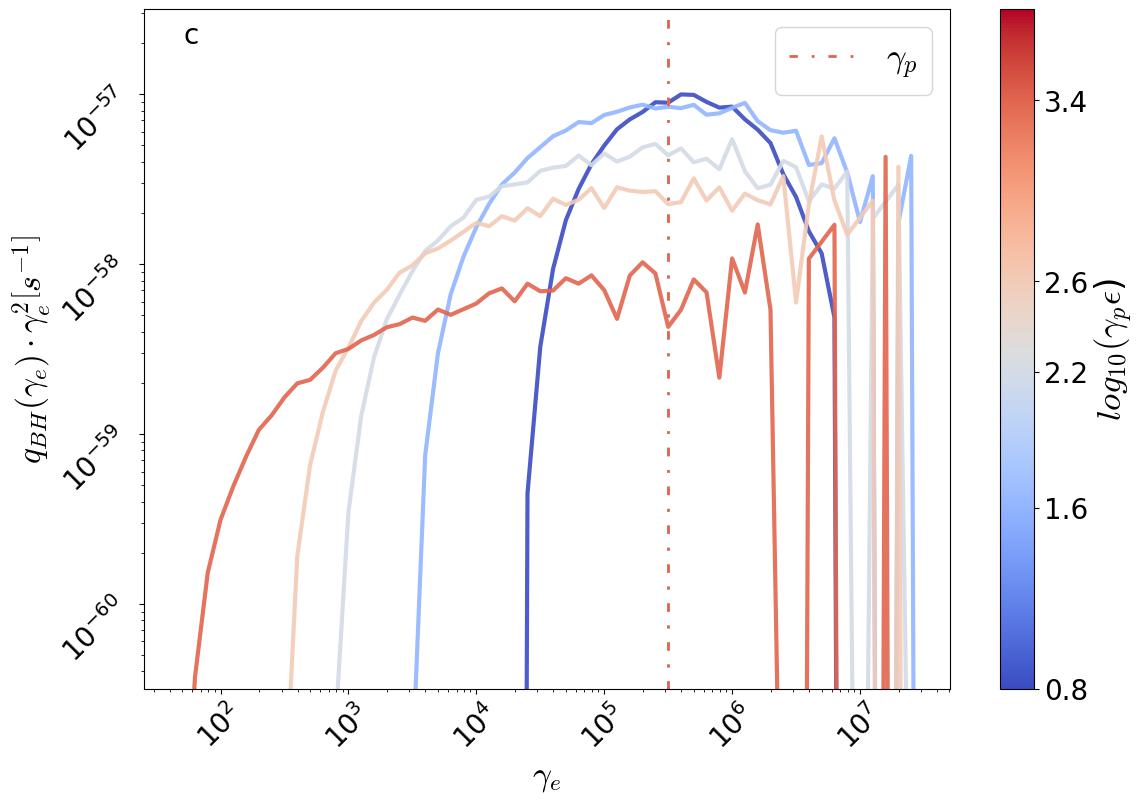}
        \includegraphics[width=0.48\textwidth]{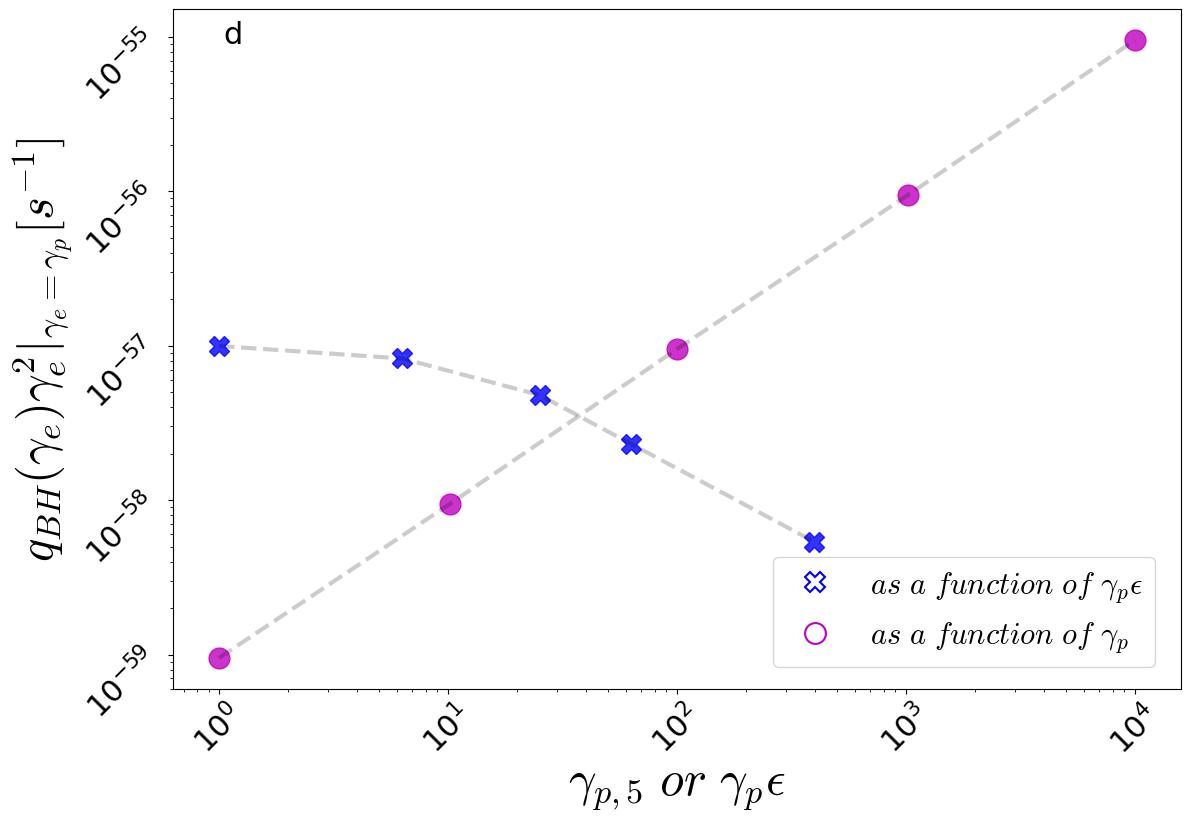}
\caption{Injection spectra, produced by \code \ \citep{mastichiadis_spectral_2005, refId0}, of Bethe-Heitler pairs produced by interactions of (i) a single proton with $\gamma_p = 10^{5.5}$ with a single photon of different energies (see color bar in panels a and c) and (ii) a single proton of a changing Lorentz factor (see color bar in panel b) with a single photon of energy adjusted, in each case, in order to keep the interaction energy constant. Panel (d) shows the maximum value of the energy injection spectra from panels (b) and (c) as a function of the proton Lorentz factor (normalized to $10^5$) and the interaction energy, respectively.}
\label{fig:BH_char}
\end{figure} 

Our goal is to describe the injection spectrum of secondaries produced by the interaction of a single proton with a single photon. In \code, where all particle energies are discretized, mono-energetic distributions are approximated by particles occupying a single bin of energy. More specifically, \code \ employs logarithmic energy grids with resolution of 0.1 for protons and pairs, and 0.2 for photons. For example, a mono-energetic proton distribution with Lorentz factor $10^5$ in \code \  is described by a power-law distribution of slope $s_p$ in the logarithmic bin $[5.0, 5.1]$. Furthermore, a single mono-energetic proton in \code \ is described by the proton distribution in the selected energy bin divided by the total number of protons occupying this bin.

Since \code \ is used to benchmark the empirical Bethe-Heitler injection spectrum, the latter function is bound to inherit the limitations of its numerical implementation. For instance, the Monte Carlo simulations performed by \cite{PJ96}, and used in tabulated form in \code, \ account for interaction energies up to $10^4$ (i.e., $\gamma_p \epsilon \le 10^4$). Furthermore, the high-energy tail of the pair production spectrum ($\gamma_e \gtrsim \gamma_p$) has saw-like edges due to low-number statistics of the Monte Carlo simulations (e.g., see figure \ref{fig:BH_char}c).
The aforementioned edges make it difficult to model the spectrum and introduce uncertainties in the description of the high-energy cutoff. Finally, we benchmark the empirical functions for interactions involving protons with $\gamma_p > 10$. The reason is that the self-similarity of Bethe-Heitler injection spectra for near-threshold interactions breaks down for protons with $\gamma_p < 10$.
All the above shape the validity regime of our  empirical function $q_{BH}(\gamma_e)$.

We first present pair production spectra computed with \code \ (figure \ref{fig:BH_char}) in order to highlight the main features that we aimed at capturing with the analytical function. We observe that the differential production rate peaks at a Lorentz factor $\gamma_{e, \rm pk}\approx\epsilon^{-1}$, with $\epsilon$ being the target photon energy (see panel a in figure \ref{fig:BH_char}). Meanwhile, the position of the aforementioned injection rate peak is not affected by the proton Lorentz factor (see figure \ref{fig:gpE_diff_comp} in Appendix \ref{App:BH_Analyt}). However, this behavior changes when we look at the energy distribution of the produced pairs, $\gamma_e^2 q_{\rm BH}(\gamma_e)$. When the interaction takes place near the threshold (i.e. $\gamma_p \epsilon \gtrsim 2$), most of energy is transferred to pairs with $\gamma_e \approx \gamma_p$ (see panel b). Both the peak Lorentz factor and the peak of the energy injection rate scale linearly with $\gamma_p$ for interactions close to the threshold (see also panel d). On the contrary, the pair energy distribution from interactions taking place away from the threshold tends to spread over a wider range of pair Lorentz factors without having a well-defined peak (panel c). Instead, these interactions seem to transfer roughly equal amounts of energy into pairs that are produced with a wider distribution of Lorentz factors (see redder curve in panel c). 
For interactions happening away from the energy threshold, the energy injection spectra show large fluctuations at high Lorentz factors due to low event statistics of the Monte Carlo simulations that are provided, in a tabular form, as input to the code. As a result, the high-energy tail of the analytical spectrum will be more uncertain (see also Appendix \ref{App:BH_Analyt}).

The empirical function we constructed reads,
    {\large
    \begin{gather}
        \centering
         q_{\rm BH}(\gamma_e)= A(\gamma_p,  \epsilon) \cdot \exp \left[ -\frac{\left[\log_{10}\left(\frac{\gamma_e}{\gamma_{e,\rm pk}}\right)\right]^{ p(\gamma_p \epsilon)}}{2 {a_1}^2}- a_2^2 \left(\frac{\gamma_{e, \rm pk}}{\gamma_e}-1\right)^2 -a_3 \frac{\gamma_e}{\gamma_{e, \rm cr}}  \right] 
        \label{Q_inj_approx}
    \end{gather} 
    }
where $\gamma_{e, \rm pk}= (1.23 \epsilon)^{-1}$ is the Lorentz factor where the number of injected pairs peaks (i.e. where $\gamma_e q_{\rm BH}(\gamma_e)$ has a maximum), $a_1, a_2, a_3$ and $\gamma_{e, \rm cr}$ are summarized in table~\ref{tab:inj-params} and 
\begin{gather}
    p(\gamma_p \epsilon) =\left \{
    \begin{array}{cc}
        \scalebox{1.1}{$a \left [\log_{10}(\gamma_p \epsilon)\right ]^{-s} e^{-\frac{\log_{10}(\gamma_p \epsilon)}{x_0}} + b \left[ \log_{10}(\gamma_p \epsilon)\right]^{s_2} + c$,}  & \hspace{0.1cm} \gamma_p \epsilon \geq 10^{x_0} \hspace{0.1cm} {\rm and} \hspace{0.1cm} \gamma_e  \geq \gamma_{e, \rm pk} \\ \\
        2, & \gamma_p \epsilon \leq 10^{x_0} \hspace{0.1cm} {\rm or} \hspace{0.1cm} \gamma_e  \leq \gamma_{e, \rm pk}
    \end{array}
    \right.
    \label{Q_inj_slope} 
\end{gather}
where $a=0.99,~b=-0.06,~c=1.56,~s=0.65,~s_2=0.94$, and $x_0=0.6586$. A more detailed discussion on how the parameters were determined is presented in Appendix \ref{App:BH_Analyt}. 

\begin{table}[h!]
    \centering
    \begin{tabular}{c c|c|c|c|c}
        {}&$a_1$ & $a_2$ & $a_3$ & $\gamma_{e, \rm cr}$ &  conditions \\ \hline
        (1) &0.47 & 0.95 & 0.1 & $\gamma_{e, \rm pk}$ & $\gamma_e \leq \gamma_{e, \rm pk} \hspace{0.1cm} \rm and \hspace{0.1cm} \forall \gamma_p \epsilon$  \\ 
        (2) & 0.468 & 1.0 & $0.2863-0.0665\gamma_p \epsilon$ & $\gamma_{e, \rm pk}$ & $\gamma_{e, \rm pk} \leq \gamma_{e} \hspace{0.1cm} \rm and \hspace{0.1cm} \gamma_p \epsilon \leq 4.2 \hspace{0.1cm}$\\
        (3) &0.468 & 1.0 &$0.007$ & $\gamma_{e, \rm pk}$ & $\gamma_{e, \rm pk} \leq \gamma_{e} \hspace{0.1cm} \rm and \hspace{0.1cm} \gamma_p \epsilon \leq 8.5 \hspace{0.1cm}$\\
        (4) &0.468 & 1.0 & 0.0 & $\gamma_{e, \rm pk}$ & $ \gamma_{e,\rm pk} \leq \gamma_e \leq 15 \gamma_p  \hspace{0.1cm} \rm and \hspace{0.1cm}  8.5 \leq \gamma_p \epsilon$  \\
         (5) &0.465 & 1.0 & 0.25 & $15\gamma_p$ & $15 \gamma_p \leq \gamma_e \leq \frac{m_p}{m_e}\gamma_p \hspace{0.1cm} \rm and \hspace{0.1cm} 8.5 \leq \gamma_p \epsilon $\\
         \hline
    \end{tabular}
    \caption{Values of parameters appearing in the empirical function of Eq.~(\ref{Q_inj_approx}).}
    \label{tab:inj-params}
\end{table}

To ensure continuity of $q_{\rm BH}$ at $\gamma_e = 15 \gamma_p$ we multiply Eq.~(\ref{Q_inj_approx}) with
\begin{gather}
    e^{0.03\left(\frac{\log_{10}(15 \gamma_p)}{\gamma_{e,pk}} \right)^{p(\gamma_p \epsilon)}}
\end{gather}
only if $8.5 \leq \gamma_p \epsilon \hspace{0.1cm} \& \hspace{0.1cm} 15 \gamma_p \leq \gamma_e \leq \frac{m_p}{m_e}\gamma_p$. Branches (2), (3) and (4) of table \ref{tab:inj-params} control the sharpness of the high-energy cutoff of the injection function, $q_{BH}$.

The normalization of the injection rate $A(\gamma_p, \epsilon)$ is given by (in units of s$^{-1}$)

\begin{eqnarray}
    \log_{10}[A(\gamma_p, \epsilon)]&=&De^{-d\left[\log_{10}(\gamma_p \epsilon)\right]^{2}}+E +F \log_{10}(\gamma_p \epsilon) + G \left[\log_{10}(\gamma_p)\right]^{g}  \\ \nonumber
    &+& H \log_{10}(\gamma_p \epsilon)^{h_1} e^{h_2\log_{10}(\gamma_p \epsilon)} + J \log_{10}(\epsilon) -3\log_{10}(R_{15})
    \label{Q_inj_norm}
\end{eqnarray}
where the parameters are listed in table \ref{tab:A_params} and $R^{\prime}_{b, 15}$ is the radius of the spherical blob normalized to $10^{15}$~cm.
\begin{table}[h]
    \centering
    \begin{tabular}{c|c|c|c|c|c|c|c|c|c|c|c|c}
       $D$ & $E$ & $F$ & $G$ & $H$ & $J$ & $d$ & $g$ & $h_1$ & $h_2$\\
       \hline
       -2.12 & -111.9 & -0.05 & 0.13 & 17.50 & 1.23  & 0.89 & 1.22 & 1.80 & -3.68
    \end{tabular}
    \caption{Fitted parameters for the normalization of the function of Eq. \ref{Q_inj_approx}. The posterior distributions of the parameters are shown in figure~\ref{fig:Q_inj_norm_corner} (appendix~\ref{App:BH_Analyt}).}
    \label{tab:A_params}
\end{table}
Additionally, $p(\gamma_p \epsilon)$ is the power-law index appearing in the first term of the exponential function in Eq.~(\ref{Q_inj_approx}), which reads

Comparison of the analytical function against \code \ for all the bench-marking cases we used can also be found in Appendix \ref{App:BH_Analyt}.
We present next the performance of the analytical function against numerical results for a more realistic case that involves interactions between protons and photons with extended power-law distributions.  The pair production spectrum in this case is a convolution of $q_{\rm BH}$ with the two particle distributions,

\begin{gather}
    Q(\gamma_e) = \iint N_p(\gamma_p) N_{ph}(\epsilon) q_{\rm BH}(\gamma_p, \epsilon, \gamma_{e}) \d \gamma_{p} \d \epsilon,
    \label{Q_tot}
\end{gather}
with $N_p(\gamma_p)=\d N_p/\d\gamma_p$ and $N_{ph}(\epsilon) = \d N_{ph}/\d \epsilon$ being the differential number distributions of protons and photons respectively. 

We consider a power-law proton distribution $N_p(\gamma_p) \propto \gamma_p^{-2}$ extending from $\gamma_{p,\min}=10^4$ to $\gamma_{p, \max}=10^8$ interacting with a power-law photon distribution, $N_{ph}(\epsilon)  \propto \epsilon^{-s_{ph}}$. We adopt five different photon indices, $s_{ph}=-1,0,1,2,3$, that cover a wide range of potentially relevant astrophysical sources. For example, non-thermal photon spectra from high-energy astrophysical sources are usually soft with photon indices $s_{ph}\ge 2$, while the Rayleigh-Jeans part of a black-body spectrum has $s_{ph} = -1 $. The photon spectrum extends from $\epsilon_{\min}=10^{-8}$ to $\epsilon_{\max}=10^{-4}$. With the adopted energy ranges the highest energy protons will ``see'' as targets the whole power-law photon distribution, while the lowest energy protons will be able to interact only with the highest energy photons of the distribution. In all cases, the target photon compactness, defined as 
\begin{equation} 
\ell_{ph} = \frac{\sigma_T R^{\prime}_b u_{ph}}{3 m_e c^2} = \frac{\sigma_T}{4 \pi R^{\prime, 2}_b}\int_{\epsilon_{\min}}^{\epsilon_{\max}} \d \epsilon \, \epsilon \, N_{ph}(\epsilon),
\end{equation}
is fixed and equal to $0.1$. Similarly, the proton compactness is $\ell_p=10^{-6}$ and $R^{\prime}_b=10^{15}$~cm.

\begin{figure}
    \centering
    \includegraphics[width=0.49\textwidth]{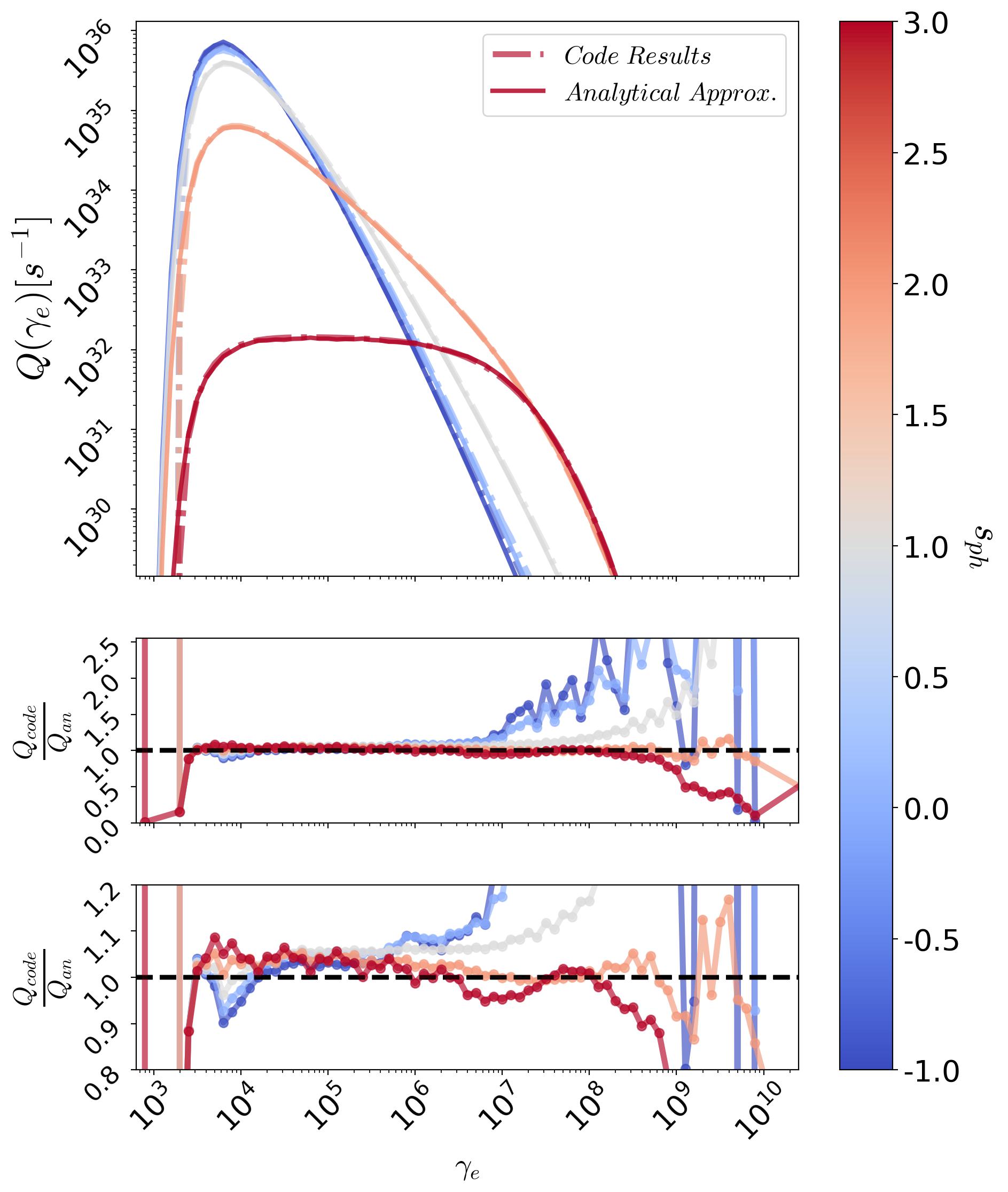}
    \hfill
    \includegraphics[width=0.49\textwidth]{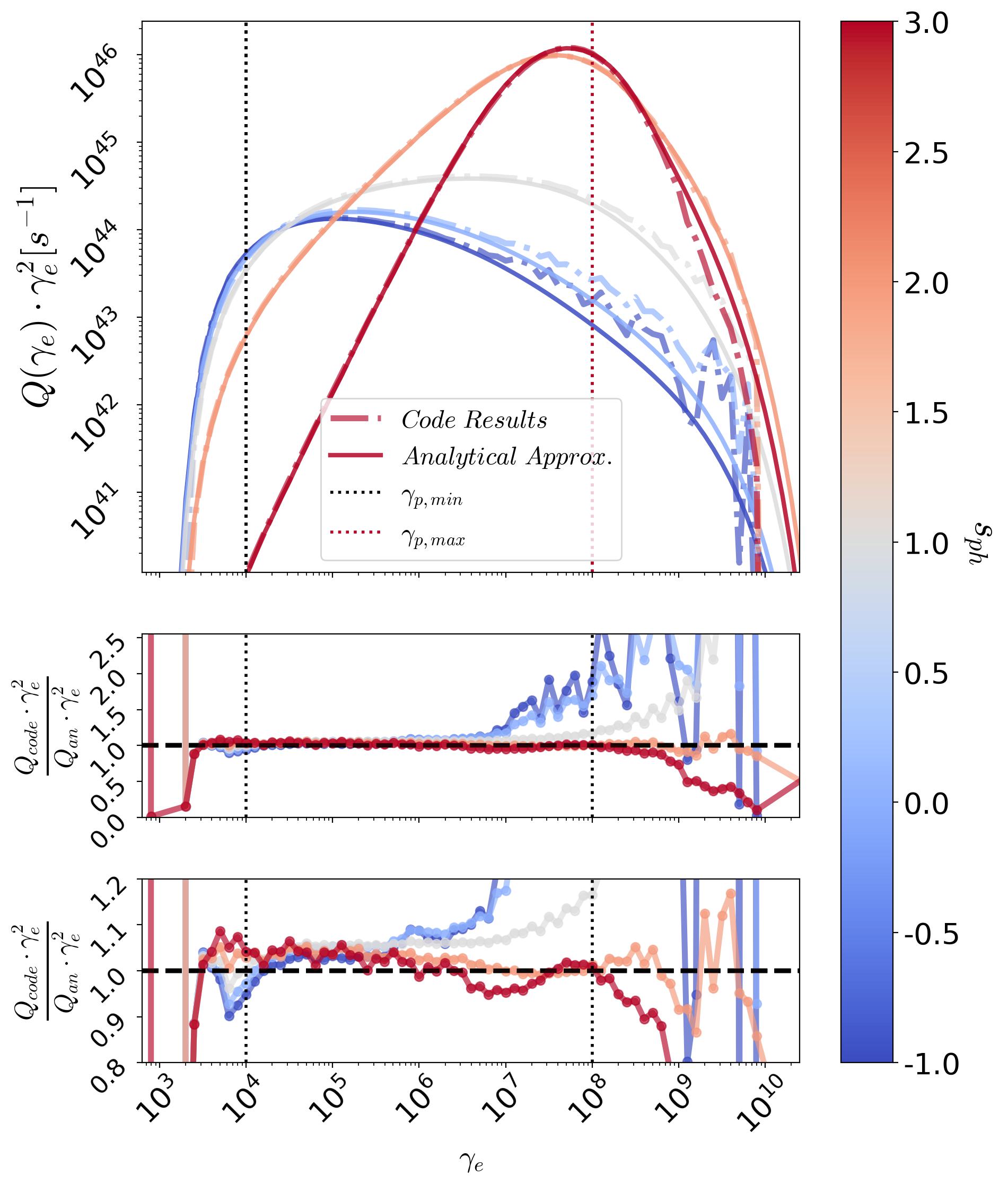}
\caption{Injection spectra of pairs produced by interactions of power-law protons with slope $s_p=2$ and $\gamma_p \in [10^4, 10^8]$ interacting with power-law photons with $\epsilon \in [10^{-8}, 10^{-4}]$ and different photon indices (see color bar). Solid lines represent numerical results from the  \code \ code \cite{mastichiadis_spectral_2005, refId0} and dashed lines represent results obtained with the analytical approximation given by Eq.~(\ref{Q_inj_approx}). To facilitate the comparison of the two solutions, a ratio plot is added below the main panels, with the lower panel zooming into values in the range of 0.8-1.2}.
\label{fig:PLs}
\end{figure}

Our results are presented in figure~\ref{fig:PLs}. The kernel of the Bethe-Heitler pair production spectrum, $q_{\rm BH}$, when convolved with the particle power-law distributions yields pair spectra that are in good agreement with the numerical results of \code \ (see left panel). We observe that the peak position and the peak value of the differential pair injection rate are well reproduced. The shape of the curve for a wide range of pair Lorentz factors is overall well described for all values of the photon index we explored. Only at the high-energy tail of the distribution we start seeing a discrepancy between the numerical results and semi-analytical spectra. The maximum difference we find is about factor of 2.5 (see ratio plot in the left panel of \ref{fig:PLs}), but the production rate at these Lorentz factors is many orders of magnitude lower than its peak value. These differences at high energies become more pronounced when comparing the energy distributions of produced pairs -- see right panel of figure \ref{fig:PLs}. For soft target photon distributions ($s_{ph}=3$), the semi-analytical result overestimates by a factor of almost 3, at most, the energy injection rate at the high-energy tail of the distribution. On the contrary, it slightly underestimates the energy injection rate by up to a similar factor when the target photon spectra are very hard ($s_{ph}\le 0$). Nonetheless, the analytical function can reproduce the shape of the energy distribution function for a wide range of Lorentz factors, including the peak position and its peak value.  
 
\subsection{Key features of the pair production spectrum and synchrotron emission}\label{sec:features}

The empirical function of Eq.~(\ref{Q_inj_approx}) describes well the general features of the numerically computed pair production spectrum. In this paragraph, we therefore use the empirical function to study key features of the pair injection spectrum produced by interacting power-law (PL) distributions (section \ref{sec:pairs}), as well as the resulting synchrotron spectra that are a common observable in non-thermal astrophysical sources (section \ref{sec:syn}). 

\begin{figure}
    \centering
        \includegraphics[width=0.99\textwidth]{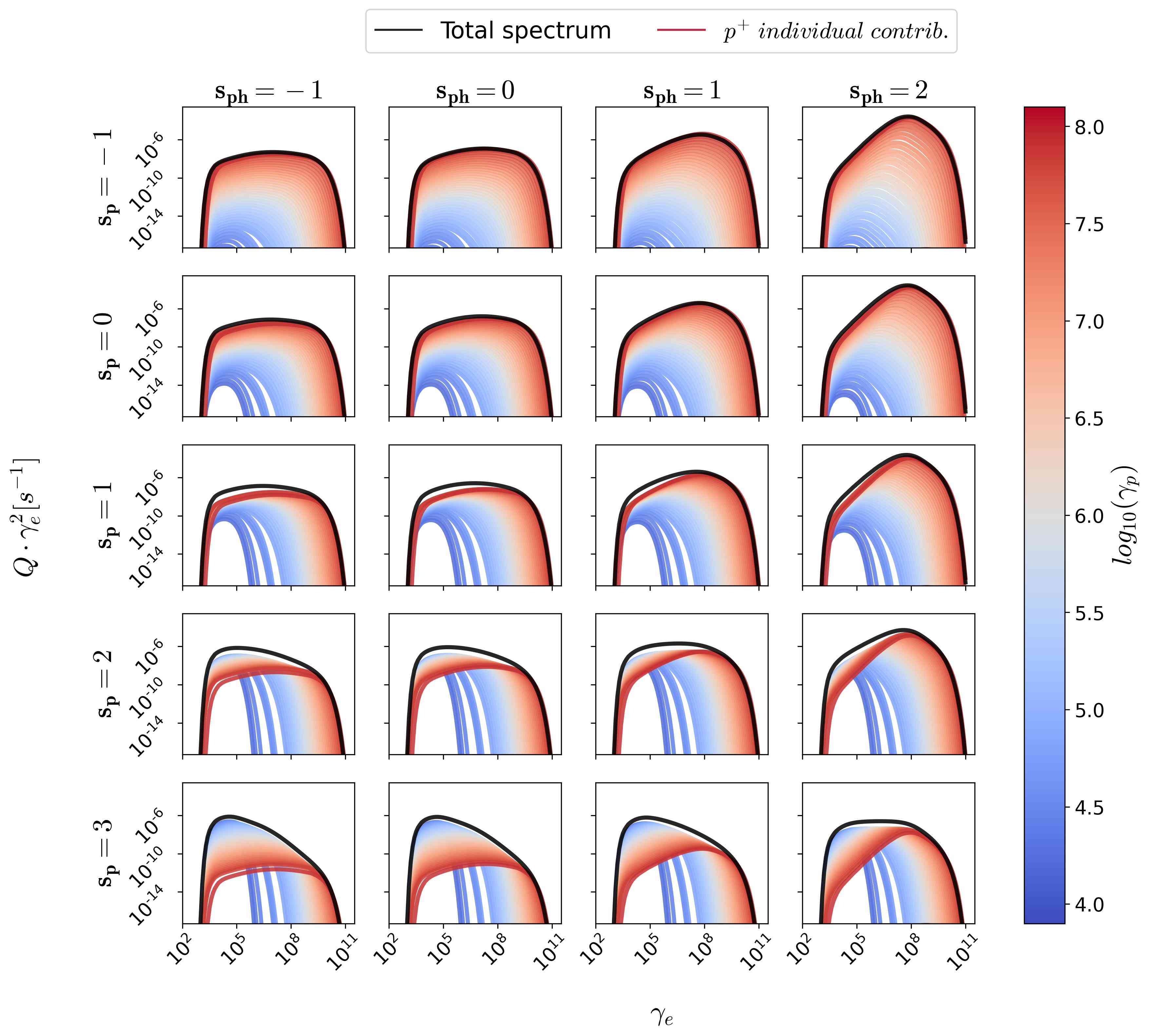}
\caption{Energy injection spectra of pairs produced in Bethe-Heitler interactions between protons with a power-law~(PL) distribution for different slopes $s_p$, and PL photon distributions with different photon indices, $s_{ph}$.  All other parameters are the same as in figure~\ref{fig:PLs}. Black solid lines represent the total injection spectrum obtained after convolving the empirical function of  Eq.~(\ref{Q_inj_approx}) with the proton and photon distributions. Colored solid lines represent the contribution of single-energy protons (see color bar) to the total spectrum. When $s_{ph}-s_{p}>1$ ($<1$) the peak position and value of the energy injection spectrum is determined by the highest (lowest) energy protons of the distribution.}
\label{fig:p_contr}
\end{figure}

\subsubsection{Injection spectra of Bethe-Heitler pairs}\label{sec:pairs}
In figure \ref{fig:PLs} (left panel) one may notice that the differential injection rate peaks at $\gamma_{e, \rm pk} \approx 1/\epsilon_{\max} \approx  \gamma_{p, \min}$ regardless of the photon index, with the exception of the $s_{ph}=3$ case where the pair injection spectrum appears almost flat for $\gamma_e \in [\epsilon_{max}^{-1}, \epsilon_{min}^{-1}]$. On the contrary, the peak position of the energy injection spectra depends on the photon index as shown in the right panel of figure \ref{fig:PLs}. 

Here, we explore the dependence of the energy injection spectrum on both $s_p$ and $s_{ph}$ by considering cases with $s_p=-1, 0, 1, 2, 3$ and $s_{ph}=-1, 0, 1, 2$. In all examples, we use $\ell_{ph}= 10^{-1}, \ell_p=10^{-6}$, and $R^{\prime}_b=10^{15}$~cm. In figure \ref{fig:p_contr} we present a decomposition of the energy injection spectrum into separate components produced by protons of specific Lorentz factors. For proton distributions with $s_p \leq 1$ (where the number of protons, $\gamma_p \frac{\d N_p}{\d \gamma_p} \propto \gamma_p^{-s_p+1}$, peaks at the highest energies for $s_p <1$ or protons are equally distributed in number across all energies ($s_p=1$)), we notice that the pair energy distribution is determined by the most energetic protons regardless of $s_{ph}$. However, for a fixed proton slope, changes in the photon index affect the shape of the produced distribution (see e.g. second row from the top). For hard photon spectra (e.g. $s_{ph}=-1, 0$) the spectrum appears to be almost flat. The reason for this behavior is that most photons are concentrated in the high-energy tail of the distribution, thus providing far-from threshold targets for the higher energy protons, which are also more numerous. The aforementioned conclusion also agrees with the shape of the pair energy distribution (for $s_{ph}=-1, 0$ and $s_{p}=-1, 0, 1$) which coincides with the away-from-threshold spectral shape (redder curves); see also spectra for $\gamma_p \epsilon \gg 2$ in figure \ref{fig:BH_char}. Similarly, for softer photon spectra (i.e. larger values of $s_{ph}$) most photons are found in the low-energy part of the distribution. As a result, near-threshold interactions of the high-energy protons become in this case more important than those taking place away from threshold. The latter leads to a harder spectrum with a well defined maximum (see e.g. panel for $s_p=-1$ and $s_{ph}=2$).

Figure \ref{fig:p_contr} also shows that when low-energy protons dominate in number (i.e. $s_p>1$) they can determine the maximum of the pair energy distribution for certain choices of the photon index: when target photons are equally distributed across the energy spectrum ($s_{ph}=1$) or concentrated near the high-energy tail of the photon distribution ($s_{ph}=-1, 0$). In these cases threshold interactions of the low-energy protons are the ones being dominant (bluer lines). Furthermore, we notice that pair production by more energetic protons is also important in those scenarios, since they are the ones creating the power-law segment of the distribution beyond its peak. Nonetheless, for a fixed value of the proton index (e.g. $s_p=2$ or $s_p=3$), we observe that as the photon index increases the contribution of the high-energy protons (redder lines) becomes more significant, thus dominating the distribution and defining its peak (cases with $s_p=2$ and $s_{ph}=2$) or balancing the low-energy proton contributions (cases with $s_p=3$ and $s_{ph}=2$). It is noteworthy that the contributions of the high-energy and low-energy protons to the total pair energy spectrum are comparable when $s_p-s_{ph} \sim 1$. 

\subsubsection{Synchrotron spectra from Bethe-Heitler pairs}\label{sec:syn}
Here we use the analytical approximation of the Bethe-Heitler injection rate to compute the photon energy distribution emitted by the pairs. After all, the photon spectrum is the main observable of an astrophysical source. Given that synchrotron radiation is relevant to the majority of non-thermal emitting sources, we will compare the synchrotron spectrum emitted by Bethe-Heitler pairs as obtained from \code \, and by our empirical function. To do so, we calculate the steady-state distribution of pairs, as described in \cite{Inoue_Takahara}:
  \begin{gather}
        N_e(\gamma_e) = e^{- \frac{\gamma_{br}}{\gamma_e}} \frac{\gamma_{br} \tau_{esc}}{\gamma_e^2} \int_{\gamma_e}^{\infty} \d \gamma_e^{\prime}  \, Q(\gamma_e^{\prime}) \,  {\rm e}^{\frac {\gamma_{br}}{\gamma_e^\prime}} \label{steady_state_integral}
    \end{gather}
    with $\tau_{esc} \sim R_b^{\prime}/c$ being the escape timescale of pairs from the source and $Q$ is computed from Eq.~(\ref{Q_tot}). Furthermore, $\gamma_{br}$ represents the cooling break Lorentz factor, which in the case of synchrotron radiative losses, is defined as
        \begin{gather}
        \gamma_{br} = \frac{3 m_e c^2}{4 u_B^{\prime} \sigma_T R_b^{\prime}}
        \label{syn_cool}
    \end{gather}
    with $u_B^{\prime}=B^{\prime 2}/(8\pi)$ being the energy density of the magnetic field. Finally, we compute the synchrotron spectrum arising from the steady state distribution of Eq.~(\ref{steady_state_integral}) using the full expression for the synchrotron emissivity \citep{Ribicki&Lightman}.
    
\begin{figure}
    \centering
        \includegraphics[width=0.49\textwidth]{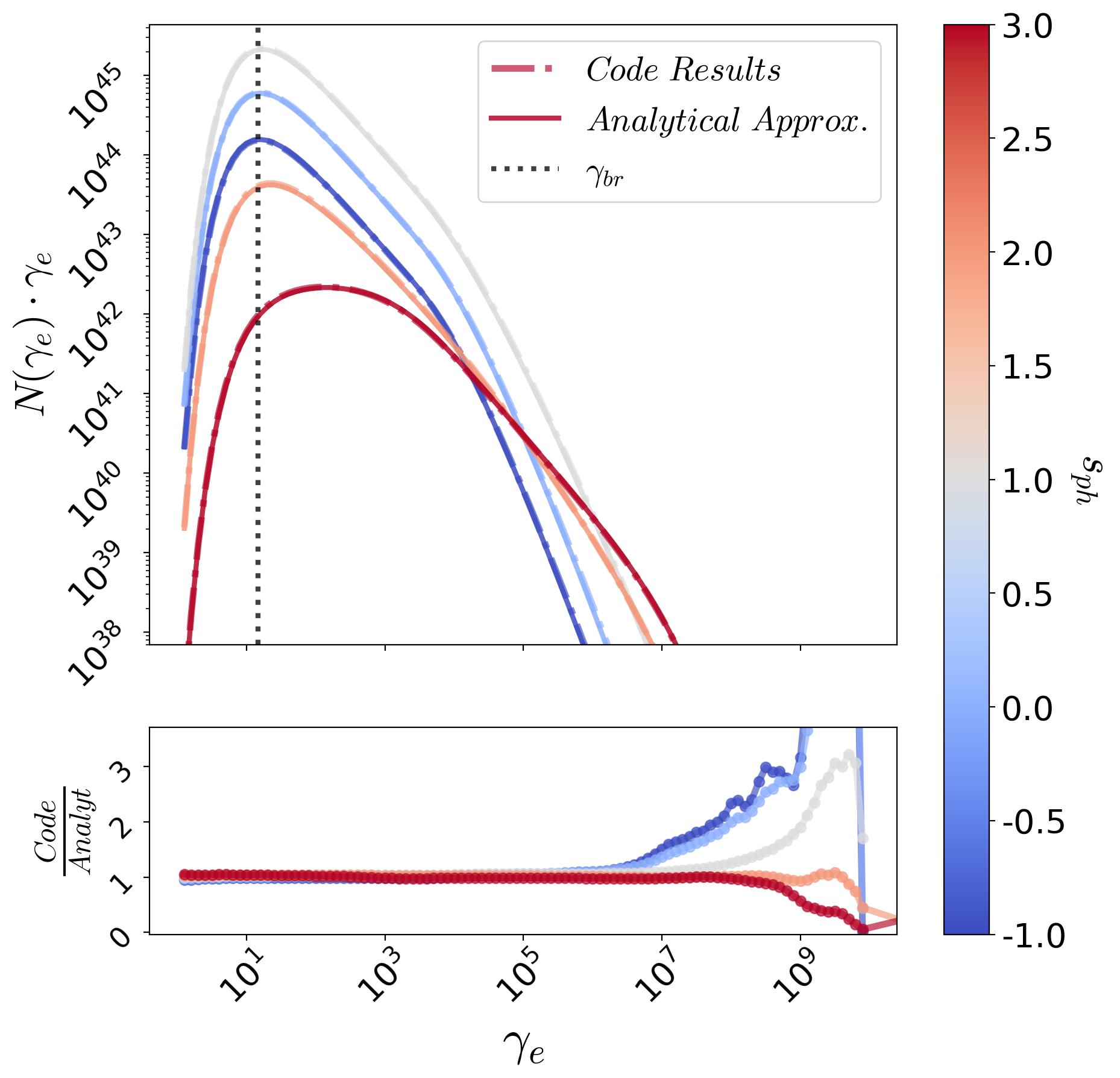}
        \includegraphics[width=0.49\textwidth]{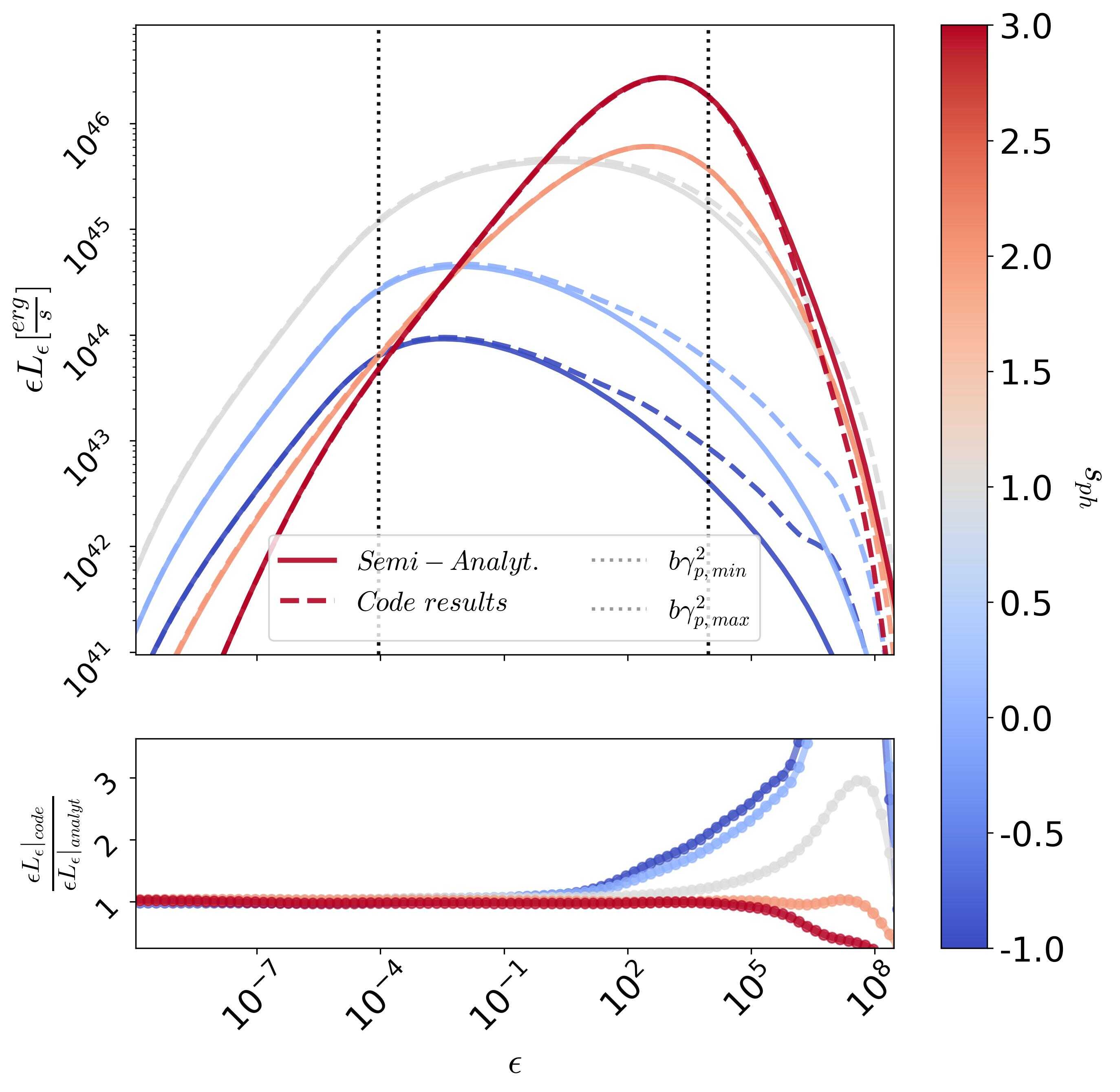}
    \caption{Steady-state number distribution of Bethe-Heitler pairs and their emitted synchrotron spectrum for the same parameters as those used in figure \ref{fig:PLs} and $B'=40$~G. Each curve, starting from the bluer one, is multiplied by a factor of  [1,  4,  16, 1, 4] respectively to avoid overlap. While the cooling Lorentz factor determines the peak of the steady-state pair distribution regardless of $s_{ph}$, the peak energy of the emitted synchrotron energy spectra  depends on the target photon index.}
    \label{fig:PL_syn} 
\end{figure}

The results presented in figure~\ref{fig:PL_syn} are obtained for the same parameters as those used for figure~\ref{fig:PLs} and magnetic field strength $B^\prime=40$ G; additional cases for different magnetic field strengths are presented in Appendix \ref{App:Bs}. The distribution of cooled pairs peaks (in number) at $\gamma_{e} \approx \gamma_{br}$, yet the synchrotron spectrum, $\epsilon L_{\epsilon}$, peaks at an energy that is defined by the proton Lorentz factor. Depending on the slope of the target-photon population, the synchrotron energy spectrum is dominated by the emission of the highest energy protons, i.e. $\propto B^{\prime}\gamma_{p,\max}^2$ (e.g. for $s_{ph}=2$ most target-photons are of low energy providing threshold targets for the highest energy protons of the distribution), or the emission of the lowest energy protons, i.e. $\propto B^{\prime}\gamma_{p,\min}^2$ (e.g. for $s_{ph}=-1$ most of the target-photons have high energies providing threshold targets for the low-energy protons).

\subsubsection{General remarks}
Before closing this section we summarize our main findings about the Bethe-Heitler pair distributions. 

For monoenergetic protons interacting with monoenergetic photons we find that:
\begin{itemize}
    \item most pairs are produced with a Lorentz factor $\sim \epsilon^{-1}$; this was also pointed out by \cite{mastichiadis_spectral_2005}. 
    \item the injection rate reaches a peak value when the protons interact with target photons near the threshold. Thus, the pair population will be dominated by near-threshold interactions.
    \item most of the injected energy is transferred to pairs that have a Lorentz factor equal to that of the proton, i.e. $\gamma_{e}\approx \gamma_p$, for near-threshold interactions. In far-from-threshold interactions roughly equal amounts of energy per logarithmic decade are transferred to pairs.  
\end{itemize}
For PL protons distributions interacting with a PL photon distribution, we find that  (see figure \ref{fig:p_contr}):
\begin{itemize}
    \item the produced pair energy spectrum, $\gamma_e^2 Q(\gamma)$,  is determined by the interactions of the most energetic protons of the distribution, when these dominate in number (i.e., $s_p \leq 1$). More specifically,
    \begin{itemize}
        \item when most of the PL photons are interacting far from threshold with the high-energy protons, then the pair energy distribution is determined by the away-from threshold interactions resulting in an almost flat distribution.
        \item when most of the PL photons are interacting near the threshold with the high-energy protons, then the shape of the pair energy distribution is determined by interactions close to the threshold, and is described by a PL component with a well defined peak at $ \gamma_e \approx \gamma_{p, \max}$.
    \end{itemize}
        \item when the lowest energy protons of the distribution dominate in number ($s_p \geq 1$), the produced pair energy spectrum is:       
        \begin{itemize}
            \item determined by the low-energy proton produced pairs when $s_{p}-s_{ph} \gtrsim 1$,
            \item determined by the high-energy proton produced pairs when $s_{p}-s_{ph} \lesssim 1$, 
            \item almost flat when $s_{p}-s_{ph}\simeq 1$.
        \end{itemize}
\end{itemize}

In astrophysical sources, such as AGN, the target photon fields are typically soft distributions with $s_{ph} \gtrsim 1$, originating by non-thermal mechanisms that cover a wide energy range starting e.g. from a few eV and extending to a few MeV. These extended photon fields provide near-threshold targets for protons of different Lorentz factors. As a result the electromagnetic component originating by Bethe-Heitler synchrotron emission is expected to have a peak value determined by those protons of the distribution that carry most of the energy. 

\section{Bethe-Heitler $\gamma$-ray emission in blazars}\label{sec:application}
Various processes have been invoked to explain the $\gamma$-ray emission from blazars, including electron inverse Compton scattering on synchrotron photons \citep[e.g.][]{1992ApJ...397L...5M,1996ApJ...461..657B, 1997A&A...320...19M} or external photons \citep[e.g.][]{1994ApJ...421..153S, 1992A&A...256L..27D}, proton synchrotron radiation \citep{2000NewA....5..377A, 2001APh....15..121M}, hadronic-initiated cascade emission~\citep{1992A&A...253L..21M} or synchrotron radiation of secondaries from charged pion decays \citep{2015MNRAS.448.2412P, petropoulou_bethe-heitler_2015}. In this next section, we investigate if synchrotron emission of Bethe-Heitler produced pairs in blazar jets can produce a dominant $\gamma$-ray spectral component.

\subsection{Analytical considerations} \label{sec:analyt}
We present analytical approximations for the locations and the luminosities of the two spectral peaks as they appear in a plot of $\log_{10} (\nu\ F(\nu))$ versus $\log_{10}(\nu)$. We also discuss the relative importance of photopion interactions and photon-photon pair production. We assume that all particle populations are effectively monoenergetic, namely they have distributions that are energetically dominated by particles of a characteristic energy. Moreover, a particle with a given Lorentz factor is assumed to produce photons at the characteristic synchrotron frequency \citep{Ribicki&Lightman}.

 \begin{table}
    \centering
    \scalebox{1.}{
    \begin{tabular}{c|c|c|c}
    \hline
        {}& LSP & ISP & HSP \\ \hline
        $\nu_{S}$ (Hz) & $[10^{12}, 10^{14}]$ & $[10^{14}, 3 \cdot 10^{15}]$ & $ [3 \cdot 10^{15}, 10^{17}]$ \\ 
        $\nu_{\gamma}$ (Hz) & $[10^{20}, 10^{22}]$ & $[10^{22}, 10^{25}]$ & $[10^{25}, 10^{28}]$\\
        $\epsilon_{S}$ & $ [8 \cdot 10^{-9}, 8 \cdot 10^{-7}]$ & $[8 \cdot 10^{-7}, 3 \cdot 10^{-5}]$ & $[3 \cdot 10^{-5}, 8 \cdot 10^{-4}]$ \\ 
        $\epsilon_{\gamma}$ & $[8 \cdot 10^{-1}, 8 \cdot 10^{1}]$ & $ [8 \cdot 10^{1}, 8 \cdot 10^{4}]$ & $[8 \cdot 10^{4}, 8 \cdot 10^{7}]$ \\ 
        \hline
    \end{tabular}
    }
\caption{Typical values of the peak frequencies (and corresponding photon energies in units of $m_e c^2$) of the two SED humps for LSPs, ISPs and HSPs \citep{abdo_spectral_2010}.}
\label{table:source_freqs}
\end{table}

\subsubsection{Characteristic photon energies} 
The low-energy peak of the SED appears at an observed frequency $\nu_S$ (see Table~\ref{table:source_freqs}), which is related to the characteristic Lorentz factor, 
$\gamma_{e}^{\prime}$, of accelerated electrons in the jet (henceforth, primaries) as
    \begin{gather}
        \frac{3}{2}\delta b \gamma_{e}^{\prime 2} (1+z)^{-1} = \frac{h \nu_S}{m_e c^2} \equiv \epsilon_S
        \label{e_syn}
    \end{gather} 
    where $b \equiv B^{\prime} \left( \frac{m_e^2 c^3}{ \hbar e} \right)^{-1}$ is a dimensionless measure of the co-moving magnetic field strength $B'$. Similarly, the high-energy peak of the SED, which appears at a frequency $\nu_{\gamma}$ (see Table~\ref{table:source_freqs}), is created by the synchrotron radiation of Bethe-Heitler pairs with characteristic Lorentz factor $\gamma'_{BH}$,
    \begin{gather}
        \frac{3}{2} \delta b \gamma_{BH}^{\prime 2} (1+z)^{-1} = \frac{h \nu_\gamma}{m_e c^2} \equiv \epsilon_{\gamma}.
    \end{gather}
   In section \ref{sec:BH-spec} we showed that the energy distribution of pairs created via Bethe-Heitler pair production peaks at a Lorentz factor almost equal to the Lorentz factor of the parent proton, i.e. $\gamma'_{BH} \approx \gamma'_p$. These are the pairs that will contribute to the peak of the emitted synchrotron spectrum (see section~\ref{sec:features}). Therefore, the peak of the high-energy component will appear at 
    \begin{gather}
       \epsilon_{\gamma} \sim \frac{3}{2} \delta b \gamma_p^{\prime 2} (1+z)^{-1} \label{BH_syn}
    \end{gather}
   The ratio of the SED peak energies is then simply given by 
    \begin{gather}
    \frac{\epsilon_\gamma}{\epsilon_S} \sim \left(\frac{\gamma'_p}{\gamma'_e}\right)^2 \label{pk_pos_comp},
    \end{gather}
    similarly to the proton-synchrotron scenario but without the multiplying factor $m_e/m_p$ \citep{2001APh....15..121M, 2016ApJ...825L..11P}. Moreover, the primary electron and proton Lorentz factors are not always independent model parameters, like in the proton synchrotron model. More specifically, assuming that protons with Lorentz factor $\gamma'_p$ interact with synchrotron photons of energy $\epsilon'_S$ at the threshold for Bethe-Heitler pair production, then the following condition has to be met,
    \begin{gather}
         \delta^{-1} \epsilon_S \gamma_{p}^{\prime} \sim 2 \label{BH_thres_obs} \rightarrow 
         \frac{3}{2} b \gamma_{e}^{\prime 2} \gamma_{p}^{\prime} \sim 2
    \end{gather} 

\begin{figure}
    \centering
    \includegraphics[width=0.6\textwidth]{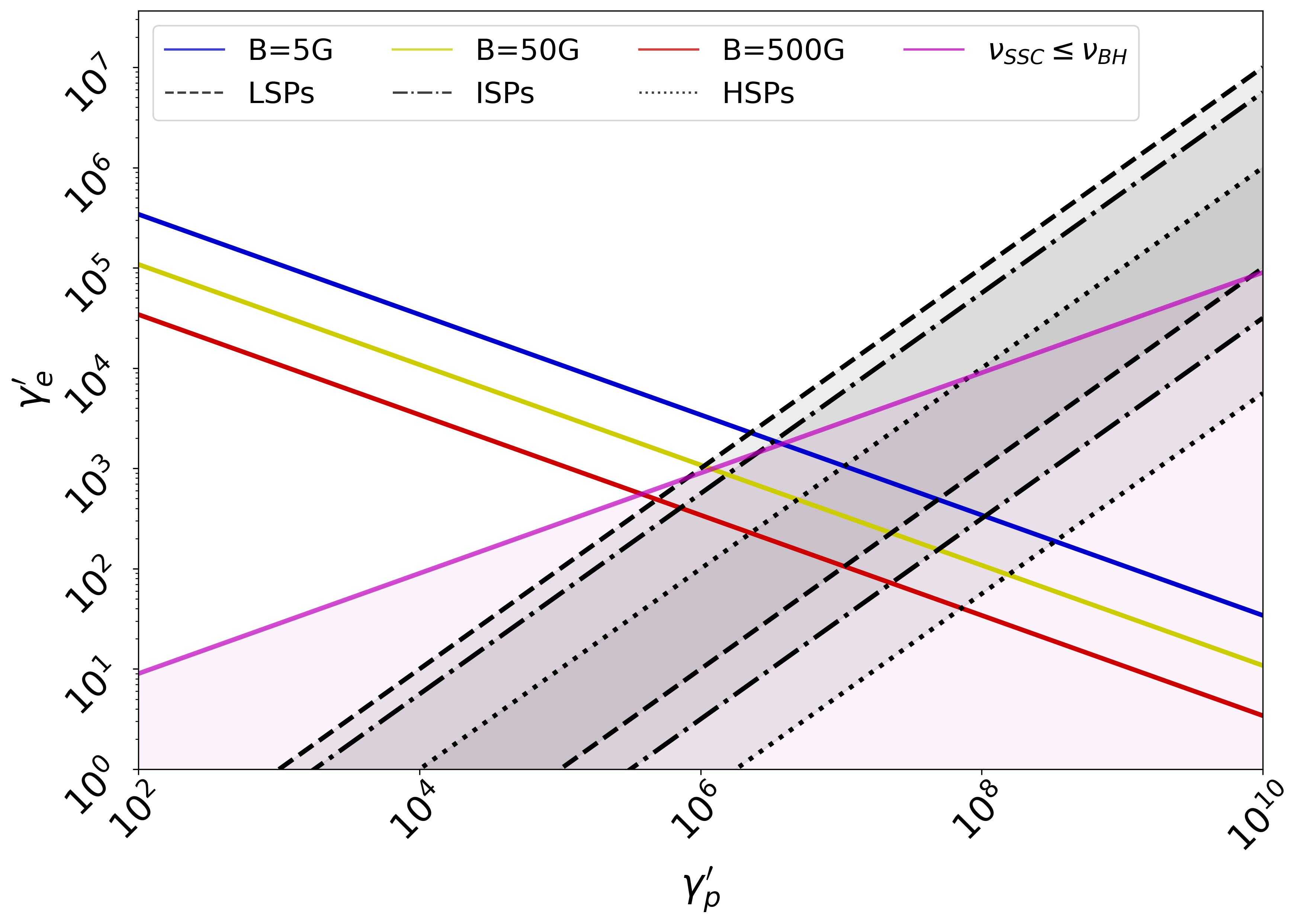}
    \caption{Analytical constraints for $\gamma$-ray production by Bethe-Heitler pairs following from Eq.~(\ref{pk_pos_comp}) (shaded regions) and Eq.~(\ref{BH_thres_obs}) (colored lines). Protons interact at the threshold with electron synchrotron photons, and produce pairs that emit synchrotron radiation within the observed $\gamma$-ray limits (see Table~\ref{table:source_freqs}) for $\gamma^{\prime}_e$ and $\gamma^{\prime}_p$ values selected from the intersection of colored lines with the shaded region for each blazar subclass. The peak energy of the SSC component is lower than the peak energy of the Bethe-Heitler synchrotron component for parameter values drawn from the magenta shaded region. There are combinations of $\gamma_p^{\prime}$ and $\gamma_e^{\prime}$ leading to Bethe-Heitler synchrotron spectra peaking in $\gamma$ rays.}
    \label{fig:ge_gp}
\end{figure}

Using Eqs.~(\ref{pk_pos_comp}) and (\ref{BH_thres_obs}) we create a parameter space for  $\gamma^{\prime}_e$ and $\gamma^{\prime}_p$ for each source class and three indicative magnetic field strengths -- see figure \ref{fig:ge_gp}. Colored lines indicate the locus of points satisfying the energy threshold condition of Eq.~(\ref{BH_thres_obs}), while the shaded-grey region corresponds to Eq.~(\ref{pk_pos_comp}) for a range of $\epsilon_S$ and $\epsilon_\gamma$ values that are representative for each blazar class. For a specific magnetic field strength and source class, there is small range of  $\gamma^{\prime}_e$ and $\gamma^{\prime}_p$  values that satisfy both conditions simultaneously.

Moreover, by combining Eqs. (\ref{BH_syn}) and (\ref{BH_thres_obs}) we can express the dimensionless energy of the high-energy hump as
  \begin{gather}
      \epsilon_{\gamma}=1.7 \cdot 10^3 \, B^\prime_1 \gamma^{\prime 3}_{p,8} \epsilon_{S,-8}  (1+z)^{-1}      \end{gather}
  where we introduced the notation $f_X \equiv f/10^X$ (in cgs units, unless stated otherwise).   Interestingly, the peak energy of the proton synchrotron spectrum in this scenario is a constant fraction of $\epsilon_\gamma$, namely
    \begin{equation}
    \epsilon_{S,p} = \frac{3}{2}\delta b \gamma_p^{\prime 2}(1+z)^{-1} \frac{m_e}{m_p} = \frac{m_e}{m_p}\epsilon_\gamma.
    \label{psyn}
    \end{equation}
The synchrotron self-Compton (SSC) emission produced by primary electrons peaks at a photon energy
\begin{gather}
    \epsilon_{SSC} = \frac{4}{3} \gamma_e^{\prime 2} \epsilon_S = 2 \delta b \gamma_{e}^{\prime 4} (1+z)^{-1} \label{e_ics}
\end{gather}
where we assumed that the scatterings take place in the Thomson regime. The SSC component peaks at lower energies than the proton synchrotron spectrum, see Eq. (\ref{psyn}), when 
\begin{gather}
    \gamma_e^{\prime} \leq \left( \frac{3 m_e}{4 m_p} \right)^{\frac{1}{4}} \gamma_p^{\prime \frac{1}{2}} \lesssim 1.4 \cdot 10^3 \, \gamma^{\prime \frac{1}{2}}_{p, 8}.
\end{gather}
Additionally, the SSC component peaks below the Bethe-Heitler spectrum, if
\begin{gather}
    \gamma_e^{\prime} \leq \left( \frac{3}{4} \right)^{\frac{1}{4}} \gamma_p^{\prime \frac{1}{2}} \lesssim 9 \cdot 10^3 \, \gamma^{\prime \frac{1}{2}}_{p, 8}.
\end{gather}
This condition is satisfied for most parameters that lead to a Bethe-Heitler component peaking in $\gamma$-rays, as shown in figure~\ref{fig:ge_gp}.
  
An isotropic external thermal photon field, if present, can also provide Bethe-Heitler threshold targets for relativistic protons in the jet. We generally describe the thermal emission with a grey body of temperature $T$ in the AGN rest frame (e.g. infrared emission from dusty torus). We also assume that the jet emitting region (blob) is moving within the external radiation field, which is considered isotropic in the AGN rest frame. Hence, the temperature and energy density of the external photon field, as measured in the blob rest frame, are relativistically boosted as $T^\prime = \Gamma T $ and $u^\prime_{ext} = \Gamma^2 u_{ext}$ \citep{mandau}. The threshold condition then reads $ \Gamma \epsilon_{ext} \gamma_{p}^{\prime} \sim 2$ or
        \begin{gather}
            \Gamma \frac{3 k_B T}{m_e c^2} \gamma_{p}^{\prime} \sim 2
            \label{BH_thres_therm},
        \end{gather}     
leaving more freedom in the determination of $\gamma_e^{\prime}$ and $\gamma_p^{\prime}$. If we, now, combine Eqs. (\ref{e_syn}) and (\ref{BH_thres_therm}), we can express the temperature of the external thermal photon field as:
\begin{eqnarray}
    T  = 1350~{\rm K} \, (\delta/\Gamma) \, B^\prime_1 \gamma_{e,2}^{\prime 2} \gamma^{\prime -1}_{p,6}  \epsilon_{S,-8}^{-1} (1+z)^{-1}, 
\end{eqnarray} 
where we considered a typical synchrotron energy for LSP blazars.

\subsubsection{Peak photon luminosities}
We continue with an estimation of the peak luminosities of the SED humps. Under the assumption of monoenergetic particle population and the $\delta$-function approximation for the synchrotron emissivity, the low-energy peak luminosity can be written as
    \begin{gather}
        L_{e,S}^{\prime} \approx \frac{4}{3} c \sigma_T u_B^{\prime} \gamma_{e}^{\prime 2} N_e  \min[1, \frac{t_{\rm syn}(\gamma_e)c}{R^\prime_b}]
        \label{esyn_lum}
    \end{gather}
where $u^{\prime}_B=B^{\prime 2}/(8\pi)$, $N_e$ is the number of primary electrons with Lorentz factor $\gamma_e^{\prime}$ in the blob, and the factor in square brackets accounts for the cooling of monoenergetic electrons. The number $N_e$  can be related to the electron energy density $u^{\prime}_e$ as
 \begin{gather}
       N_e = \frac{4 \pi}{3} R_b^{\prime 3} u_e^{\prime} (m_e \gamma^{\prime}_{e} c^2)^{-1}.
       \label{e_number}
    \end{gather}
At this point it is useful to introduce the electron compactness  
  \begin{gather}
        \ell_e=\frac{u^\prime_e \sigma_T R_b^{\prime}}{3 m_e c^2}, \label{e_energ_dens}
    \end{gather}
    which is a dimensionless measure of the total energy density and  an input parameter of the numerical code \code, results of which will be presented in section~\ref{sec:numres}. By combining Eqs.~(\ref{esyn_lum}), (\ref{e_number}) and (\ref{e_energ_dens}), we obtain the low-energy peak luminosity (in the observer's frame)
    \begin{gather}
         L_{e, S} \approx \frac{16 \pi}{3} \delta^4 c R_b^{\prime 2} u_B^{\prime} \gamma_e^{\prime} \ell_e \min[1, \frac{t_{\rm syn}(\gamma_e)c}{R^\prime_b}].
        \label{esyn_tot_lum_obs}
    \end{gather}
    In order to compute the peak synchrotron luminosity emitted by Bethe-Heitler pairs, we will assume that the latter radiate all of their energy via synchrotron, namely the injected power in Bethe-Heitler pairs equals the power emitted in synchrotron photons $L_{BH,S}^{\prime}=L_{BH,e}^{\prime}$. The  total energy transferred to Bethe-Heitler pairs per unit time is a fraction of the relativistic proton luminosity $L^{\prime}_p$,
        \begin{gather}
            L_{BH,e}^{\prime}= f_{BH} L_p^{\prime} \approx  \hat{\sigma}_{BH}  R_b^{\prime} n_{t}^{\prime}  L_p^{\prime}
            \label{L_BH}
        \end{gather}
        where $f_{BH}\le 1$ is the Bethe-Heitler pair production efficiency, $\hat{\sigma}_{BH} \simeq 8\times10^{-31}$~cm$^2$ is the maximum effective cross section accounting for the  inelasticity of the interaction \citep{chodorowski_1992}, and $n_{t}^{\prime}$ is the target photon number density. If $\epsilon_{t}^{\prime}$ is the characteristic target photon energy, then the number density can be written as
         \begin{gather}
            n_{t}^{\prime}=\frac{u_{t}^{\prime}}{ \epsilon_{t}^{\prime} m_ec^2} = \frac{3 \ell_{t}}{\epsilon_{t}^{\prime} \sigma_T R_b^{\prime}}
            \label{n_t}
        \end{gather}
        where in the last expression we introduced the photon compactness $\ell_t$, defined in a similar manner as in Eq.~(\ref{e_energ_dens}).  We can also express $L^{\prime}_p$ in terms of the proton compactness $\ell_p$ as 
        \begin{gather}
            L_p^{\prime}=\frac{4 \pi R_b^{\prime} \ell_p m_p c^3}{\sigma_T} \cdot
        \end{gather}
        Combining the equation above with Eqs.~(\ref{L_BH}) and (\ref{n_t}) we obtain the final expression for the peak luminosity of the high-energy component of the SED (in the observer's frame)  
        \begin{gather}
            L_{BH, S} \approx \frac{ 12 \pi \hat{\sigma}_{BH} m_p c^3}{\sigma_T^2} \delta^4 R_b^{\prime}  \ell_p \ell_{t} \epsilon_{t}^{\prime -1} .
            \label{L_BH_full_obs}
        \end{gather}

         In the special case where the target photons for Bethe-Heitler pair production are the primary electron synchrotron photons, which may be relevant for HSP blazars, then $\epsilon_t^\prime = \epsilon_S (1+z)/\delta$ and the photon compactness is given by
        \begin{gather}
            \ell_{t}= \frac{u^{\prime}_{e,S} \sigma_T R^\prime_b}{3 m_e c^2} = \frac{4 \sigma_T R_b^{\prime} u_B^{\prime}}{3 m_e c^2} \gamma^{\prime}_{e} \ell_e  \min[1, \frac{t_{\rm syn}(\gamma_e)c}{R^\prime_b}]
            \label{l_esyn}
        \end{gather}
        where we also used Eq.~(\ref{esyn_lum}). The ratio of the peak luminosities, which is the equivalent of the Compton ratio in purely leptonic models, can then be written as
        \begin{gather}
        A^{(syn)}_{\gamma S} \equiv \frac{L_{BH, S}}{L_{e, S}} = 3 \frac{\hat{\sigma}_{BH}}{ \sigma_T} \frac{m_p}{m_e} \frac{\delta}{(1+z)\epsilon_S} \ell_p. 
        \label{BH_dominance_ratio_fin_syn}
        \end{gather}  
        If the target photons were provided by an external source of radiation of energy density $u^\prime_{t} \simeq \Gamma^2 u_{ext}$ and photon energy $\epsilon^\prime_t = \Gamma \epsilon_{ext}$, then the ratio of the peak luminosities would read
        \begin{gather}
        A^{(ext)}_{\gamma S} \equiv \frac{L_{BH, S}}{L_{e, S}} = 3 \frac{\hat{\sigma}_{BH}}{ \sigma_T} \frac{\ell_p m_p}{\ell_e m_e} \frac{u_{ext}}{u^\prime_B}\frac{\Gamma} {\epsilon_{ext}}\left(\frac{3 b \delta}{2 \epsilon_S (1+z)}\right)^{1/2}.
        \label{BH_dominance_ratio_fin_ext}
        \end{gather} 
   
   In the proposed scenario the peak luminosity ratio has a different dependence on $\epsilon_S$ depending on the dominant target photon field for Bethe-Heitler pair production: synchrotron jet emission or external radiation. By varying model parameters, like $\ell_p$ and $\ell_e$, it is possible to obtain a wide range of values for the peak luminosity ratio. We will compare the predictions of the fully numerical models against the distribution of observed values in Sec.~\ref{sec:numres}.

\subsubsection{Photopion and photon-photon pair production processes} \label{sec:pg-gg}

Relativistic protons may pion produce on low-energy photons, leading eventually to the production of energetic $\gamma$-ray photons, pairs, and neutrinos. It is therefore interesting to compare the expected luminosities of synchrotron-emitting pairs and neutrinos with the $\gamma$-ray luminosity of Bethe-Heitler pairs. Moreover, we discuss the role of photon-photon ($\gamma \gamma$) pair production, which is relevant for the attenuation of energetic photons from pion decay.

For simplicity, we assume that photopion interactions take place at the $\Delta^{+}$ resonance, where $\pi^{0}$ and $\pi^{+}$ pions are produced at a ratio $1:1$ (i.e. equally probable). Neutral pions will decay into two photons, while charged pions will produce positrons/electrons and neutrinos, after a series of reactions. We consider that the energy lost by the reactants is equally distributed to the products, resulting in 
$\gamma$-ray photons from $\pi^0$ decay with energy,

\begin{gather}
    \epsilon^\prime_{\gamma, \pi^0} \approx \frac{1}{2} \gamma_{\pi^0} \frac{m_{\pi^0}}{m_e} \approx \frac{1}{2} \kappa_{p \gamma} \gamma^\prime_{p} \frac{m_{p}}{m_e} \approx 0.1 \gamma^\prime_p \frac{m_p}{m_e} \approx 1.8\cdot 10^9 \, \left(\frac{\gamma^\prime_p}{10^7} \right)
\end{gather}
where we used that the inelasticity of p--$\gamma$ interactions is $\kappa_{p \gamma} \simeq 0.2$ close to the $\Delta^{+}$ resonance.  

Similarly, the neutrinos produced in the $\pi^{+}$ decay chain will acquire $3/4$ of the pion energy, leaving $1/4$ of the latter for the positron created. The energy injected per unit time in $\gamma$-ray photons and neutrinos of all flavors can be then estimated as: 
\begin{gather}
    \label{Lgamma}
    L_{\gamma}= \frac{4}{3}L_{\nu}= L_{\pi^0} = \frac{1}{2} \delta^4 \kappa_{p\gamma} \sigma_{p\gamma} R_b^{\prime} n'_{t, p\gamma} L_p^{\prime} 
\end{gather}
where $R_b^{\prime}$ is the blob radius, $L_p^{\prime}$ the proton population luminosity, $\kappa_{p\gamma} \sigma_{p\gamma}  = \hat{\sigma}_{p\gamma}\simeq 7 \cdot 10^{-29}~\rm cm^2$ is the product of the inelasticity and the cross section of the photopion interaction \citep{2009HEA_BH} and $n'_{t, p\gamma}$ is the number density of the relevant target photons.

We compare next the luminosities of secondaries from photopion interactions with the luminosity of synchrotron photons emitted by Bethe-Heitler pairs, see Eq.~(\ref{L_BH})
\begin{gather}
    \frac{4}{3} \frac{L_{\gamma}}{L_{BH, S}} =\frac{L_{\nu}}{L_{BH, S}}= \frac{3}{8} \frac{\hat{\sigma}_{p \gamma}}{\hat{\sigma}_{BH}}\frac{n'_{t, p\gamma}}{n'_{t, BH}}  \approx 33 \ \frac{L_{t, p\gamma}}{L_{t, BH}} \frac{\epsilon_{t, BH}}{\epsilon_{t, p\gamma}}
    \label{photopion_prod}
\end{gather}
In the case of monoenergetic protons, as assumed here for simplicity, the luminosity ratio depends only on the cross sections and number densities of the target photons of the interaction processes. In AGN jets where photons are produced by non thermal mechanisms, the differential number density of photons can be usually described by a power law $n^\prime_t\propto \epsilon^{\prime -s_{ph}+1}$. Therefore, the number density of target photons for each interaction will generally depend on the photon index $s_{ph}$. In the special case of $s_{ph}=1$, where the photons are equally distributed in number across the energy spectrum, we expect $n^\prime_{t, p\gamma}/n^\prime_{t, BH} = 1$ and $L_\nu/L_{BH,S}\approx 30$. For $s_{ph}\neq 1$ the target number ratio would be $n^\prime_{t,  p\gamma}/n^\prime_{t, BH} \approx 10^{2(-s_{ph}+1)}$,  since $\epsilon_{p\gamma, thr}/\epsilon_{BH, thr} \approx 140$. For example, if $s_{ph}=2$, then $n^\prime_{t,  p\gamma}/n^\prime_{t, BH}\approx0.01$ resulting in $L_\nu/L_{BH,S} \approx 0.1$. Summarizing, for photon indices in the range 1 to 2, we could expect $0.1 \lesssim L_\nu/L_{BH,S}\lesssim 10$. 

\begin{figure}[h!]
    \centering
    \includegraphics[width=0.6\textwidth]{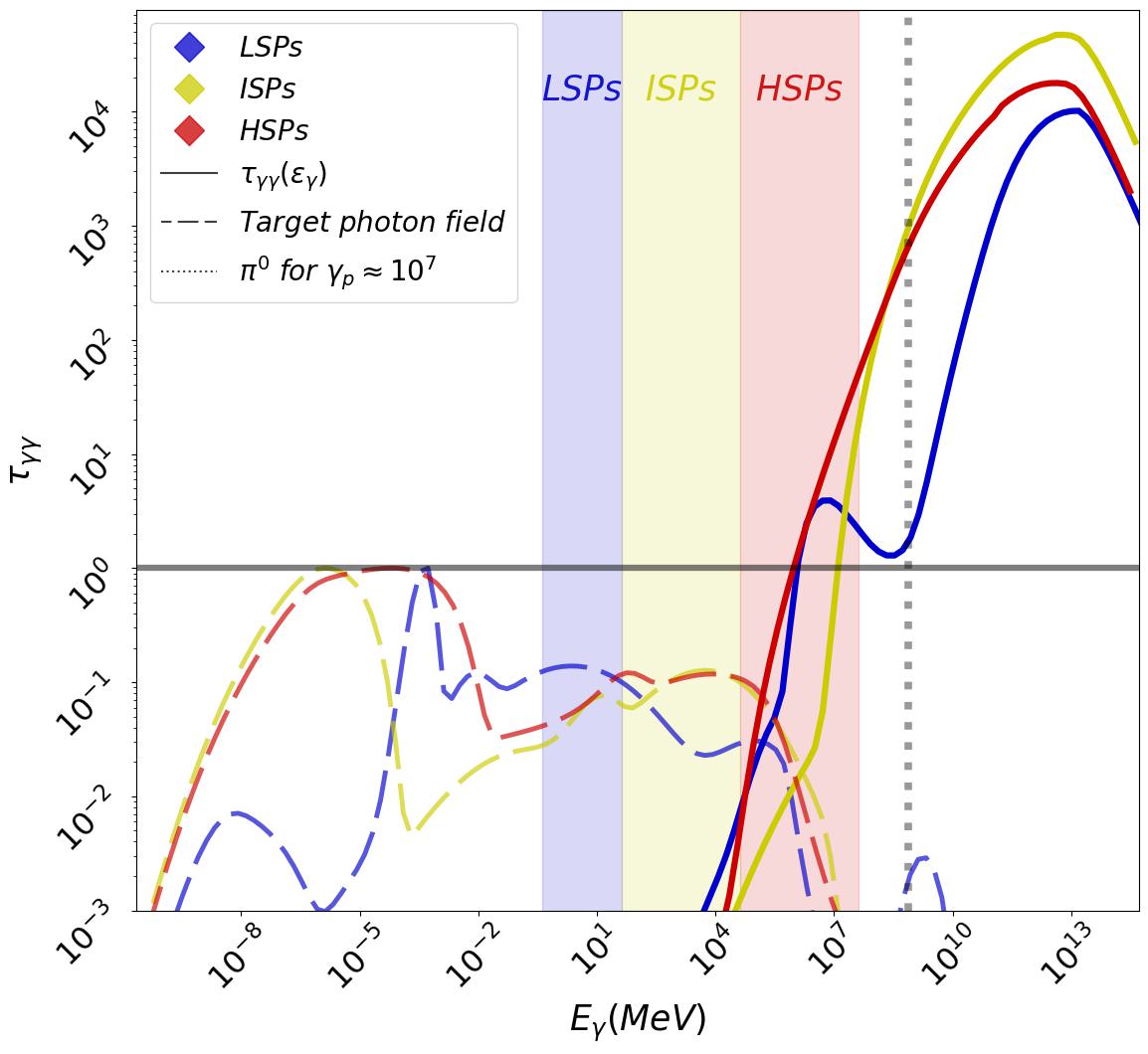}
    \caption{Graphical representation of the $\gamma \gamma$ optical depth as a function of the high-energy photon energy for different blazar classes. Colored areas indicate the typical range of observed high-energy peak energies. For each class we integrate Eq.~(\ref{tau_gg_mono}) on an indicative blazar spectrum. We use the photon spectra of figure \ref{fig:models_var_15} for $B^\prime=50$~G. The latter are also shown in the figure with dashed lines, in the blob frame and are normalized to their peak values. We also take into account the Doppler factor and the redshift of each source, as $\delta=55, 45, 27$ and $z=0.5, 0.3, 0.1$ for the LSP, the ISP and the HSP cases respectively. In all cases, the emitting region is opaque to $\gamma \gamma$ pair production for photons of energy above 10 TeV.}
    \label{fig:tau_gg}
\end{figure}

The energetic photons produced by the decay of neutral pions are expected to be attenuated by low-energy photons in the source. We can estimate the opacity of the source to $\gamma \gamma$ pair production of energetic photons with observed energy as $\epsilon_\gamma = \delta \epsilon^\prime_\gamma/(1+z)$ as 
\begin{gather}
  \tau_{\gamma \gamma}(\epsilon_\gamma ) \approx 0.625 \sigma_T R^\prime_b \int_{2 \delta/[\epsilon_\gamma (1+z)]} \d \epsilon^\prime n^\prime(\epsilon^\prime) \frac{(\epsilon^\prime_{\gamma} \epsilon^\prime_{0})^{2}-1}{(\epsilon^\prime_{\gamma} \epsilon^\prime_{0})^3} \ln(\epsilon^\prime_{\gamma} \epsilon^\prime_{0}) 
    \label{tau_gg_mono}
\end{gather}
where we used the approximate $\gamma \gamma $ cross section from Ref.~\cite{coppi_reaction_1990}. In figure~\ref{fig:tau_gg} we present indicative results for LSP, ISP, and HSP blazars. The optical depth for each case is integrated over a broadband target photon distribution\footnote{The displayed SEDs were numerically computed and will be presented later in figure \ref{fig:models_var_14} of section \ref{sec:numres}.}. Figure \ref{fig:tau_gg} shows the transition from an optically thin source ($\tau_{\gamma \gamma} \leq 1$) to an optically thick source ($\tau_{\gamma \gamma} \geq 1$). When the latter transition happens we can observe that the luminosity of the electromagnetic spectrum drops significantly due to the internal photon-photon annihilation that becomes important. For the examples considered, the source is optically thick to energetic $\gamma$-rays from pion decay.

\subsection{Numerical approach}\label{sec:numerical}
According to the analytical results presented in the previous section, it is possible to find parameter values that can lead to a high-energy SED component powered by synchrotron radiation of Bethe-Heitler pairs. This analysis is however simplified, because it considers effectively monoenergetic particle distributions. Moreover, the effects of other processes, like photomeson interactions and photon-photon pair production were evaluated in an approximate way.

In order to examine in more detail the hypothesis that Bethe-Heitler pair production can produce the $\gamma$-ray component in blazar spectra, we will use the leptohadronic code \code  \ \citep{mastichiadis1995synchrotron, refId0}. The code numerically solves the kinetic equations that describe the evolution of stable particle distributions that are injected or created inside a spherical magnetized source. More specifically, the code solves a system of coupled partial differential equations for protons, neutrons, electron-positron pairs, photons, and neutrinos. Unstable particles, such as kaons, pions, and muons, are assumed to decay instantly into secondary lighter particles, hence only their injection rates are computed. All stable particles are also allowed to escape the source on  a timescale equal to the light crossing time $R^{\prime}_b/c$. The physical processes that are included are: synchrotron emission from charged particles,  synchrotron self-absorption, electron inverse Compton scattering, photon-photon pair production, and proton-photon interactions that may lead to pion production (photomeson) or  pair production (Bethe-Heitler) -- for the numerical implementation, see \cite{refId0}.

The code takes as input the parameters describing the emitting region, namely radius and magnetic field strength,
and the characteristics of the power-law distributions of electrons and protons injected in the blob (i.e. minimum and maximum Lorentz factors, slope and compactness). Given these input parameters, we evolve the system long enough to reach a steady state (i.e. $\sim 10\ R'_b/c$ for most cases). Based on the steady-state photon number density, we compute the escaping photon flux in the observer's frame after performing the appropriate Doppler boosting \citep{Ghisellini_HEA}.

We numerically investigate the parameter space of the model in search for parameters that produce a high-energy SED component dominated by synchrotron emission of Bethe-Heitler pairs for different blazar classes (LSPs, ISPS, and HSPs). As indicative values for the peak photon energy of each subclass we use those listed in Table~\ref{table:source_freqs}.

We start by selecting 3 indicative values for the blob radius, $R^{\prime}_b \in \{10^{14}, 10^{15}, 10^{16}\}~\rm cm$. For each value we then use 3 magnetic field strengths, $B' \in \{0.5, 5, 50, 500\}~\rm G$, in order to cover a wide range of physically plausible values. Then, for each pair of $(R'_b, B')$ values and for each spectral subclass with fixed $(\epsilon_S, \epsilon_\gamma)$, we
determine $\gamma'_e$ and $\gamma'_p$ using  the analytical expressions of the previous section as a guide (see e.g. Eqs. \ref{BH_syn}, \ref{BH_thres_obs} and \ref{BH_thres_therm} and figure~\ref{fig:ge_gp}). More precisely, the distributions of injected particles are modeled as power laws that energetically peak close to the analytically predicted values. We then vary the normalization of the particle distributions, or equivalently $\ell_e$ and  $\ell_p$, to produce a range of peak luminosities that fall within the range of observed values.

For LSPs we also consider the presence of an external thermal photon field with a grey-body spectrum. The temperature is chosen so that the thermal photons from the peak of the distribution fulfil the Bethe-Heitler threshold condition in the blob rest frame, see Eq. (\ref{BH_thres_therm}). The compactness (or comoving density) of the external photon field is adjusted so that it leads to values of $A_{\gamma S}^{(ext)} > 1$, as usually observed in LSP spectra \citep{abdo_spectral_2010}. The plausibility of the chosen values will be discussed in section \ref{sec:ext_ph_fields}.

\subsection{Numerical results}\label{sec:numres}
\begin{figure}[h!]
    \centering
    \includegraphics[width=0.75\textwidth]{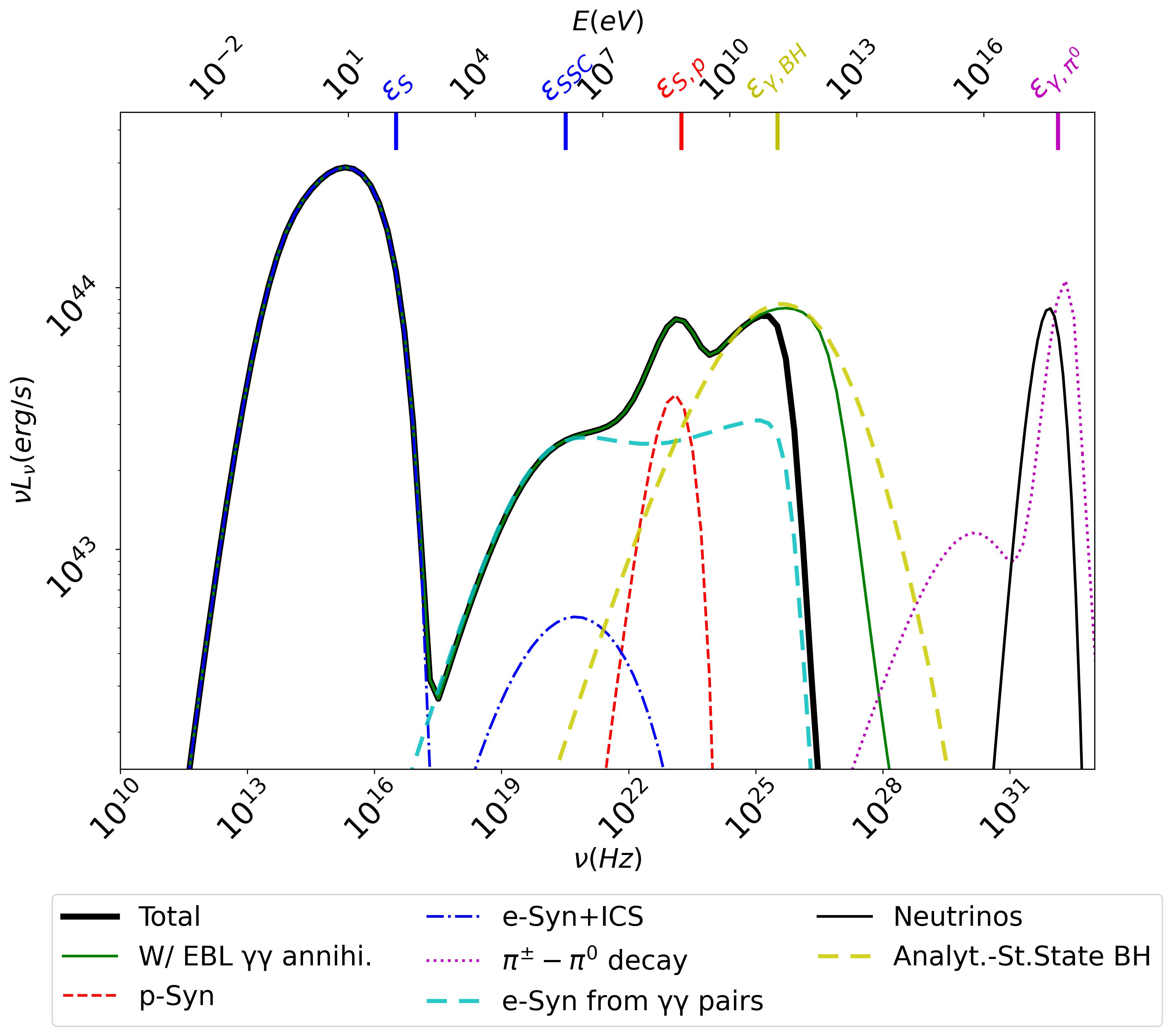}
    \caption{Decomposition of an indicative spectral energy distribution (SED) from an ISP-like blazar into various spectral components indicated by different types of lines. The all-flavor neutrino spectrum is also plotted (black solid line). Colored markers on the upper horizontal axis indicate analytical estimates of characteristic energies presented in section \ref{sec:analyt}. For the parameter values used, see Table~\ref{tab:model_params} under ISPs with $R^{\prime}_b=10^{14}$~cm and $B'=50$~G.} 
    \label{fig:spectr_components}
\end{figure}

In figure~\ref{fig:spectr_components} we present a ISP-like blazar spectrum with different spectral components highlighted (the parameters used are the same as those listed in Table \ref{tab:model_params} for the ISP case with $B^{\prime}=50$~G and $R_b^{\prime}=10^{14}$~cm). Starting from low energies, we identify first the synchrotron spectrum of primary electrons, Their distribution is steeper than the injected power law due to synchrotron cooling. The analytical expectation for the peak of the synchrotron spectrum in this case would be $\propto B\gamma_{e,\max}^2$, but this overestimates the position of the numerical spectrum by a factor of $\sim 7$. The reason for the discrepancy is that the cooled electron distribution falls off faster than the asymptotic power law with $-s_e-1$ due to the narrow dynamic range of the injected power law. As a result, the electrons that contribute most to the peak of the radiated synchrotron spectrum have $\gamma \sim \gamma_{e,\max}/2.5$. The primary synchrotron component is followed by a subdominant, in this example, synchrotron self-Compton bump (dash-dotted blue line); its peak energy agrees with the estimate of Eq.~(\ref{e_ics}). At slightly higher frequencies, we observe a distinct bump due to proton synchrotron radiation (dashed red curve), which peaks at energy given by Eq.~(\ref{psyn}). 
At $\sim 10-100$~GeV energies we observe the synchrotron emission of Bethe-Heitler produced pairs (thick dashed yellow line) when $\gamma \gamma$ absorption is neglected; the spectrum is computed using the analytical function for the pair injection, see Eq.~(\ref{Q_inj_approx}).  A two-component spectrum (dotted magenta line) emerges at the highest energies, which consists of synchrotron emission of pairs from charged pion decays and photon emission from neutral pion decay. These energetic photons are attenuated, however, inside the source (see e.g. yellow lines in figure~\ref{fig:tau_gg}), producing electron-positron pairs that, for the specific parameters, emit synchrotron photons across a wide range of energies between 10~keV and 100 GeV. The synchrotron spectrum of the $\gamma \gamma$-produced pairs is shown as a separate component marked with cyan color (with attenuation included). For the selected parameters is less luminous than the Bethe-Heitler and proton synchrotron components, but emerges as a separate component between 10 keV and 100~keV. Finally, the black line shows the spectrum after accounting for attenuation due to the extragalactic background light (EBL). For this example, we considered a source at redshift $z=0.3$ and used the EBL model of Franceschini~\cite{franceschini_extragalactic_2008}. Due to the combined attenuation, only part of the Bethe-Heitler component would be visible. The magenta line in the high-energy part of the spectrum describes the energy distribution of neutrinos and anti-neutrinos of all flavors. The peak neutrino luminosity is comparable to the peak luminosity of the Bethe-Heitler component before attenuation, a result that agrees with Eq.~(\ref{photopion_prod}). 

\subsubsection{Blazar SEDs}
\begin{figure*}
    \centering
    \includegraphics[width=0.99\textwidth]{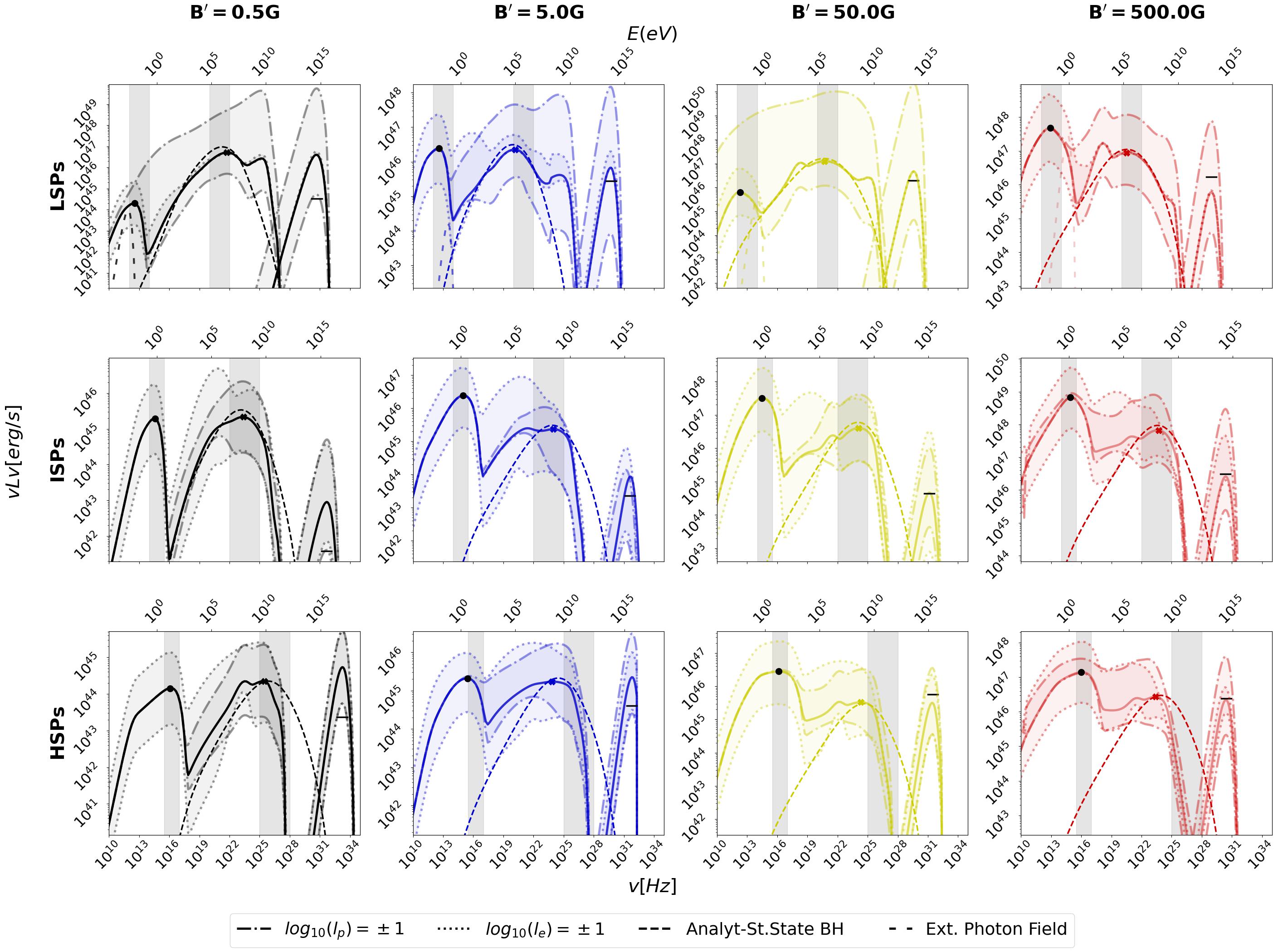}
    \caption{Theoretical spectral energy distributions (SEDs) for LSP-, ISP-, and HSP-like blazars (thick solid lines) computed for an emitting region with radius $R^{\prime}_b=10^{15}$~cm and different values of the (comoving) magnetic field strength (all other parameter values are listed in Table \ref{tab:model_params}). In all panels, dashed lines represent the synchrotron spectrum of Bethe-Heitler pairs computed using the analytical function for the pair injection, see Eq.~(\ref{Q_inj_approx}). Dash-dotted and dotted lines represent spectra obtained by varying the proton and electron compactness, respectively, by two orders of magnitude with respect to their nominal values (solid lines). The energies of the low- and high-energy humps for the baseline model (solid lines) are indicated with symbols. Long-dashed lines in the LSP graphs represent the external photon field.  
    Gray shaded regions indicate the observational range of peak frequencies for the low- and high-energy SED components. Black horizontal lines at $\sim 1-10$~PeV represent the neutrino flux prediction according to Eq.~(\ref{photopion_prod}). }
    \label{fig:models_var_15}
\end{figure*}

In figure \ref{fig:models_var_15} we present an ensemble of theoretical SEDs that were built as outlined in Sec.~\ref{sec:numerical}, assuming an emitting region with radius $R^\prime_b = 10^{15}$~cm; all other parameter values  are listed in Table~\ref{tab:model_params} of appendix \ref{App:Params}. We find parameter sets for all blazar subclasses  resulting in a dominant BH-synchrotron component in the $\gamma$-ray regime (see thick solid lines in each panel) while producing broadband spectra with similar features as the observed ones (e.g. peak photon energies and luminosities). Changes in $\ell_{e}$ and/or $\ell_p$ are commonly invoked to reproduce flux variability \citep[e.g.][]{2013MNRAS.434.2684M, 2014A&A...571A..83P, 2021MNRAS.505.6103P}. Therefore, for each case, we demonstrate the effect that different values of $\ell_e$ (dotted lines) and $\ell_p$ (dash-dotted lines) have on the SED morphology. In all panels, we also include the expected all-flavor neutrino spectrum for completeness. 

In the first row of figure \ref{fig:models_var_15} we show spectra that resemble those of LSPs. The high-energy peak remains unaffected by changes in $\ell_e$ because protons pair produce close to the threshold with external photons that have higher number densities than the electron synchrotron photons in the comoving frame. Changes in $\ell_p$, however, have a linear effect on the peak luminosity of the Bethe-Heitler synchrotron spectrum  as long as the target density remains fixed (see Eq.~\ref{L_BH_full_obs}). As $\ell_p$ increases, the third bump peaking at $\sim 0.1$~TeV, which originates from pions created by the proton population interacting with the low-energy photons of the electromagnetic spectrum, becomes progressively more luminous that the Bethe-Heitler component (see cases for $B'=5$~G and $B'=50$~G). However, in smaller sources (e.g. $R^\prime_b=10^{14}$~cm) with stronger magnetic fields, such as $B'=50-500$~G, the same increase in $\ell_p$ leads to drastically different spectra that are almost featureless, characterized by a MeV peak with much higher luminosity (see e.g. dash-dotted lines in middle and right panels of the top row in figure \ref{fig:models_var_14}). These spectra are characteristic of a non-linear cascade developed in the source as protons lose efficiently their energy through photo-hadronic interactions on target photons that are not any more fixed, but are indirectly related to the proton population: attenuation of energetic photons produced in photo-hadronic interactions leads to the injection of relativistic pairs that produce low-energy target photons \citep[for more details on hadronic supercriticalities see][]{2014MNRAS.444.2186P, 2020MNRAS.495.2458M}. Larger sources, like those presented in figures \ref{fig:models_var_15} and \ref{fig:models_var_16} are less compact for the adopted values, hence we do not observe featureless spectra peaking in MeV energies (with the exception of the LSP-like case with $B'=50$~G in figure \ref{fig:models_var_15}). Lower $\ell_p$ values still produce a two-component spectrum, but with lower $\gamma$-ray luminosity than the default spectra displayed with solid curves. In this case, higher bolometric luminosities may be achieved for higher values of the Doppler factor. 

In the ISP- and HSP-like cases, the primary electron synchrotron photons are the main targets for Bethe-Heitler pair production, thus changing the dependence of the BH-synchrotron component on $\ell_e$. The target photon density is proportional to $\ell_e$ while the electron inverse Compton luminosity scales as $\ell_e^2$. Therefore, there are cases where an increase of $\ell_e$ can produce an inverse Compton component that is more luminous than the Bethe-Heitler synchrotron  -- see e.g. upper dotted curves for $B^\prime=0.5$~G and $B^\prime=5$~G. We also note that the peak energy of the SSC component is typically lower than the Bethe-Heitler synchrotron peak energy (see Eqs. (\ref{psyn}) and (\ref{e_ics})), which was chosen to fall within the typical range of $\epsilon_\gamma$ values for ISP and HSPs. Lower values of $\ell_e$ value will result in the suppression of inverse Compton emission from primary electrons. Still, a lower limit on $\ell_e$ is needed to produce a Bethe-Heitler synchrotron component that is more luminous than the proton synchrotron, contrary to the cases displayed by the lower dotted curves in the panels for ISPs and HSPs with $B^\prime=50$~G and $B^\prime=500$~G. For the ISP and HSP cases we examined, we find a linear dependence of the Bethe-Heitler synchrotron emission on $\ell_p$. Because of the lower values of $\ell_p$ used to produce the ISP/HSP-like spectra, the system does not enter into the non-linear regime by increasing $\ell_p$, as shown in upper panels for the LSPs of figure \ref{fig:models_var_14} in Appendix \ref{App:SEDs}. Finally, we note that photon-photon pair production severely attenuates the Bethe-Heitler synchrotron component in most HSP-like cases. As a result, the high-energy SED component does not peak within the range of observed peak energies, especially for cases with $B^\prime \ge 5$~G. In larger sources (e.g. $R^{\prime}_b=10^{15}$~cm and $R^{\prime}_b=10^{16}$~cm) with $B^\prime = 0.5$~G the photon density of the synchrotron-produced photons is lower and the attenuation due to $\gamma \gamma$ pair production not being as intense as for stronger magnetic fields.

In figure \ref{fig:models_var_15} we also observe that the neutrino luminosity is in most cases comparable to the Bethe-Heitler synchrotron peak luminosity, since the targets for BH pair production are comparable to those for photopion production. In such cases the analytical prediction of Eq.~(\ref{photopion_prod}) states that the Bethe-Heitler synchrotron and the photopion-related luminosities will differ from each other no more than an order of magnitude. In ISP- and HSP-like models we require the primary electron synchrotron photons to provide on-threshold targets for Bethe-Heitler interactions of protons.  However, the on-threshold target photons for photomeson interactions are $\sim150$ times more energetic than the threshold photons for Bethe-Heitler interactions. In some models the photomeson and Bethe-Heitler target photons are both part of the power-law synchrotron spectrum of primary electrons. In these cases (e.g. HSP-like models in figure \ref{fig:models_var_15}), the luminosity ratio of Bethe-Heitler synchrotron photons and neutrinos may range from 0.1 to 10 depending on the photon index of the target photon spectrum (see also section~\ref{sec:pg-gg}). Otherwise, if the photomeson targets are part of the exponential cut-off of the electron synchrotron spectrum, their number density is much smaller than the density of photons responsible for Bethe-Heitler pair production, resulting in a much lower neutrino luminosity than the one of the Bethe-Heitler component (e.g. ISP-like cases in figure \ref{fig:models_var_15}).

\begin{figure}
    \centering
    \includegraphics[width=0.49\textwidth]{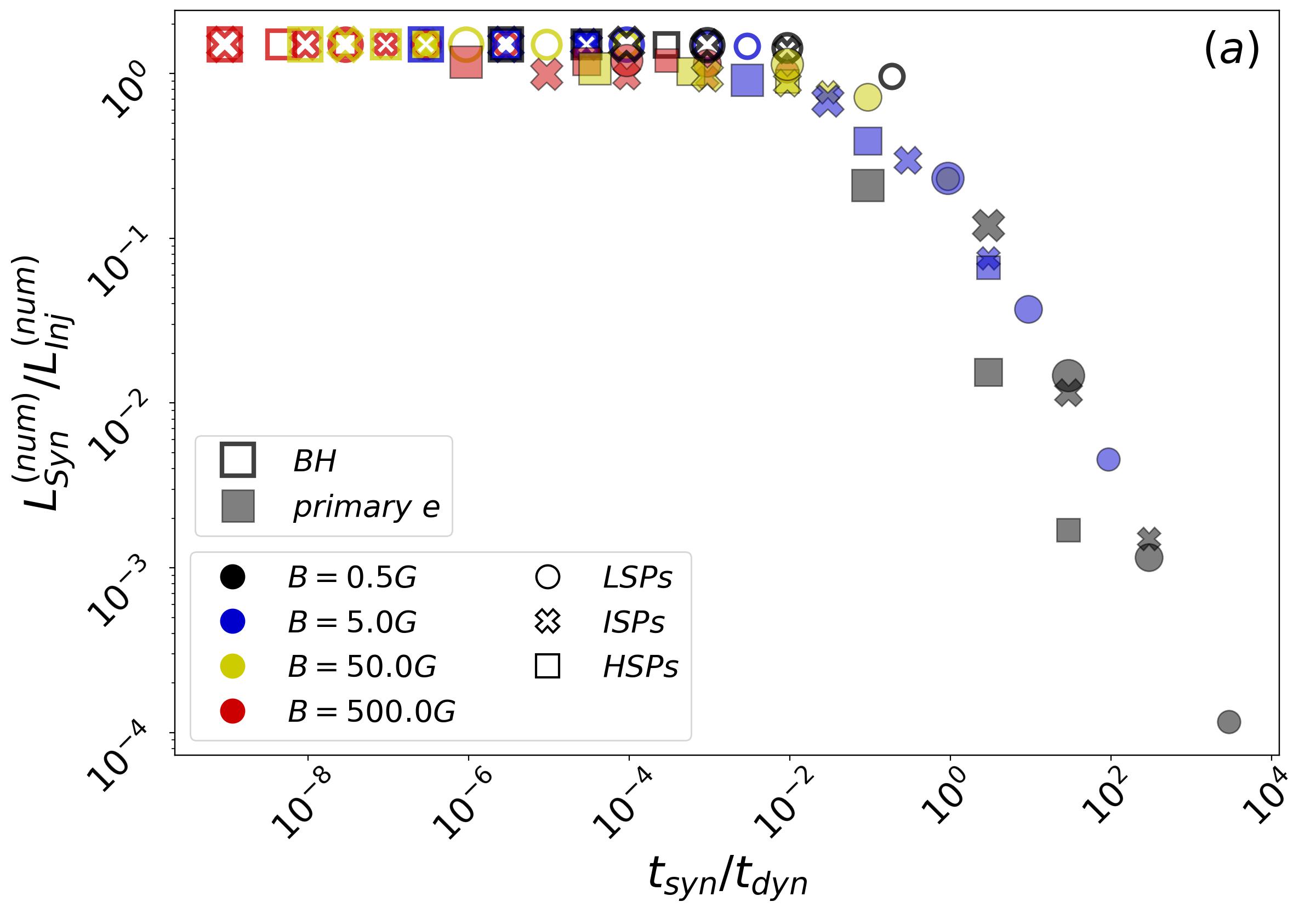}
    \vfill
    \includegraphics[width=0.49\textwidth]{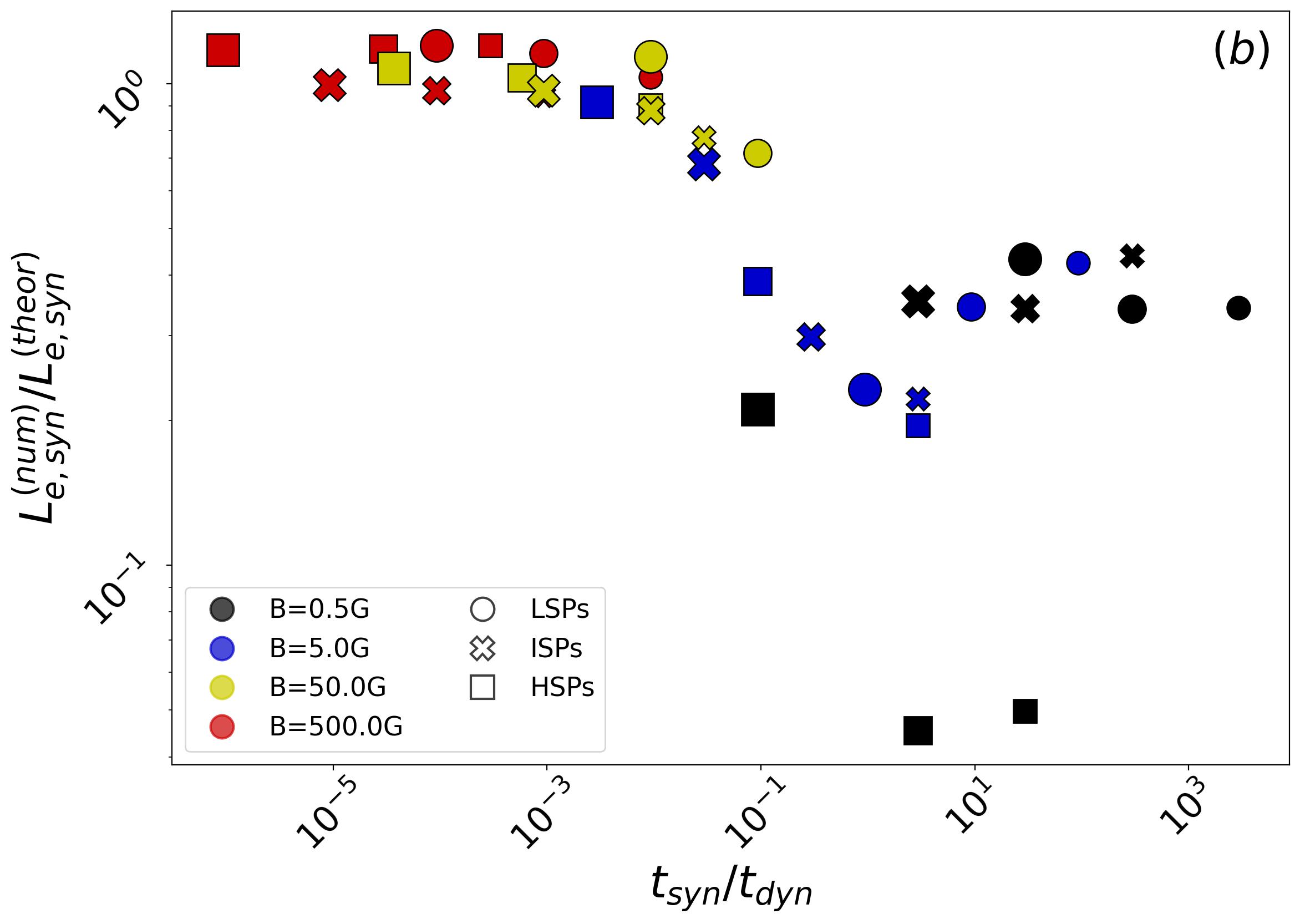}
    \hfill
    \includegraphics[width=0.49\textwidth]{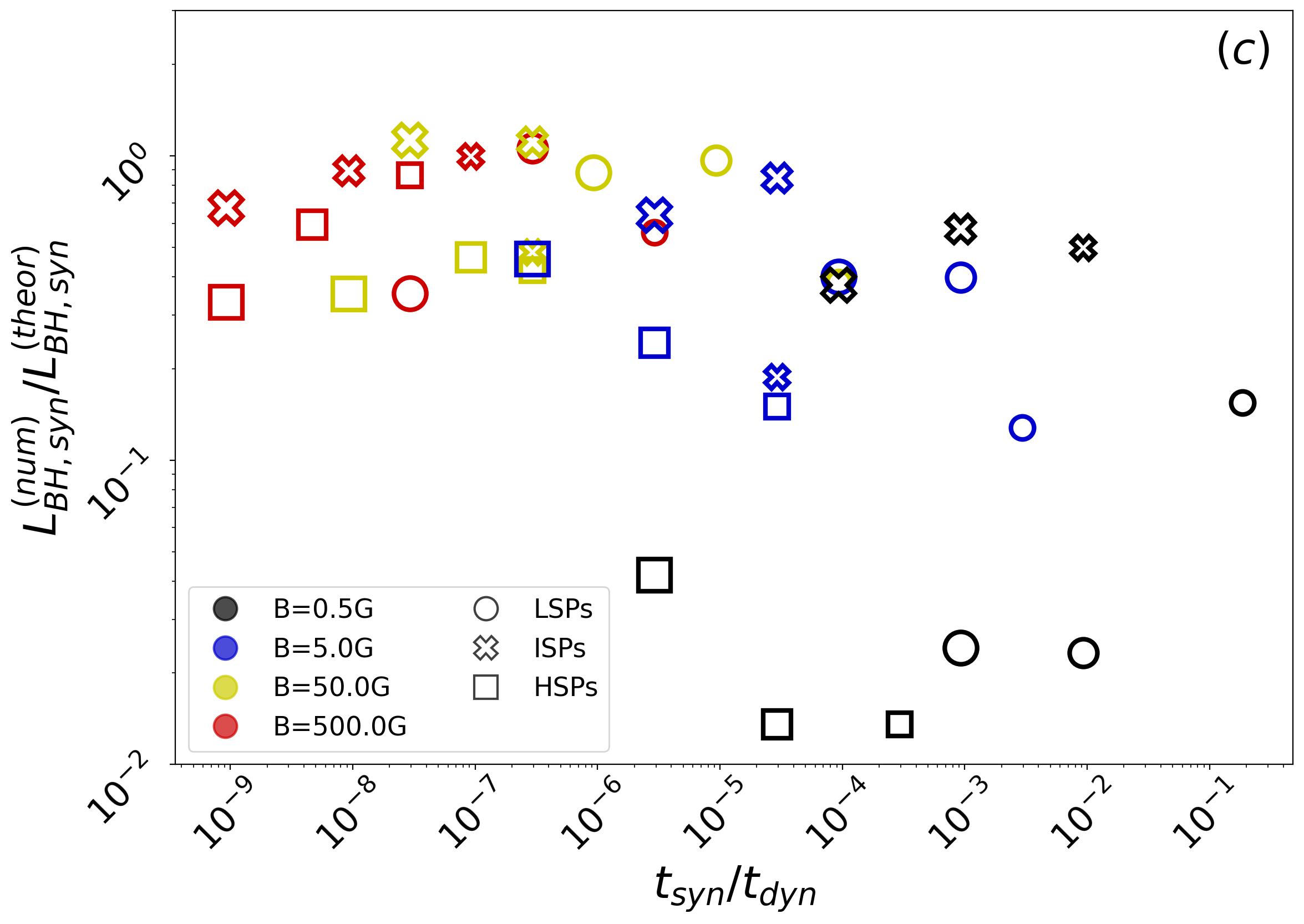}
    
    \caption{Bolometric luminosities of the primary electrons and Bethe-Heitler injected pairs and their emitted synchrotron radiation. Panel (a) shows the ratio of the synchrotron luminosity to the injected luminosity derived numerically for primary electrons (filled symbols) and secondary pairs (open symbols). Panel (b) displays the ratio of the primary electron synchrotron luminosity computed numerically to the analytical estimate of using Eq.~(\ref{esyn_tot_lum_obs}). Panel (c) shows the same ratio as in panel (b) but for Bethe-Heitler injected pairs, with the analytical estimate given by Eq.~(\ref{L_BH_full_obs}). All the ratios are plotted against the synchrotron cooling timescale of particles injected with Lorentz factor $\gamma_e$ to the timescale $R^\prime_b/c$.}
    \label{fig:anlyt_comp}
\end{figure}

At this point it would be interesting to compare the accuracy of the analytical estimations presented in section \ref{sec:analyt} about the injected and synchrotron emitted luminosities of primary electrons and Bethe-Heitler pairs against our numerical results that are not limited by simplifying assumptions. Figure \ref{fig:anlyt_comp} shows the ratio of the synchrotron radiated luminosity over the injected one for both the primary and secondary electrons (panel a), the ratio of the numerically calculated synchrotron luminosity over the analytical expectation for the primary electrons (panel b), and the same ratio for the Bethe-Heitler pairs (panel c). In all panels, the ratios are plotted against  $t_{syn}/t_{dyn}$, where $t_{dyn}=R'_b/c$, which can be considered as a proxy for the synchrotron cooling efficiency of particles injected with Lorentz factor $\gamma_e$. Panel (a) shows that Bethe-Heitler pairs are fast cooling due to synchrotron radiation for all models, since $t_{syn}/t_{dyn} \leq 1$, while for the primary electrons this is true only for models with $B^\prime \geq 50$~G (regardless of the source size). 
Our analytical formulas (see e.g., Eqs.~\ref{esyn_tot_lum_obs} and \ref{L_BH_full_obs}) were derived assuming that all the injected luminosity of the electrons is radiated through synchrotron, which is not always the case for the primary population. As a result, we expect that our analytical predictions for the primary electrons will not match the numerical results  when electrons do not cool efficiently due to synchrotron (i.e., $t_{syn}/t_{dyn} > 1$). The latter can, indeed, be observed in panel (b) where the ratio of the numerical and the predicted synchrotron luminosities is less than unity for models with $B^\prime \le 5$~G (i.e., the analytical expression in this regime overestimates the primary electron synchrotron luminosity).

To analytically calculate the Bethe-Heitler synchrotron luminosity (Eq.~\ref{L_BH_full_obs}) we assumed that (i) pairs radiate all their energy due to synchrotron and (ii) the target photons for pair production in ISP- and HSP-like case are the primary electron synchrotron photons  with a luminosity given by \ref{esyn_tot_lum_obs}. While the first assumption is always fulfilled, as shown in panel (a) of figure \ref{fig:anlyt_comp}, the second is not always true; the primary synchrotron luminosity is not always equal to the theoretical value given by \ref{esyn_tot_lum_obs} due to inefficient synchrotron cooling, as shown by panel (b) (black and blue markers). For the LSP-like cases the dominant targets for Bethe-Heitler pair production are assumed to be of external origin with a thermal energy distribution. In these cases, the available targets might also be of lower luminosity value than the theoretical one, being the grey body bolometric luminosity. However, since the protons interact with the near-peak photons of the spectrum and not with the low-energy tail of the grey body distribution, not all the grey body luminosity is available for pair creation. Panel (c) shows that Eq. \ref{L_BH_full_obs} is good proxy of the Bethe-Heitler synchrotron luminosity (within factors of 2-3) when the primary electron luminosity is radiated through synchrotron (red and yellow markers), but it can overestimate by a factor of 10-30 the luminosity when primary electrons are not fast synchrotron cooling (blue and black markers).

\subsubsection{Compton Dominance}
\begin{figure}[h!]
    \centering
    \includegraphics[width=0.48\textwidth]{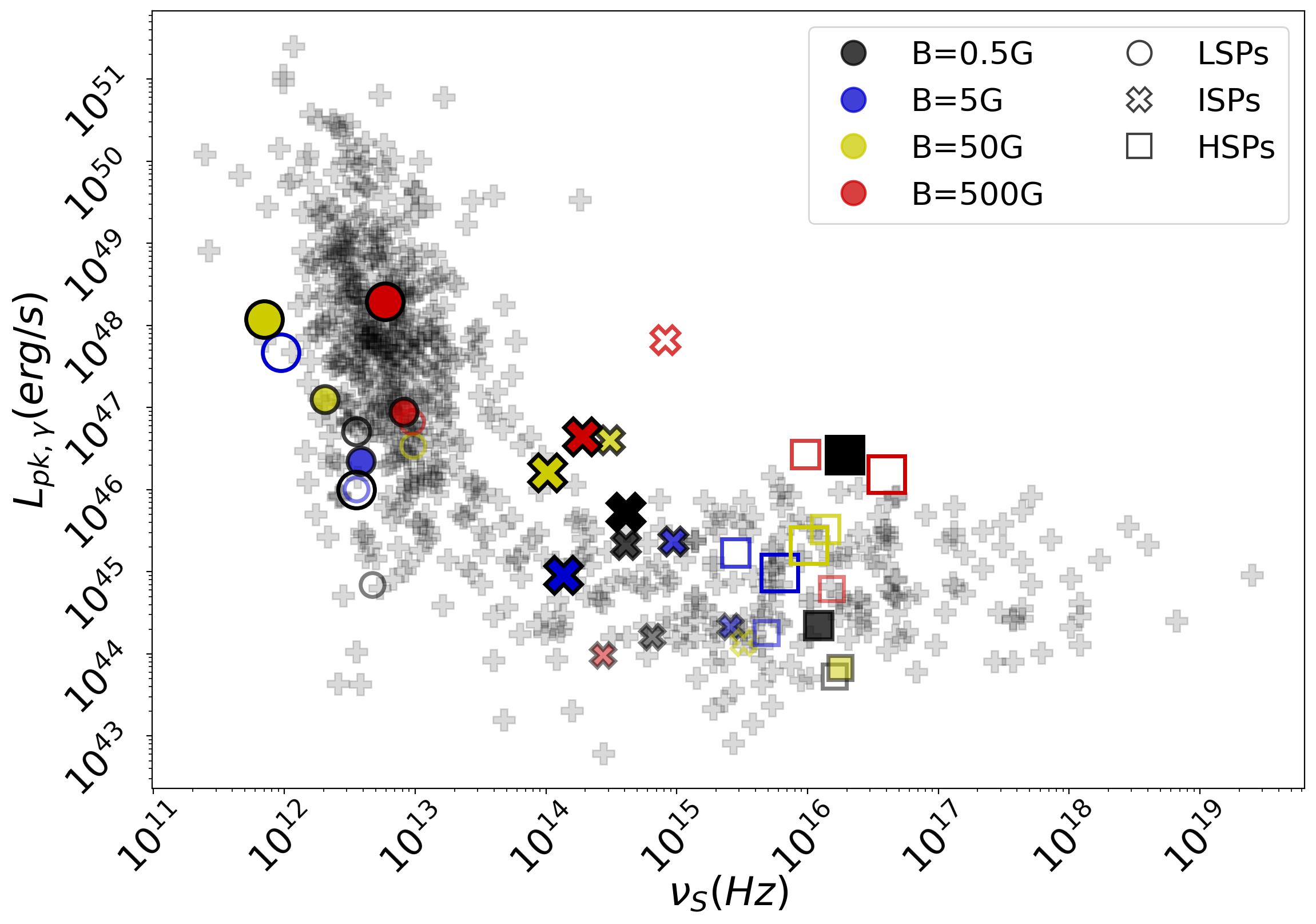}
    \includegraphics[width=0.494\textwidth]{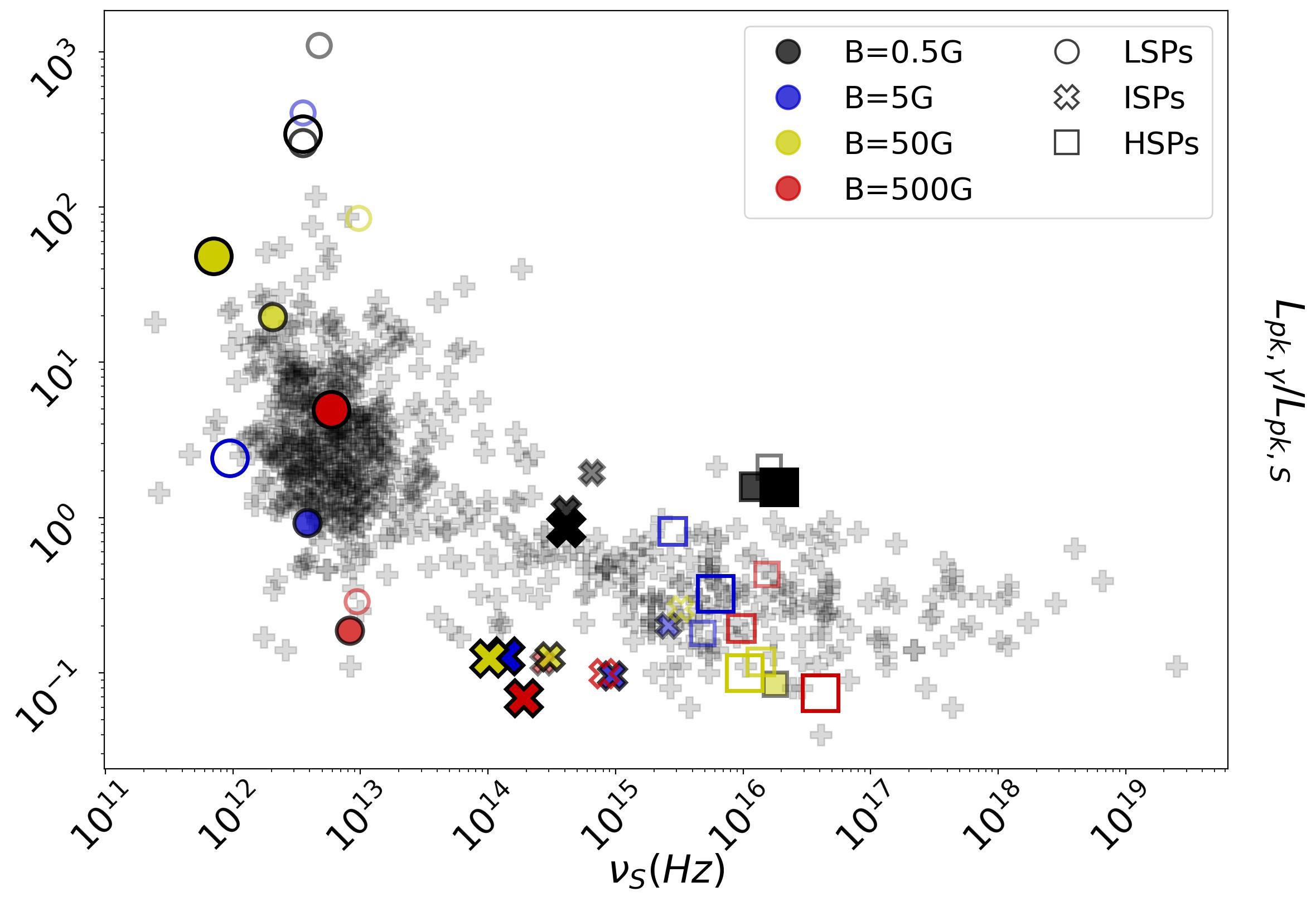}
     \caption{Peak synchrotron luminosity (left panel) and Compton dominance (right panel) versus the peak synchrotron frequency. Our model results are plotted with colored markers (see inset legends) on top of observationally inferred values (gray crosses) from \cite{chen_curvature_2023}. The size of colored markers indicates the size of the blob, with the smallest markers corresponding to $10^{14}$~cm and the largest to $10^{16}$~cm. Open markers indicate models that (i) fall outside the locus of observed points, or (ii) have $\gamma$-ray peak frequencies outside the typical range of values for each subclass (see table \ref{table:source_freqs}), or (iii) require non-physical combinations of temperature and luminosity of the external photon field (see section \ref{sec:ext_ph_fields}).}
     \label{fig:Lgamma_obs}
\end{figure}

Next we investigate how our theoretical SEDs compare to observations of \textit{Fermi}-detected blazars. In addition to the SED models shown in figure~\ref{fig:models_var_15} we compute photon spectra for $R^{\prime}_b=10^{14}$~cm and $10^{16}$~cm (see  figures~\ref{fig:models_var_14} and \ref{fig:models_var_16}, and Table~\ref{tab:model_params} for a complete list of parameter values). For each model we calculate the peak synchrotron luminosity and peak frequency ($L_{pk, S}$ and $\nu_S$), and the peak luminosity $L_{pk, \gamma}$ of the high-energy component. We then define the Compton dominance as the ratio of the two peak luminosities,  $A_{\gamma S} = L_{pk, \gamma}/L_{pk, S}$. In figure \ref{fig:Lgamma_obs} we compare our results against those derived from a sample of 781 \textit{Fermi}-detected AGN, which includes 504 FSRQs and 277 BL Lac objects \cite{chen_curvature_2023}.

Most theoretical models fall within the range of observational values and follow the same trend as the latter. The scatter in our models is attributed to different sizes of the emitting region and Doppler factor values. While this agreement with the data is achieved partially by construction (i.e., we selected our parameter values in order to reproduce peak luminosities and frequencies, as explained in section~\ref{sec:application}), it was not clear a priori if the Bethe-Heitler emission from pairs could dominate the high-energy emission in different blazar subclasses. This might be an indication that systems where Bethe-Heitler pairs play an important part in the production of the high-energy component of the blazar SED can exist and be hidden amongst the already observed AGN spectra.

However, some models can be rejected if they do not pass certain criteria, which we describe below. \textit{Criterion (i):} if a model falls well outside the observational locus of points in the peak $\gamma$-ray luminosity (or Compton dominance) versus peak synchrotron frequency diagrams, then we characterize it as non-plausible (e.g., ISP-like model with $B^\prime=500$~G and $R^{\prime}_b=10^{15}$~cm). \textit{Criterion (ii):} models might not qualify as plausible candidates for a certain blazar subclass, if the peak frequency of the high-energy hump does not fall into the typical range of values for that subclass (see table~\ref{table:source_freqs} and grey-shaded areas in figure \ref{fig:models_var_15}). \textit{Criterion (iii):} LSP-like models that require non-physical combinations of temperature and luminosity for the external photon field are also excluded (for more details see the next section). Rejected models are indicated with open markers in figure~\ref{fig:Lgamma_obs}.

\begin{figure}[h!]
    \centering
    \includegraphics[width=0.6\textwidth]{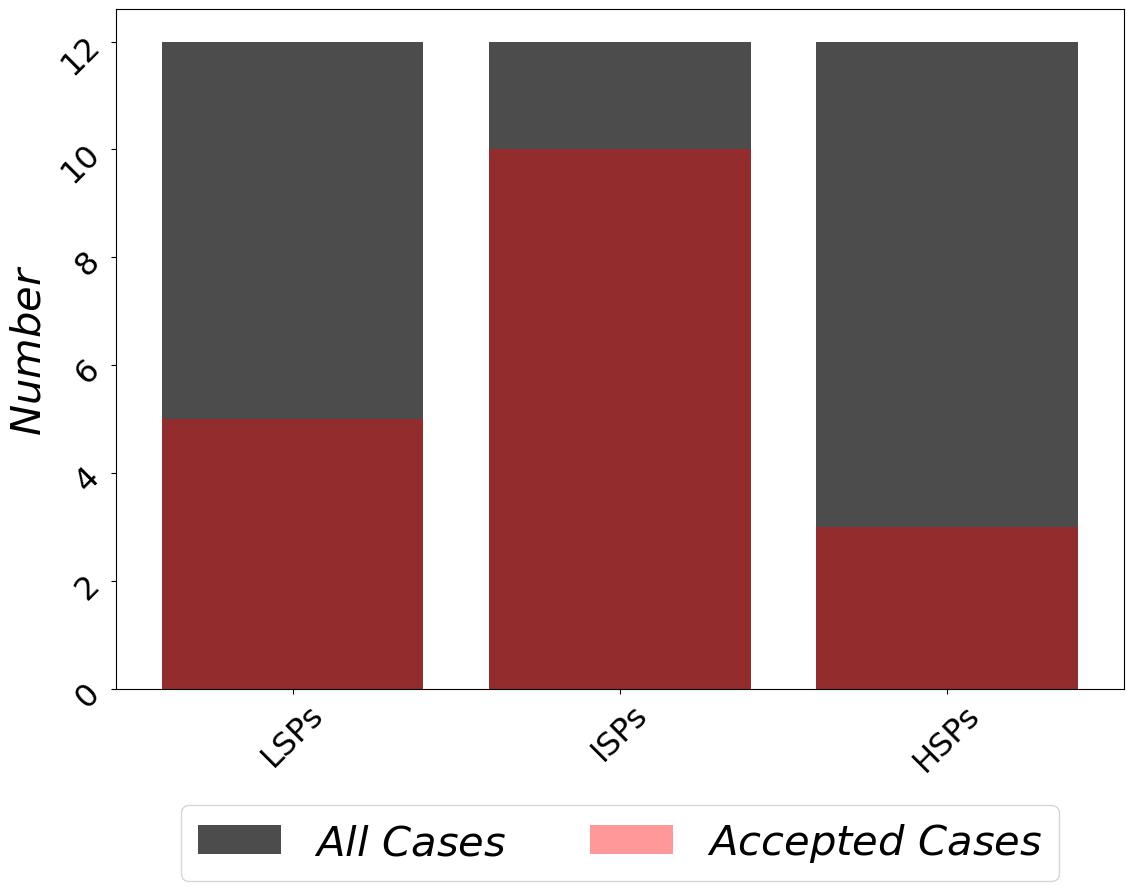}
    \caption{Chart of number of total models (black) and valid models (red), categorized in respect of their electromagnetic spectrum type (LSP-, ISP-, HSP-like).}
    \label{fig:chart}
\end{figure}

Figure \ref{fig:chart} shows both the total number of models we computed, 12 for each type (LSP-, ISP-, HSP-like), and the number of the accepted models  based on the criteria described in the previous paragraph. In summary, we found that almost all HSP-like models (9/12) can be excluded solely by criterion (ii). Most LSP-like models (7/12)  can be excluded based on criterion (iii), while ISP-like models (2/12) can be excluded either on criterion (i) or (ii). Our results suggest that if we were to observe blazars with a Bethe-Heitler-dominated $\gamma$-ray spectrum, most of them (55\%) would be ISPs, a few LSPs (28\%) and, only a a minority would be HSPs (17\%).

\subsubsection{External photon fields}\label{sec:ext_ph_fields}

To reproduce LSP-like spectra within our scenario we invoked arbitrary external radiation fields to compensate for the low energy of jet synchrotron photons. More specifically, these photons are not energetic enough (for typical values in LSPs) to provide near-threshold targets for Bethe-Heitler interactions, which are needed to produce a luminous high-energy component. Therefore, a grey-body photon field with effective temperature $T^\prime \approx 10^2-10^5$~K in the blob frame was introduced. Considering a Doppler factor of $\delta \in [40-80]$ and $\Gamma = \delta/2$ (see table \ref{tab:model_params} in appendix \ref{App:Params}), the aforementioned temperatures translate to $T = T^\prime/\Gamma \approx 10-10^3$~K in the AGN rest frame. The energy density of the thermal field (in the AGN rest frame) is related to the photon compactness $\ell_{ext}$ as $u_{ext}= 3 \ell_{ext} m_e c^2 /\Gamma^2 \sigma_T R^\prime_b$. By requiring $u_{ext} < u_{BB}$, where $u_{BB} = \alpha T^4$ is the energy density of a black body with effective temperature $T$ and $\alpha$ is the radiation density constant, we can set an upper limit on $\ell_{ext}$, 

\begin{equation}
    l_{ext} \leq \frac{\alpha \sigma_T}{3 m_e c^2} R^\prime_b \Gamma^{-2} T^{\prime 4} \lesssim 2 \times 10^{-4} R^\prime_{b, 15} \Gamma^{-2}_1 T^{\prime 4}_4.
\end{equation}

In most LSP-like sources, the invoked thermal field does not satisfy the limit above (see table~\ref{tab:model_params} in appendix \ref{App:Params}), making this a non physical choice. An attempt to lower $\ell_{ext}$, while trying to produce the same $L_{\gamma}$ through Bethe-Heitler synchrotron radiation, would require a higher proton compactness (see Eq.~\ref{L_BH_full_obs}). This, in turn, would increase the energetic requirements of the model (see also section~\ref{sec:power}) and change the relative ratio of the proton synchrotron and pair synchrotron peak luminosities, leading to dual-component $\gamma$-ray spectra (see e.g. solid red line in the top right panel of figure~\ref{fig:models_var_15}). From this analysis, only LSP-like models with $R^\prime=10^{15}-10^{16}$~cm and $B^\prime=5-500$~G are viable.

\subsubsection{Jet power}\label{sec:power}
The power of a two-sided blazar jet, ignoring the contribution of cold protons, can be expressed as
\citep[e.g.][]{2014Natur.515..376G, 2015MNRAS.450L..21Z, petropoulou_properties_2016}:
\begin{gather}
\label{eq:jetpower}
    P_{jet} = 2 \pi R^{\prime 2}_b \beta \Gamma^2 c \sum_{i=B,e,p,ph} (u^{\prime}_i+P^{\prime}_i) 
\end{gather}
where $R^\prime_b$ is the radius of the region that produces the steady emission, and is assumed to be equal to the cross-sectional radius of the jet, $\Gamma$ is the jet Lorentz factor, $\beta \approx 1$ is the jet speed (in units of $c$),  $u^{\prime}_i$ and  $P^{\prime}_i= \frac{1}{3}u^{\prime}_i$  are the energy density and the pressure of the magnetic field and of the relativistic particles in the emitting region.

\begin{figure*}[h!]
    \centering
    \includegraphics[width=0.95\textwidth]{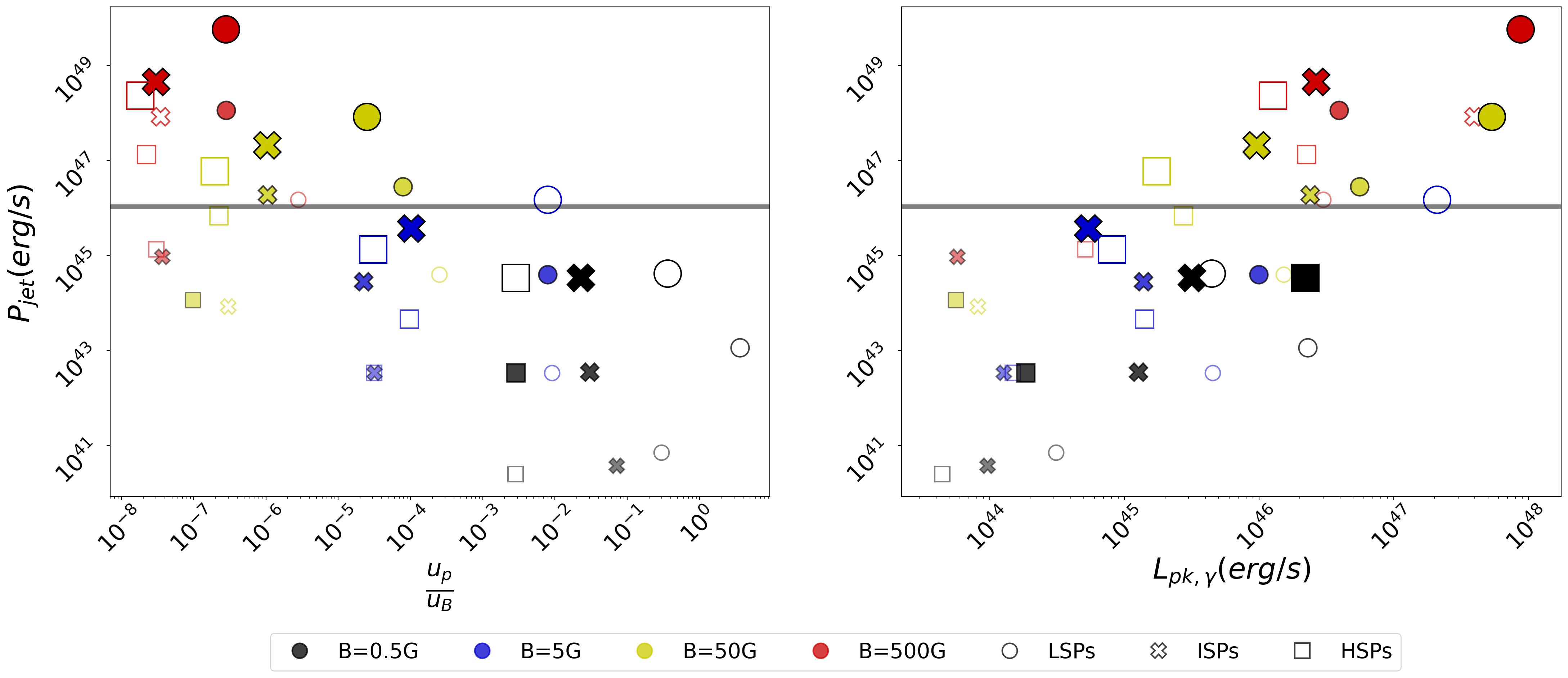}
    \caption{Jet power (Eq.~\ref{eq:jetpower}) as a function of the ratio of the relativistic proton and magnetic field energy densities (left panel), and the peak $\gamma$-ray luminosity of the source (right panel). Smaller size and more transparent markers indicate smaller emitting regions. The horizontal line marks the Eddington luminosity of a $10^8~M_\odot$ black hole.}
    \label{fig:Ljet}
\end{figure*}

In figure \ref{fig:Ljet} we plot the jet power of our models against the proton-to-magnetic energy density ratio (left panel) and the $\gamma$-ray peak luminosity (right panel). Note that the jet Lorentz factor is not the same for all cases (see Table~\ref{tab:model_params}). Models with large emitting regions and strong magnetic fields are characterized by high powers that surpass the Eddington luminosity for a $10^8~M_{\odot}$ black hole mass (horizontal line) by several orders of magnitude. While the Eddington luminosity of the accreting black hole is not a strict physical limit to the power of the jet, such high values are not expected in sub-Eddington accreting black holes. The jet power in an accreting system can be written as $P_{jet}=\eta_j \dot{M}c^2$ , where $\dot{M}$ is the accretion rate onto the black hole and $\eta_j$ is the jet-formation efficiency, which can be $\sim 1.5$ at most for
magnetically arrested accretion disks \citep{2003PASJ...55L..69N, 2011MNRAS.418L..79T}. Thus, the jet power can be as high as $P_{jet} \lesssim 15 \dot{M}_{\rm Edd} c^2 / \eta_{-1} \simeq 1.5 \times 10^{47}~M_8/\eta_{-1}$~erg s$^{-1}$, where $\eta$ is the radiative efficiency of the accretion flow. 
Furthermore, the displayed powers are actually lower limits (due to the assumed very narrow proton energy distribution and the neglect of the non-relativistic plasma component). Therefore, models with $P_{jet}\gg 10^{47}$~erg s$^{-1}$ can be discarded based on energetic grounds. Moreover, blazar jets are thought to be strongly magnetized and narrow near their base, with weaker magnetic fields on larger scales \citep{1979ApJ...232...34B, 2004ApJ...605..656V}. It is therefore difficult to physically explain the formation of large emitting regions (at pc scales) with hundreds of Gauss (comoving) magnetic field strengths. These arguments also disfavor the models with extreme jet powers. From this analysis, the physically plausible models are those with: $R_b^{\prime}=10^{14}$~cm and $B^\prime\le100$~G, $R_b^{\prime}=10^{15}$~cm and $B^\prime<100$~G, and $R_b^{\prime}=10^{16}$~cm with $B^\prime \lesssim 5$~G.

\section{Summary and Discussion}\label{sec:discussion} 

In this paper we took a closer look at the characteristics of the distribution of relativistic electrons and positrons produced in Bethe-Heitler interactions of relativistic protons with low-energy photons. While most pairs are created with a Lorentz factor $\gamma_e \approx \epsilon^{-1}$, most of the energy is transferred to particles with $\gamma_e \approx  \gamma_p$ for near-threshold interactions between protons and photons of single energy $\epsilon$. Far-from-threshold interactions lead to more extended pair distributions, and tend to distribute almost equal amounts of energy per logarithmic decade in $\gamma_e$. In astrophysical environments where interactions of protons take typically place on extended target photon distributions, we find that the energy distribution of pairs is usually determined by near-threshold interactions, and has a well-defined peak energy. 

Additionally, we provide an empirical function that is able to describe the Bethe-Heitler pair distribution satisfactorily. The empirical function can be implemented in numerical codes and replace the time-consuming calculation of double integrals. Our function, however, has not been constructed to model pair distributions for interaction energies $\gamma_p \epsilon < 2$. The contribution of these interactions to the total pair production spectrum has been found to be smaller by an order of magnitude than the contribution of interactions with $\gamma_p \epsilon \approx 5$ (very close to the threshold, the effective cross section of the process is very small -- see e.g. figure 1 in \cite{mastichiadis_spectral_2005}). As a result, when target photons for Bethe-Heitler are described by an extended distribution, $\gamma_p \epsilon <2$ interactions will not be important in shaping the overall secondary pair distribution. 

Furthermore, we examined if it is possible to produce $\gamma$-rays from energetic Bethe-Heitler pairs in blazar jets, and produce SEDs that resemble the observed ones. We used analytical arguments (under several simplifying assumptions) 
to determine the relevant parameter space, and created 12 numerical SED models for each blazar subclass (36 in total) using the code \code. As the numerical models do not suffer from the simplifying assumptions used in our analytical estimates, they were also used to cross check the validity of the latter. We found that
the parameter space leading to a dominant Bethe-Heitler $\gamma$-ray component is very constrained, especially when the jet photons are the targets  for Bethe-Heitler interactions  (see e.g. figure~\ref{fig:ge_gp}). Moreover, small changes in parameters like the energy injected to accelerated electrons and protons, can favor other processes, like SSC or proton-synchrotron radiation, and eventually hide the Bethe-Heitler $\gamma$-ray emission (figure~\ref{fig:models_var_14}). The latter attribute can motivate small adjustments of the injected proton and electron luminosities in order to fine tune the proton synchrotron and Bethe-Heitler synchrotron peak luminosities. There are cases (see e.g. figure \ref{fig:spectr_components}) where the aforementioned components are both distinguishable with comparable luminosities. But by changing the injection luminosities by a factor of 0.1-0.2 in logarithm, one could suppress the proton synchrotron emission and acquire a spectrum with a broad $\gamma$-ray component, the Bethe-Heitler synchrotron one. Given the limited number of models we created, we did not use the presence of additional spectral components in the broadband SED to exclude said models. However, we implemented three other criteria to determine whether a model is accepted as a possible candidate for a blazar subclass or not. First, we checked if the peak position of the $\gamma$-ray component fell inside the frequency range set in table \ref{table:source_freqs} for each blazar type. Second, we checked  whether the numerical models fell within the locus of observed points in figure \ref{fig:Lgamma_obs}. Lastly, for LSP-like models, we checked if the combination of temperature and density for the external photon field was physical or violated the black body limit. Interestingly, after enforcing the three criteria, we found that most HSP-like models can be excluded because of the strong attenuation of $\gamma$-rays above a few tens of GeV (see e.g. figures \ref{fig:tau_gg} and \ref{fig:models_var_14}). Nevertheless, a few LSP-like and most ISP-like models passed all criteria, suggesting that sources with Bethe-Heitler $\gamma$-ray emission could exist in the blazar population, but they would not be the most common ones. Interestingly, in an independent work where SED fitting was performed for a sample of 34 blazars, the best-fit model for 3 ISPs yielded a dominant Bethe-Heitler synchrotron component in $\gamma$-rays (Rodrigues, X. et al., in preparation). 

Our blazar models were computed using narrow power-law proton distributions, even though acceleration mechanisms create wide distributions, typically starting from the proton rest mass energy. Therefore, questions about the contribution of lower energy protons to the total spectra and jet power may arise. If the proton population is hard ($s_p < 2$), we find that the low-energy protons do not significantly contribute to Bethe-Heitler pair production and the total jet power, since most of the proton population energy is carried by protons at the high-energy tail of the distribution. These energetic protons are responsible for the production of pairs emitting in $\gamma$-rays. When we extend the proton distributions down to $\gamma^\prime_p \approx 1$, the compactness is slightly affected (by $\sim0.7$ in logarithm). For soft power laws ($s_p > 2$) extending down to $\gamma^\prime_p\sim 1$, the required jet power would be order of magnitude higher than the values displayed in figure~\ref{fig:Ljet}, to compensate for the ``inactive'' lower energy protons of the distribution that would not contribute to pair production. Alternatively, protons can accelerate into a broken power law with a low energy branch having a slope $s_{p, \rm lo} \leq 2$ from $\gamma^\prime_p \sim 1$ up to $\gamma^\prime_{p,\min}$, and $s_p>2$ for $\gamma^\prime_p > \gamma^\prime_{p, \min}$. When we extend the proton distributions down to $\gamma^\prime_p \approx 1$, assuming $s_{p,\rm lo}=2$ or 1, the compactness is increased by a factor $\approx 18$ or 2, respectively.  As a result, the main conclusions presented in section \ref{sec:application} remain valid, with the assumption of a proton distribution following a broken power law distribution with a slope of $< 2$ for $\gamma^\prime_p < \gamma^\prime_{p,min}$. 

The flat spectrum radio quasar 3C 279 is one of the brightest $\gamma$-ray blazars, also classified as an LSP source, located at redshift  $z = 0.536$ \citep{1965ApJ...142.1667L}. In June 2015, the Large Area Telescope (LAT) on board of the \textit{Fermi} satellite recorded a luminous $\gamma$-ray flare in GeV energies ($L_\gamma \sim 10^{49}$~erg s$^{-1}$) with 5-minute timescale variability  \citep{ackermann_minute-timescale_2016}. The short variability timescale, the large amount of energy released in GeV $\gamma$-rays, and the large Compton ratio (about a 100) of the flare make it difficult to explain this event with standard leptonic or lepto-hadronic models \citep[for details, see][]{ackermann_minute-timescale_2016, Petropoulou_2017}. 
Ackermann et. al \citep{ackermann_minute-timescale_2016} proposed an alternative leptonic scenario in which the $\gamma$-ray component of the event is produced by a second, energetic electron population, with no specified origin, emitting through synchrotron. This proposal motivated us to investigate whether Bethe-Heitler pair production was able to produce such a relativistic leptonic population. We found that is possible to explain the SED of the 2015 flare if certain conditions are met. First, the shape of the flaring $\gamma$-ray spectrum (see figure 4 in \cite{ackermann_minute-timescale_2016}) is not very broad around its peak, thus requiring near-threshold Bethe-Heitler interactions to dominate the pair production. Second, a strong magnetic field, e.g. $B^\prime=150$~G, is needed for boosting the luminosity of the pair synchrotron spectrum. However, even such strong magnetic fields were not enough to reproduce the observed luminosity, so a high Doppler factor value, $\delta \approx 190$, was needed, as well. For an adopted $B^\prime$ value, the proton Lorentz factor can be determined so that the pair distribution peaks at the observed $\gamma$-ray energy (see Eq. \ref{BH_syn}). We can then determine the target photon energy that satisfies the Bethe-Heitler threshold condition (see Eq. \ref{BH_thres_therm}). Because of the very high $\delta$ values needed, the target photon energy translates to a very low grey body temperature in the AGN rest frame, $T \approx 12$~K. Typical temperatures of AGN dust torii are $\sim 300$~K to $\sim 1500$~K \citep{2006NewAR..50..728E}. Evidence for colder dust with T$\sim 25$~K has been reported in blazar Ap Librae, but on hundreds of parsec scales \citep{roychowdhury_circumnuclear_2022}. Moreover, the invoked target photon density exceeds that of a black body field with such low temperature. In conclusion, explaining the SED of the 3C 279 2015 flare using a second electron population originating by Bethe-Heitler pair production would require extreme physical conditions, and is therefore deemed unrealistic. 

In conclusion, Bethe-Heitler pair production in blazar jets can generate a secondary relativistic lepton population that can emit in $\gamma$-rays via synchrotron radiation. Because of the extended proton and photon distributions involved,  Bethe-Heitler interactions can happen both near and away from threshold. Nonetheless, their synchrotron spectrum is determined by pairs created by near-threshold interactions. Bethe-Heitler synchrotron radiation can be the dominant $\gamma$-ray emission mechanism in low- and intermediate-peaked blazars, but for a limited part of the parameter space.  

\acknowledgments
We would like to thank the anonymous referee for their very detailed and constructive report that helped us improve our manuscript. We would also like to thank Stamatios I. Stathopoulos for facilitating the comparison against the numerical results from the code {\tt LeHaMoC}. D. K. and M.P. acknowledge support from the Hellenic Foundation for Research and Innovation (H.F.R.I.) under the ``2nd call for H.F.R.I. Research Projects to support Faculty members and Researchers'' through the project UNTRAPHOB (Project ID 3013).

\paragraph{Note added.} The empirical function for the Bethe-Hetler pair production spectrum is available in a python-class form in the available link:\\
\url{https://github.com/Des0053/Bethe-Heitler-Injection-Rate-Analytical-Approximation}








\bibliographystyle{JHEP}
\bibliography{bibliography}

\appendix 
\section{An empirical function for the Bethe-Heitler pair production spectrum} \label{App:BH_Analyt}
Our goal is to provide an analytical function for the Bethe-Heitler pair production spectrum that could be easily implemented in numerical codes. Our function is benchmarked with numerical results from \code \ that are based on Monte-Carlo simulations of proton-photon interactions performed by \cite{PJ96}. 

\subsection{Building an empirical function}

Equation (\ref{Q_inj_approx}) presents our heuristic function for the pair injection spectrum produced per unit time by interactions of monoenergetic protons and photons. Our goal was not to find a unique mathematical representation of the pair production spectrum. Instead, our function was conceptualized by examining the shape of the energy injection spectrum as described below.

\begin{figure}[h!]
        \centering
        \includegraphics[width=0.49\textwidth]{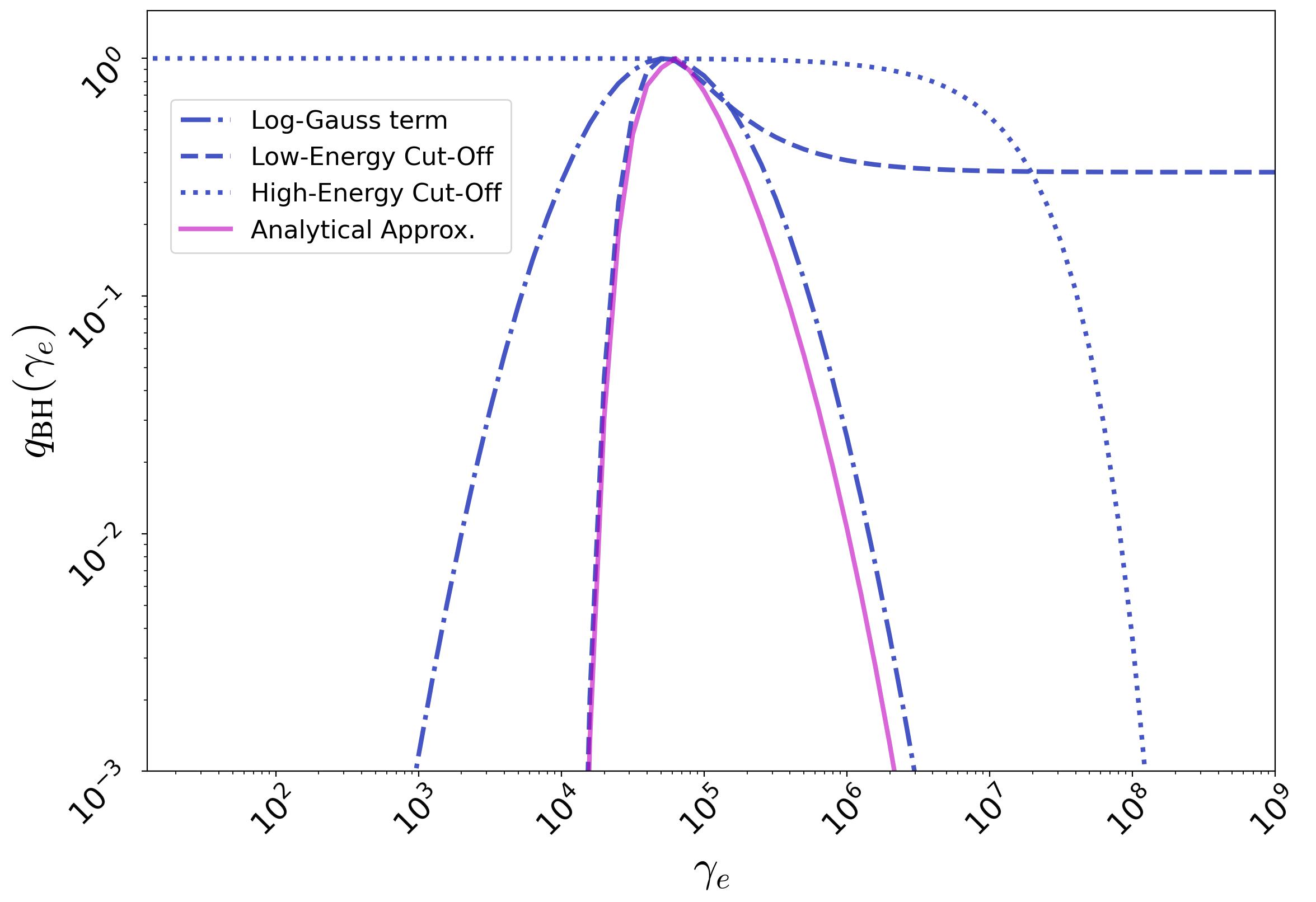}
        \hfill
        \includegraphics[width=0.49\textwidth]{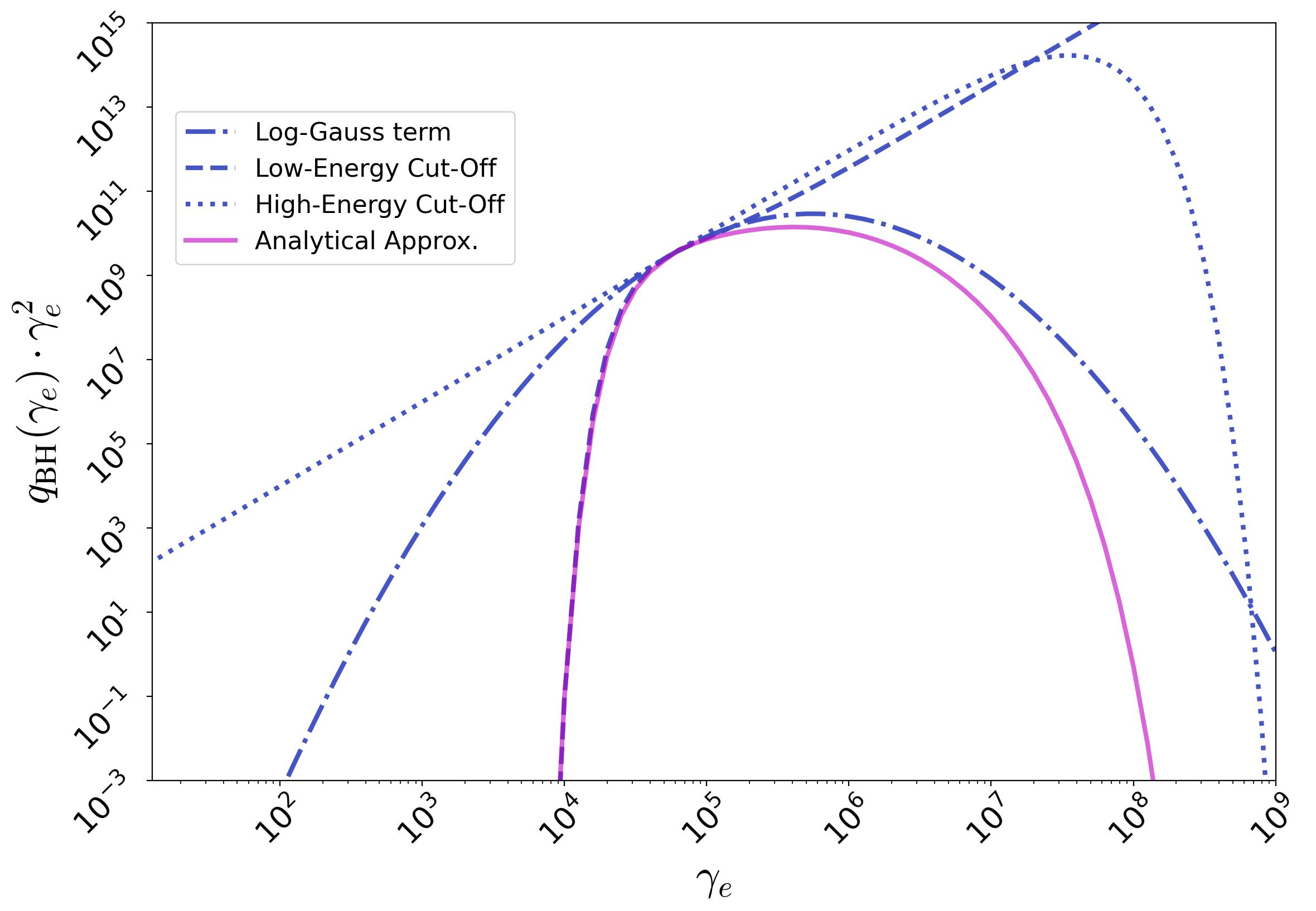}
    \caption{Graphical representation of the three exponential terms appearing in Eq.~(\ref{Q_inj_approx}) without accounting for the normalization $A$. All curves in the left panel are normalized to a peak value of one. The magenta line in both panels represents the product of the three terms.}
    \label{fig:Q_inj_comp}
\end{figure}

\begin{figure}[h!]
    \centering
     \includegraphics[width=0.49\textwidth]{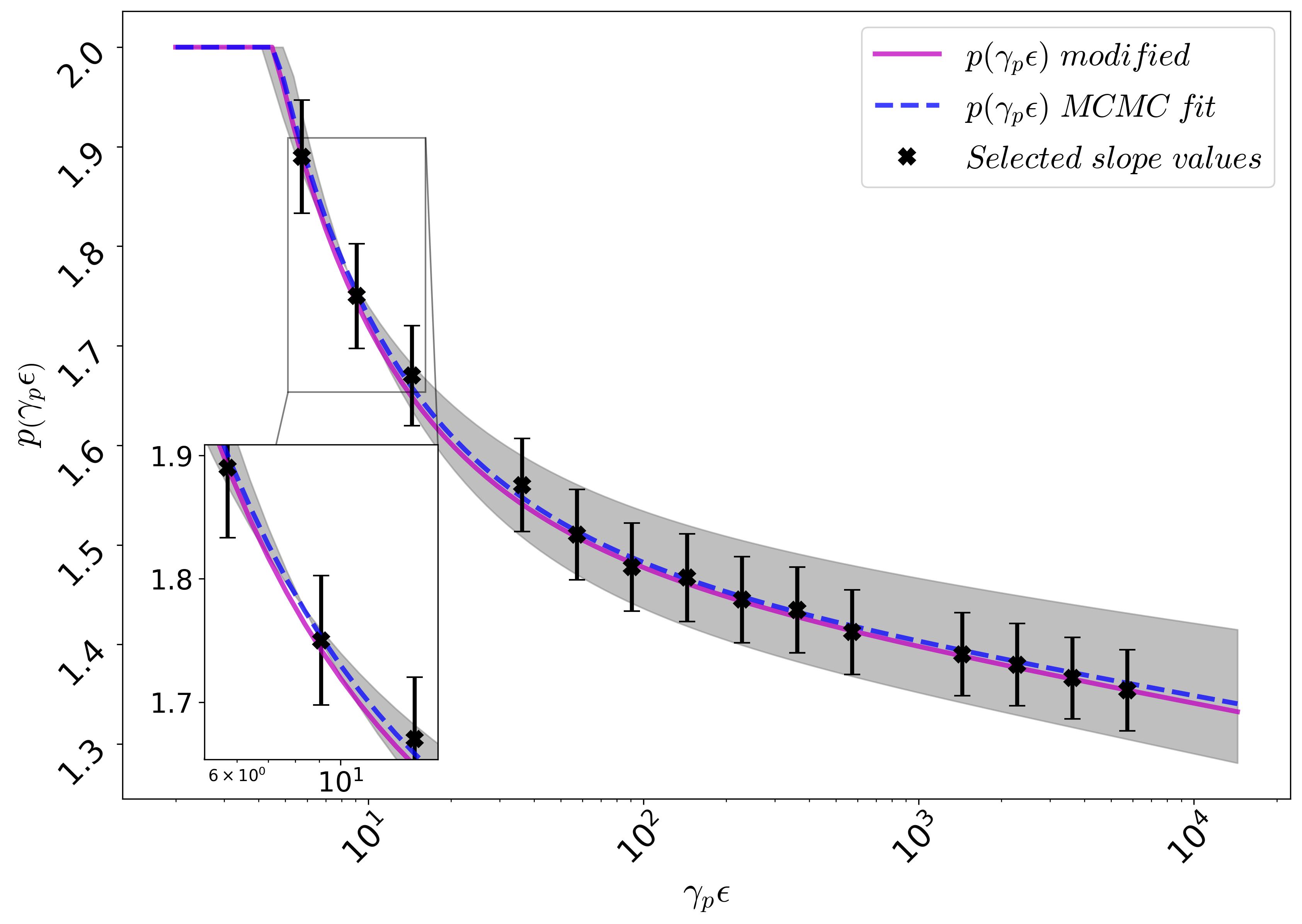}
     \includegraphics[width=0.49\textwidth]{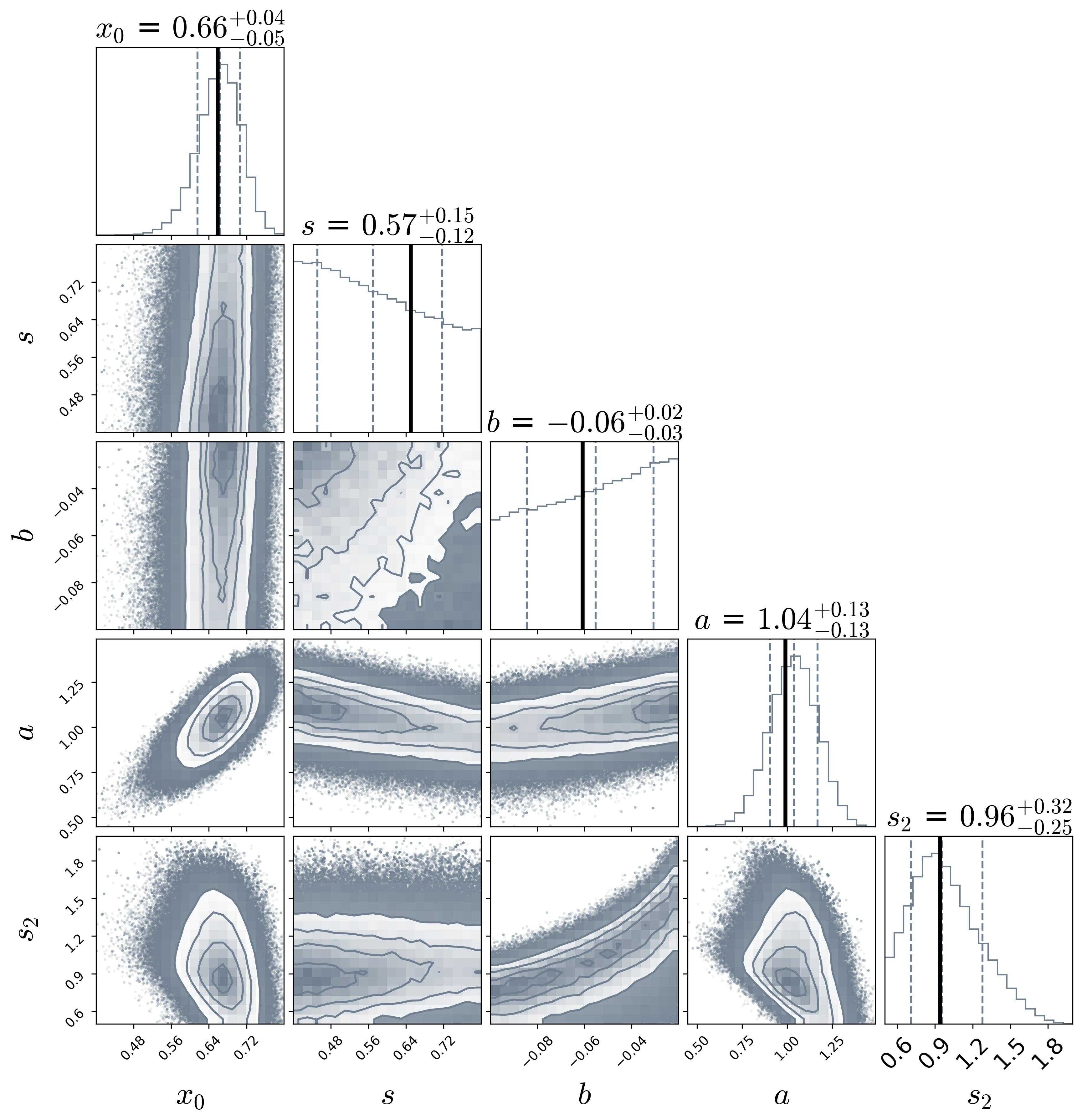}
    \caption{\textit{Left panel}: Power-law index $p$ appearing in Eq.~(\ref{Q_inj_approx}) as a function of the interaction energy $\gamma_p \epsilon$. The dashed blue line represents the MCMC fit result with the gray shaded area being the 1$\sigma$ spread of the values. The solid magenta line is another solution obtained for parameter values slightly different than the median values of the posterior distributions (see corner plot on the right panel).  \textit{Right panel}: Corner plot showing the posterior distributions of the parameters in Eq.~(\ref{eq:slope}). Dashed lines in the histograms indicate the median value and the 68\% confidence interval. Solid black lines represent the values that reproduce the magenta line of the left panel.
    }
    \label{fig:norm_n_slop}
\end{figure}

Each of the three exponential terms appearing in Eq.~\ref{Q_inj_approx} has a specific role in shaping the overall spectrum. First of all, the shape of the energy injection spectrum, namely $\gamma_e^2 q_{\rm BH}(\gamma_e)$, resembles a logarithmic Gaussian function for interactions near the interaction threshold (see figure~\ref{fig:BH_char}b), which translates to $p(\gamma_p \epsilon) \approx 2$ (see figure~\ref{fig:BH_char}c). We empirically determined that the Gaussian-like shape depends on the interaction energy $\gamma_p \epsilon$. In particular, as the interaction energy increases, the energy injection rate has no longer a well-defined peak but instead becomes flat while extending to lower energies (see e.g. figure \ref{fig:BH_char}b). This behavior can be captured by a smaller value of the power-law index $p$ (see figure \ref{fig:norm_n_slop}). Since the Gaussian-like function overestimated significantly the energy transferred to the low-energy pairs, a super-exponential term, $\exp[-a_2^2(\gamma_{e, pk}/\gamma_e-1)^2]$, was added to make the low-energy part steeper. Finally, the third exponential term controls the high-energy cutoff of the injection spectrum. This, in particular, needed to be softer than the low-energy turnover, hence it is linearly dependent on $\gamma_e/\gamma_{e, pk}$. The shape of the three exponential terms of the injection rate function is better illustrated in figure \ref{fig:Q_inj_comp}. 

In order to determine the analytical functions for the power-law index $p(\gamma_p \epsilon)$ and the normalization $A(\gamma_p, \epsilon)$, we used the numerical results presented in  figure \ref{fig:BH_char}c and 11 additional cases corresponding to different values of the interaction energy $\gamma_p \epsilon$. For each of the 15 benchmark cases we first adjusted the slope $p$ by eye so that the analytical function would match the \code \ results. The inferred values of the slope (with a 3\% error) are shown with black symbols in figure \ref{fig:norm_n_slop} (left panel). To describe the obtained trend of $p$ with $\gamma_p \epsilon$ we introduce the following function,
\begin{gather}
    p(x) = a x^{-s} e^{-\frac{x}{x_0}} + b x^{s_2}+ c 
    \label{eq:slope}
\end{gather}
where  $x \equiv \log_{10} (\gamma_p \epsilon) $ and $a, b, c, s, s_2, x_0$ are free parameters. The first term,  $\propto x^{-s}$, describes the function near the threshold values of $\gamma_p \epsilon$. It is also multiplied by an exponential term, $e^{-x/x_0}$, to suppress its contribution at large interaction energies where the linear term dominates.

To determine the free parameters\footnote{The constant parameter, $c$, is frozen during the fit at a value that ensures continuity of $p(x)$  at $x = 5$.} of the function, we fit the data (black symbols) using {\tt emcee} \citep{emcee}, a python implementation of the Affine invariant Markov chain Monte Carlo (MCMC) ensemble sampler. This allows a better estimation of the uncertainties and identification of possible degeneracies. We used uniform priors, 100 walkers and propagated each chain for 10,000 steps. Figure \ref{fig:norm_n_slop} (right panel) shows the posterior distributions of the model parameters. The blue dashed curve on the left panel is computed using the median values of model parameters and the gray-shaded area indicates the 68\% confidence interval computed from the posterior samples. We also show another solution (solid magenta curve) that is obtained for slightly different parameter values (see black vertical lines in the corner plot).

\begin{figure}
    \centering\includegraphics[width=0.49\textwidth]{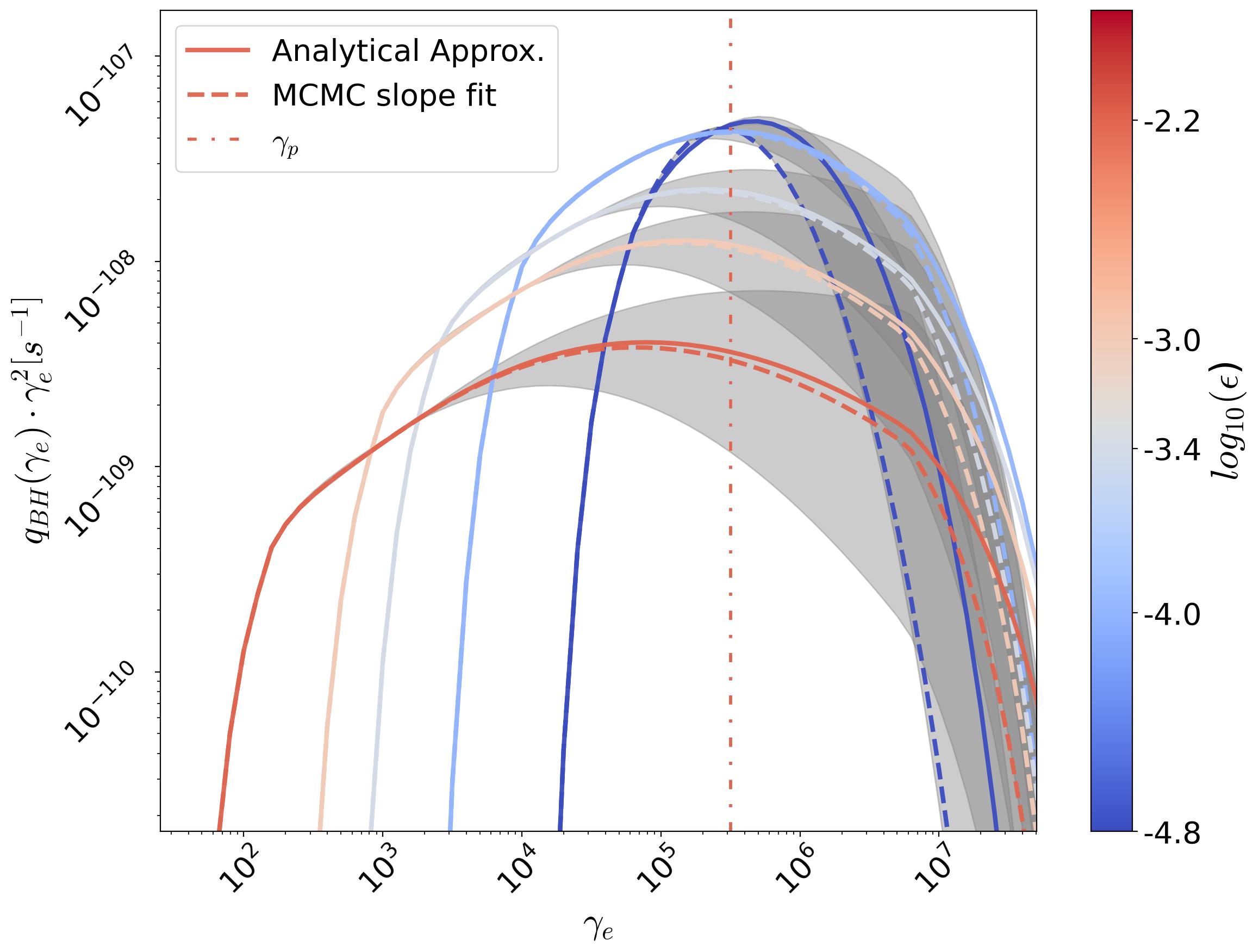}
    \hfill
    \includegraphics[width=0.49\textwidth]{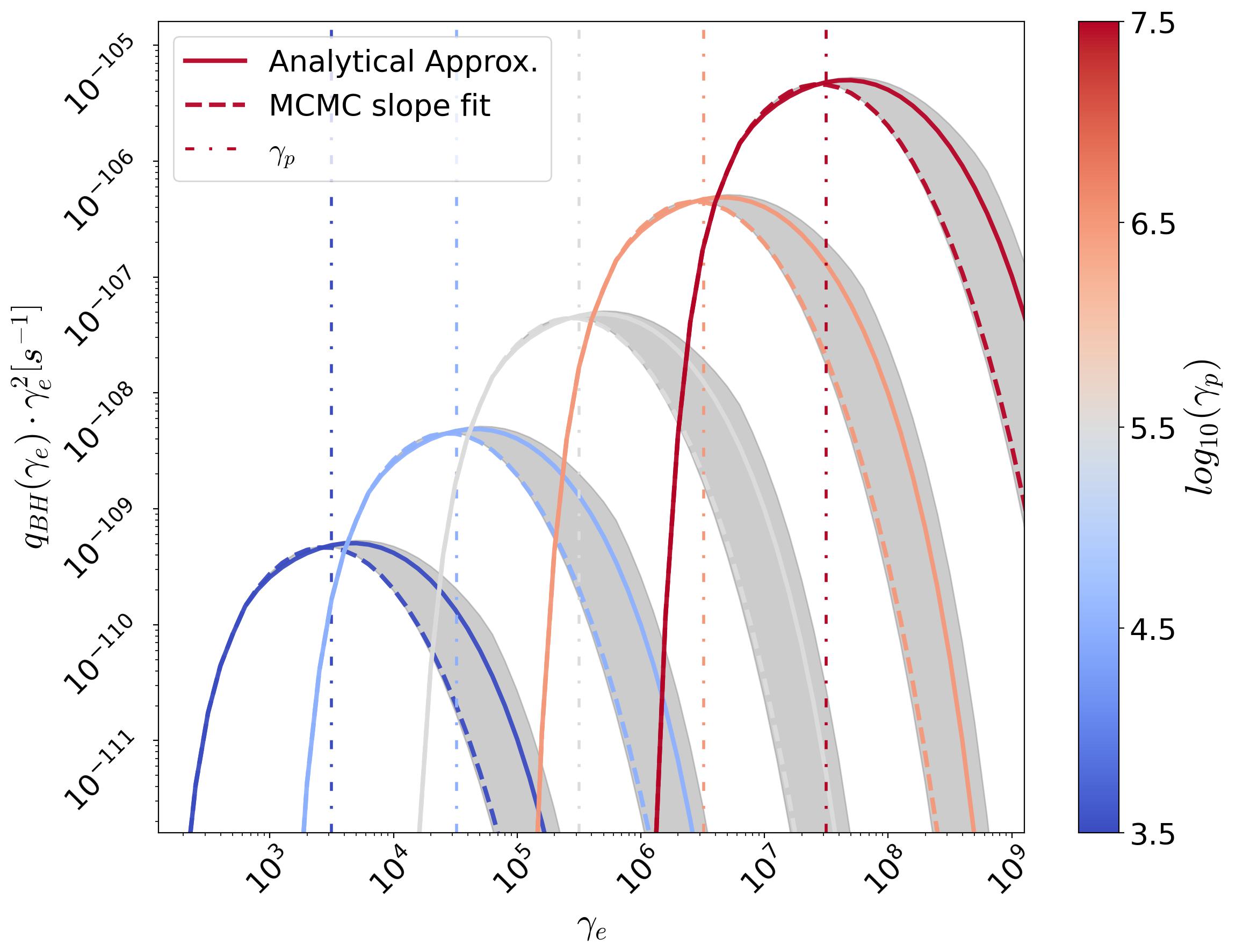}
    \caption{Energy injection spectra for the cases shown in figure \ref{fig:BH_char}. Solid and dashed lines show results when $p(\gamma_p \epsilon)$ follows the the magenta  and blue curves, respectively, shown in figure \ref{fig:norm_n_slop}. Grey-shaded areas demonstrate spectra obtained for values of $p(\gamma_p \epsilon$) drawn from the 68\%  of figure \ref{fig:norm_n_slop}.}
    \label{fig:fit_spread}
\end{figure}

We next compute the pair injection spectra using the fitted slopes and proceed with a comparison against \code \ results -- figure~\ref{fig:fit_spread}. The high-energy cutoff of the energy spectrum is extremely sensitive to the value of $p$ as indicated by the large spread in solutions. Therefore, even small changes in the adopted value of $p$ may have a  large impact on the energy injection spectra -- compare the dashed and solid lines that are obtained using the blue and magenta curves, respectively, shown on the left panel of figure \ref{fig:norm_n_slop}. Moreover, the total energy transferred to the pairs better matches the results of \code \ when the modified slope fit is used in Eq.~(\ref{Q_inj_approx}) -- see figure \ref{fig:Q_and_Qg_int}. For these reasons, we choose to work with the modified slope fit of figure \ref{fig:norm_n_slop} (solid magenta line).  

\begin{figure}[h!]
    \centering
    \includegraphics[width=0.7\textwidth]{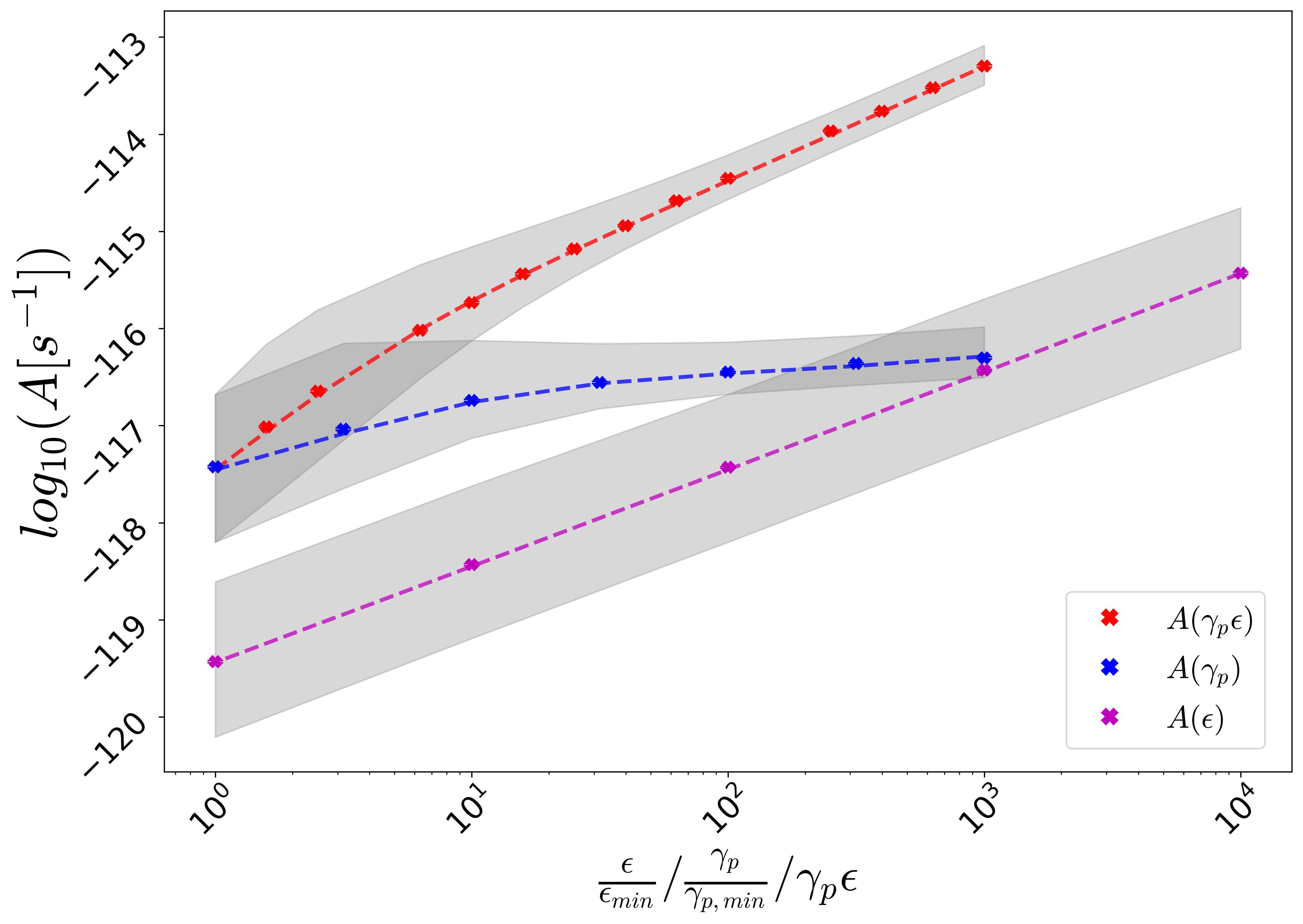}
    \caption{Normalization $A$ of the injection function (see Eq.~\ref{Q_inj_approx}) plotted against the proton Lorentz factor $\gamma_p$, or the target-photon energy $\epsilon$, or the interaction energy $\gamma_p \epsilon$. Red symbols show results when the proton Lorentz factor changes but the photon energy is $\epsilon \approx 2/\gamma_p$ (i.e. interactions happen close to the threshold). Magenta symbols show results when photons of different energies interact with protons of a fixed Lorentz factor (i.e. the interaction energy changes). Blue symbols show results when protons of different Lorentz factors interact with photons of $\epsilon=10^{-5}$. In all cases, the proton and photon distributions are normalized to their total number. The grey shaded areas represent the normalization function range in the 1$\sigma$ spread of the fitted values.}
    \label{fig:Q_inj_norm}
\end{figure}

\begin{figure}[h!]
    \centering
    \includegraphics[width=0.9\textwidth]{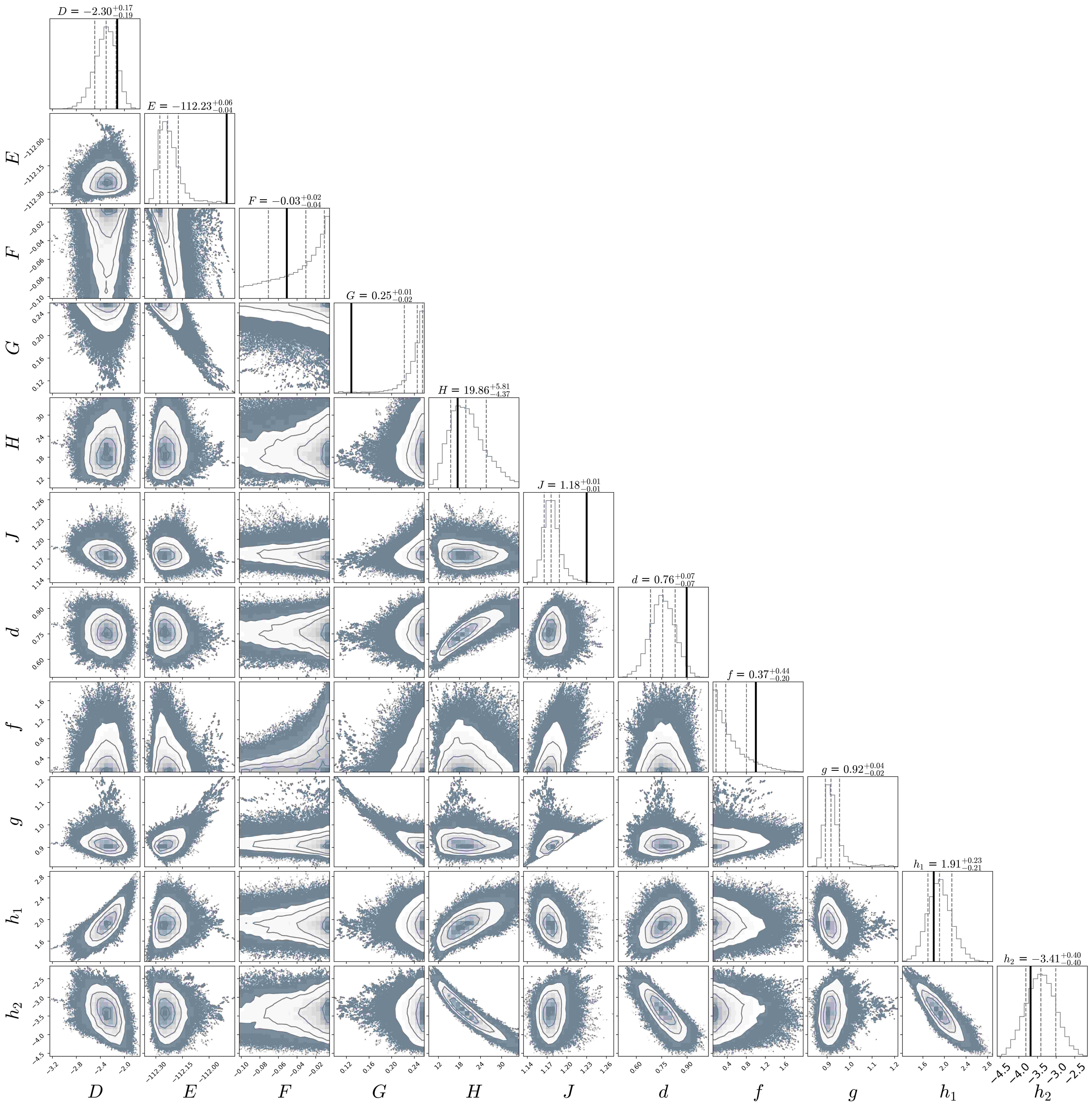}
    \caption{Corner plot showing the posterior distributions of the parameters of the normalization function $A$ displayed in Eq.~(\ref{Q_inj_norm}). Dashed lines in the histograms indicate the median value and the 68\% confidence interval. Solid black lines show the values reported in table~\ref{tab:A_params}.}
    \label{fig:Q_inj_norm_corner}
\end{figure}

After determining the shape of the Bethe-Heitler injection rate function, only the normalization $A$ (see Eq.~(\ref{Q_inj_approx})) was left to fit. We first noticed that the logarithm of the peak of the injection spectrum depends linearly on the logarithm of the target photon energy for a fixed proton Lorentz factor (see magenta markers in figure \ref{fig:Q_inj_norm}). Using the same 15 benchmark cases as in the slope fit, we were also able to determine the dependence of $\log_{10}(A)$ on the interaction energy, $\gamma_p \epsilon$ (see red markers in figure \ref{fig:Q_inj_norm}). In order to check if the normalization  depended also on $\gamma_p$ separately, we used five additional cases that are displayed in figure \ref{fig:BH_char}b. These injection spectra where computed for different values of $\gamma_p$ that still corresponded to the same interaction energy $\gamma_p \epsilon$ (by appropriately selecting the target photon energy $\epsilon$). The dependence of $\log_{10}(A)$ on $\gamma_p$ turns out to be the weakest, as shown by the blue markers in figure \ref{fig:norm_n_slop}). Based on these trends we devised the function presented in Eq.~(\ref{Q_inj_norm}), which has 11 free parameters. To determine these parameters we performed a combined fit to the three sets of points displayed in figure \ref{fig:Q_inj_norm}) using {\tt emcee}. We employed 500 walkers and each chain was propagated for 5,000 steps. We used uniform prior distributions for all parameters. The corner plot showing the posterior distributions is shown in figure~\ref{fig:Q_inj_norm_corner}, and the grey-shaded bands in figure \ref{fig:Q_inj_norm} display the 68\% confidence range. Finally, following the same process as we did for the slope $p(\gamma_p \epsilon)$, we slightly adjusted the optimal parameters we obtained from the MCMC fitting (see solid black lines in the corner plot) to ensure better agreement of the overall spectral shape with the  \code \ results .

\begin{figure}
    \centering
        \includegraphics[width=0.48\textwidth]{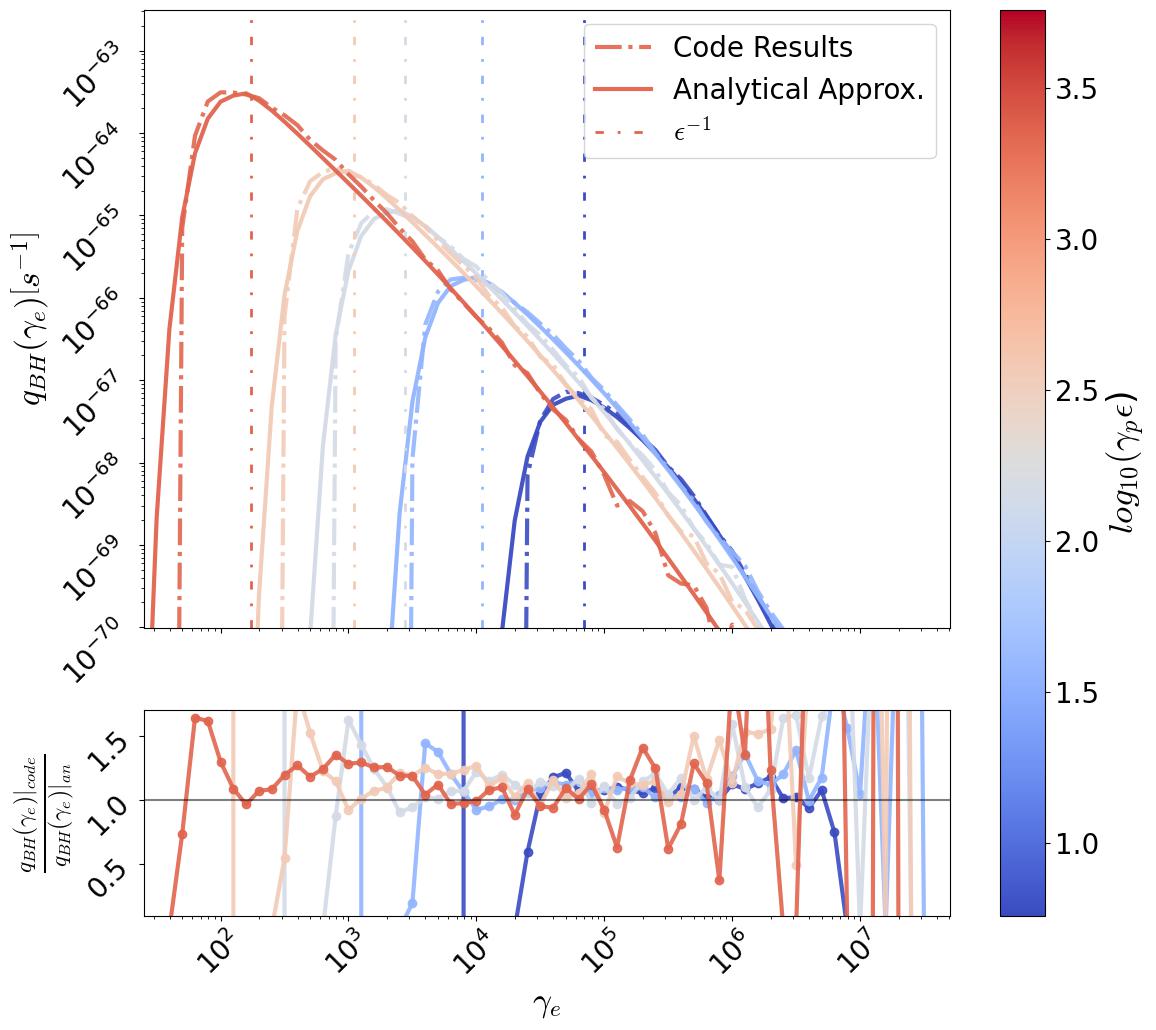}
        \hfill
        \includegraphics[width=0.48\textwidth]{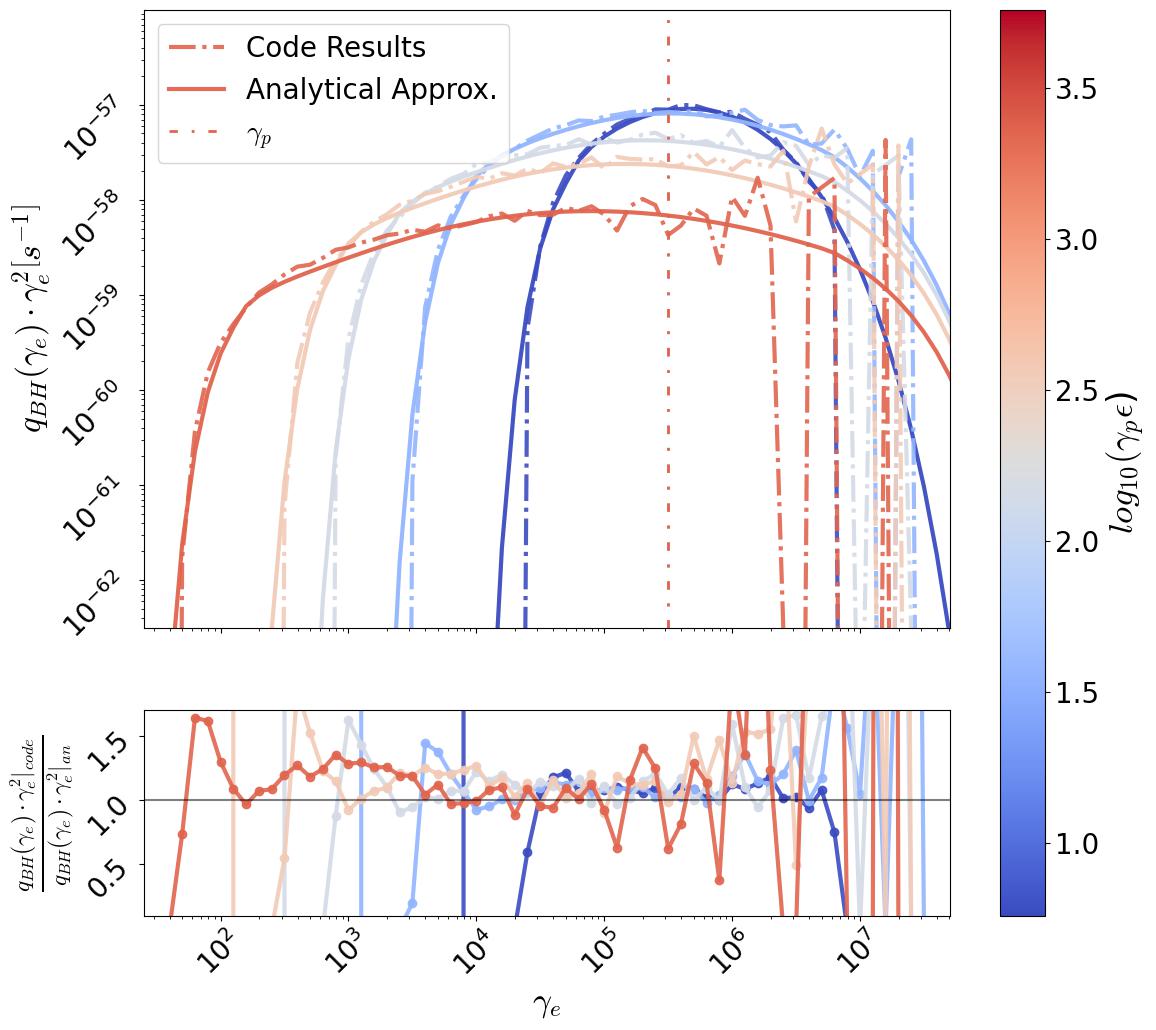}
\caption{Injection spectra of Bethe-Heitler pairs produced by interactions of a single proton with $\gamma_p \approx 3 \cdot 10^{5}$ with a single photon of different energies (see color bar) -- same cases as in panel c of figure \ref{fig:BH_char}. Solid lines represent numerical results from the \code \ code \citep{mastichiadis_spectral_2005, refId0} and dashed lines represent the empirical Bethe-Heitler spectrum of Eq.~(\ref{Q_inj_approx}). Bottom panels show the ratio of the code to analytical results.}
\label{fig:gpE_diff_comp}
\end{figure}

\begin{figure}
    \centering
    \includegraphics[width=0.49\textwidth]{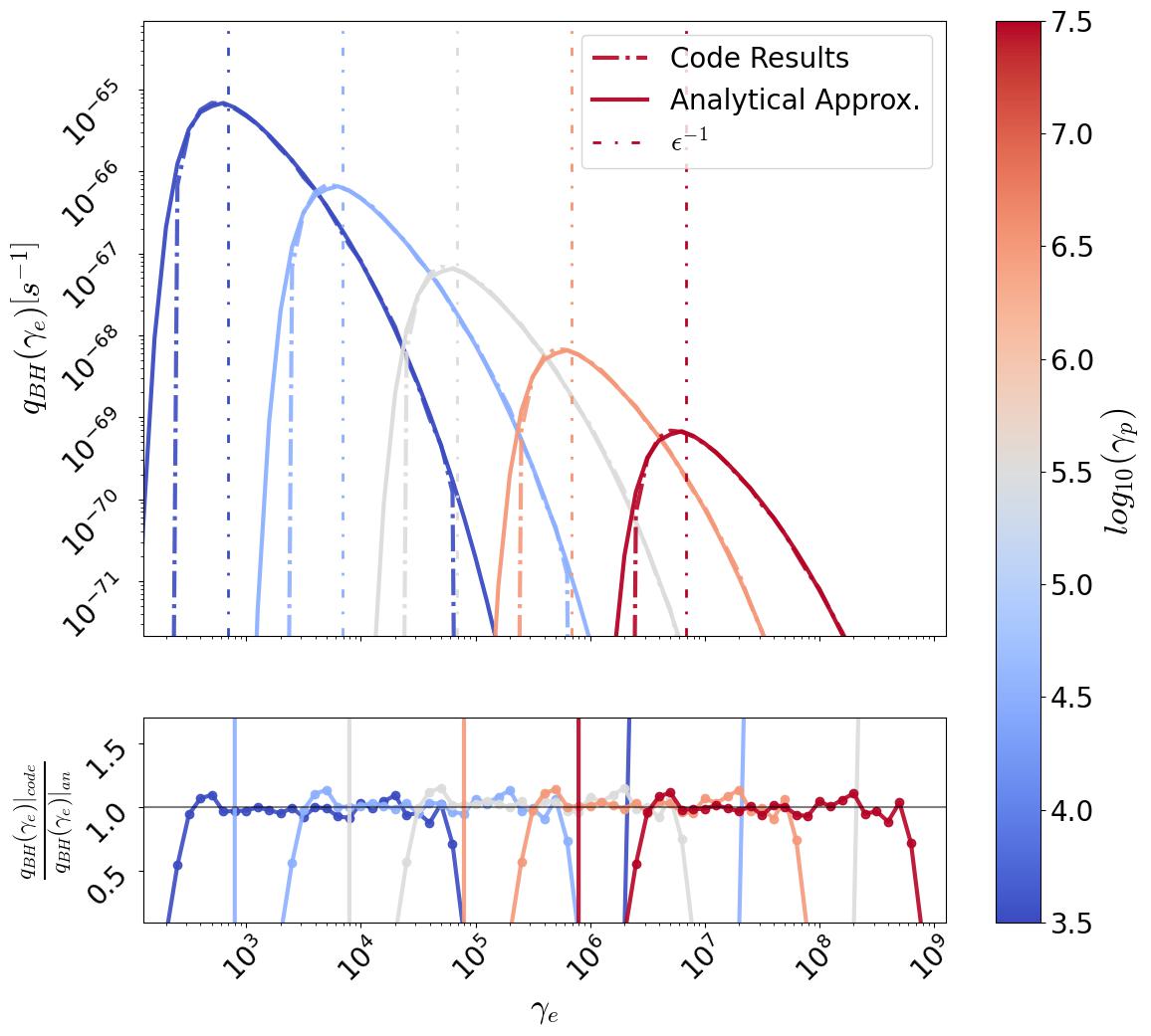}
    \hfill        
    \includegraphics[width=0.49\textwidth]{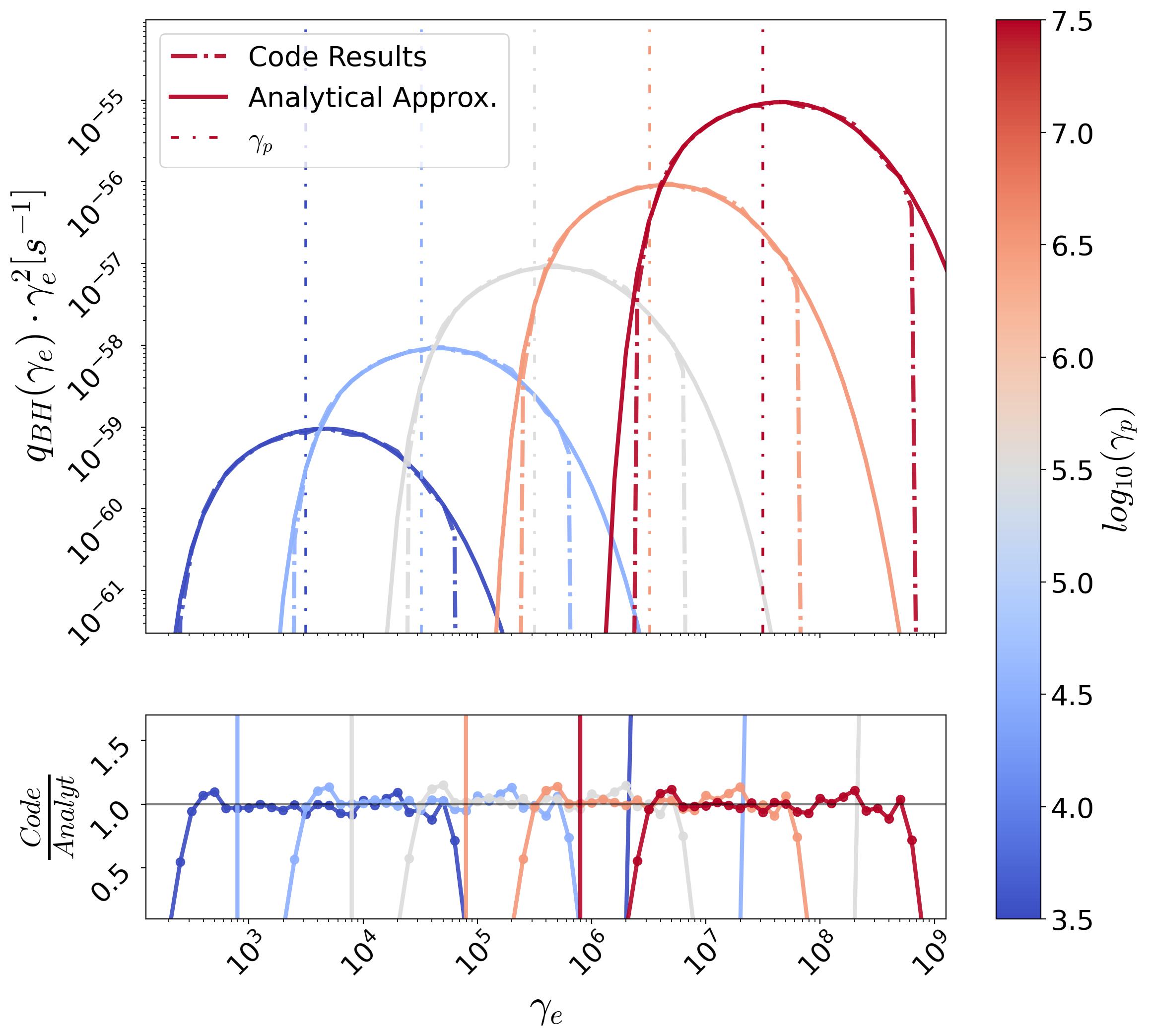}
    \caption{Same as in figure \ref{fig:gpE_diff_comp} but for different proton Lorentz factors (see color bar). The target photon energy in each case is adjusted in order to keep the interaction energy constant, i.e. $\gamma_p \epsilon \simeq 5$. }
    \label{fig:gp_diff_thres_comp}
\end{figure}

We finally compare the spectra obtained using the benchmarked/optimized empirical function against the numerical results of \code. Our results are presented in figures \ref{fig:gpE_diff_comp} and \ref{fig:gp_diff_thres_comp} for the same cases displayed in figure \ref{fig:BH_char}. The function given by Eq.~(\ref{Q_inj_approx}) describes well the peak of the numerical distribution of the Bethe-Heitler produced pairs as a function of the proton Lorentz factor (see left plot of figure \ref{fig:gpE_diff_comp}) and the energy of the target photon (see left plot of figure \ref{fig:gp_diff_thres_comp}). Similarly, the empirical function describes very well the energy distributions of pairs -- see right panels of figures \ref{fig:gpE_diff_comp} and \ref{fig:gp_diff_thres_comp} respectively. 

\subsection{Integral quantities}
In this section we test the performance of the analytical function of Eq.~\ref{Q_inj_approx} in reproducing the code results in terms of integral quantities of importance, like the total number and total energy injected into the pair distribution.

The total number of Bethe-Heitler pairs produced per unit time, in the case of monoenergetic proton and photon populations, is given by
\begin{gather}
    \left. \frac{\d N_{BH, tot}}{\d t} \right\vert_{tot} = N_p N_{ph} \int q_{\rm BH}(\gamma_{e}) \d \gamma_{e}
\end{gather}
where $N_p$ and $N_{ph}$ represent the total number of protons and photons, respectively. In a similar manner the total amount of dimensionless energy that is injected into relativistic pairs is given by
\begin{gather}
    \left. \frac{\d E_{BH, tot}}{\d t} \right\vert_{tot} = N_p N_{ph} \int q_{\rm BH}(\gamma_{e}) \gamma_{e} \d \gamma_{e} m_e c^2
\end{gather}

\begin{figure*}[h!]
\centering
        \includegraphics[width=0.98\textwidth]{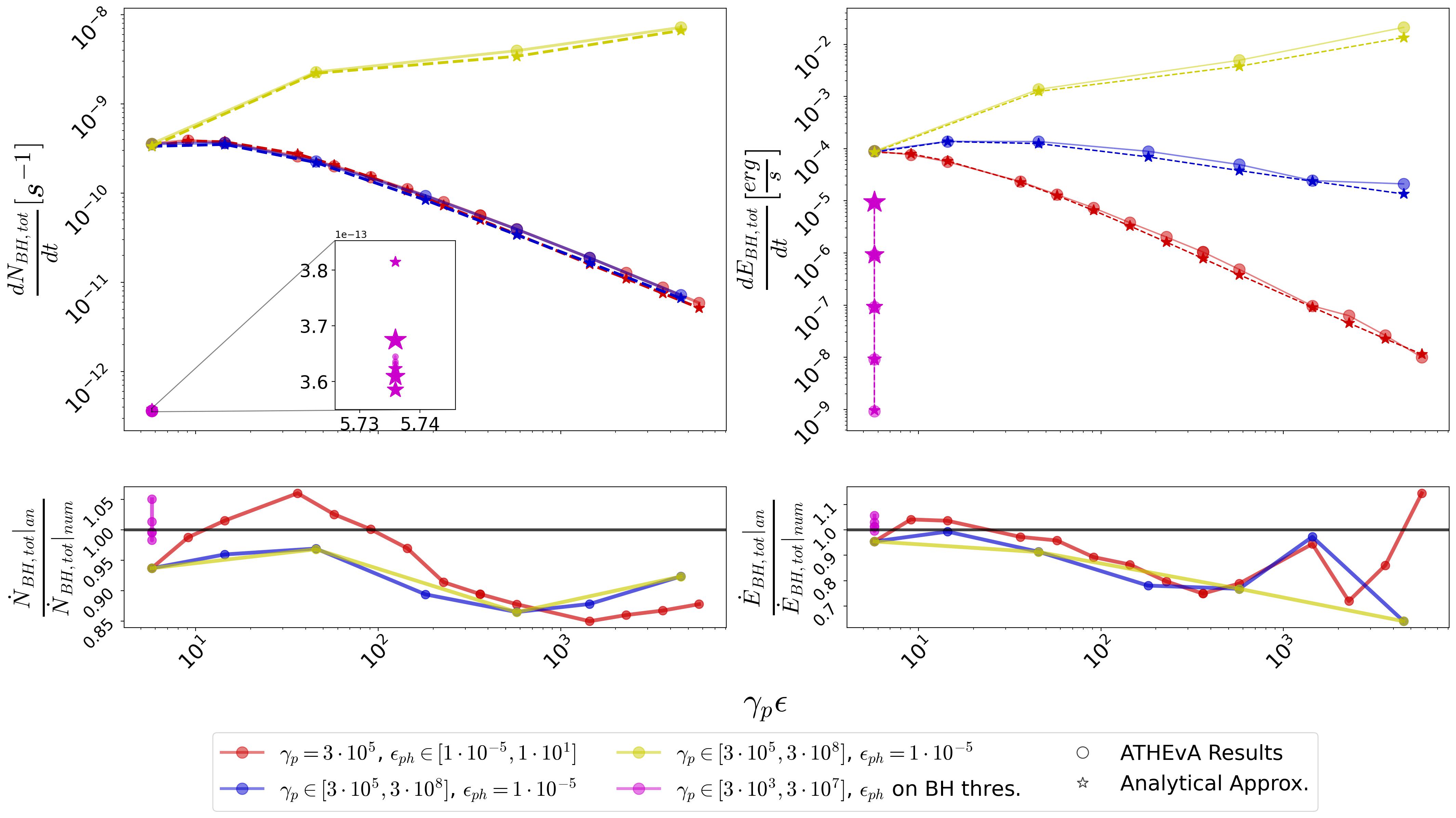}
    \caption{Integrals of total number (upper left panel) and energy (upper right panel) of injected Bethe-Heitler pairs per unit time (with residuals comparing against the \code \ results) in cases where (i) the proton Lorentz factor changes while the photon energy is $\epsilon \approx 2/\gamma_p$; larger marker size indicates higher $\gamma_p$ values (magenta), (ii) photons of different energies interact with protons of the same energy (red), (iii) protons of different Lorentz factors interact with $10^{-5}$ energy photons when the former compactness changes so the total number of the injected protons to be constant (blue) or $l_p \propto \gamma_p$ (yellow).}
    \label{fig:Q_and_Qg_int}
\end{figure*}

Inspection of both panels in figure~\ref{fig:Q_and_Qg_int} shows that the empirical function of Eq.~(\ref{Q_inj_approx}) reproduces satisfactorily the numerical results for the various test cases displayed. In particular, we recover the increasing trend in the number of injected pairs and energy with the proton Lorentz factor (yellow symbols and lines). On the contrary, when the proton Lorentz factor is fixed but the photon energy increases, both the number of pairs and the energy transferred to them drops as we move away from the threshold (see blue and red markers). 
Based on these results we can understand that the increasing trend found for the cases plotted in yellow is due to the presence of more energetic protons and not to the higher interaction energy, $\gamma_p \epsilon$, itself. This result might be explained from the fact that the energy transferred to the secondary population is a fraction of the proton energy \citep{mastichiadis_spectral_2005}, so as the latter increases, the energy transferred to the pair population will follow the same trend.

While the results depicted with yellow and blue symbols are obtained for the same proton-photon energies (see figure legend), the trend with interaction energy is different because of the number of interacting protons: the number of the injected protons remains the same for all cases shown in yellow, while for the cases shown in blue the compactness of the proton population  $\ell_p$ is kept fixed. Since $N_{p, tot} \propto l_p \gamma_p^{-1}$, the combination of a constant $\ell_p$ with an increasing $\gamma_p$ means that the total number of protons decreases. On the contrary if $N_{p, tot}$ is constant and $\gamma_p$ increases, then $\ell_p$ will also be increasing. Since the energy of the pair population is a fraction of the proton population energy, we expect that the energy transferred to pairs will be higher for cases with fixed $N_{p,tot}$ (yellow symbols).

Moreover it is interesting to compare cases of fixed proton Lorentz factor and variable target photon energy (red symbols) with cases where the proton Lorentz factor  changes and the target-photon energy is constant (blue symbols). In both examples the compactnesses of both populations are fixed. We notice, that the total number of the produced pairs as a function of the interaction energy, $\gamma_p \epsilon$, is the same for both cases, see top panel in figure \ref{fig:Q_and_Qg_int}. However, the energy transferred to the pair population is significantly higher in the case where $\gamma_p$ increases, see blue symbols in the bottom panel of figure \ref{fig:Q_and_Qg_int}. This observation is expected if we consider that the total energy the pairs acquire (i) decreases away from the threshold, and (ii) is a fraction of the total energy of the proton population. In both the red and the blue cases, the interaction energy moves away from the threshold, thus we expect the total energy transferred to the pair population to decrease as $\gamma_p \epsilon$ increases. This is, indeed, in the red cases where the photon-target energy is increased while the proton Lorentz factor remains constant. On the contrary, in the cases where $\gamma_p$ increases with $\epsilon$ being fixed, the total energy acquired by the pairs remains almost constant.

Lastly if we take a closer look at the magenta points we see that the total number of  pairs is approximately the same for interactions close to the threshold ($\gamma_p \epsilon \approx 2$), even when $\gamma_p$ and $\epsilon$ are very different. However, the total energy of the pair population increases for higher proton Lorenz factors. This is in agreement with the fact that the pairs receive a fraction of the proton energy \citep{mastichiadis_spectral_2005}, and as the latter increases, the former will too. In other words, more energetic proton populations interacting with photon fields close to the threshold will produce approximately the same number of pairs, but with higher energy on average.

\subsection{Comparison to other numerical implementations of the Bethe-Heitler injection spectrum}\label{app:KA-comparison}
In this work we have constructed an analytical function that approximates the injection rate of the particle distribution created due to Bethe-Heitler pair production. The aforementioned function was benchmarked against the \code \ code \citep{mastichiadis_spectral_2005,refId0}. However, Kelner and Aharonian (KA) \citep{kelner_energy_2008} provided with a formula that describes the Bethe-Heitler distribution of electrons or positrons produced when a single proton interacts with a photon population. This formula involves the numerical computation of three integrals, as shown in Eq.~(\ref{dNdE_full}) \cite{blumenthal_1970}. In this section, we use the leptohadronic numerical code {\tt LeHaMoC} \citep{lehamoc} that implements and solves the Kelner and Aharonian formula for Bethe-Heitler pair production, in order to compare our analytical function against the aforesaid formula.

\begin{figure}[h!]
    \centering
    \includegraphics[width=0.49\textwidth]{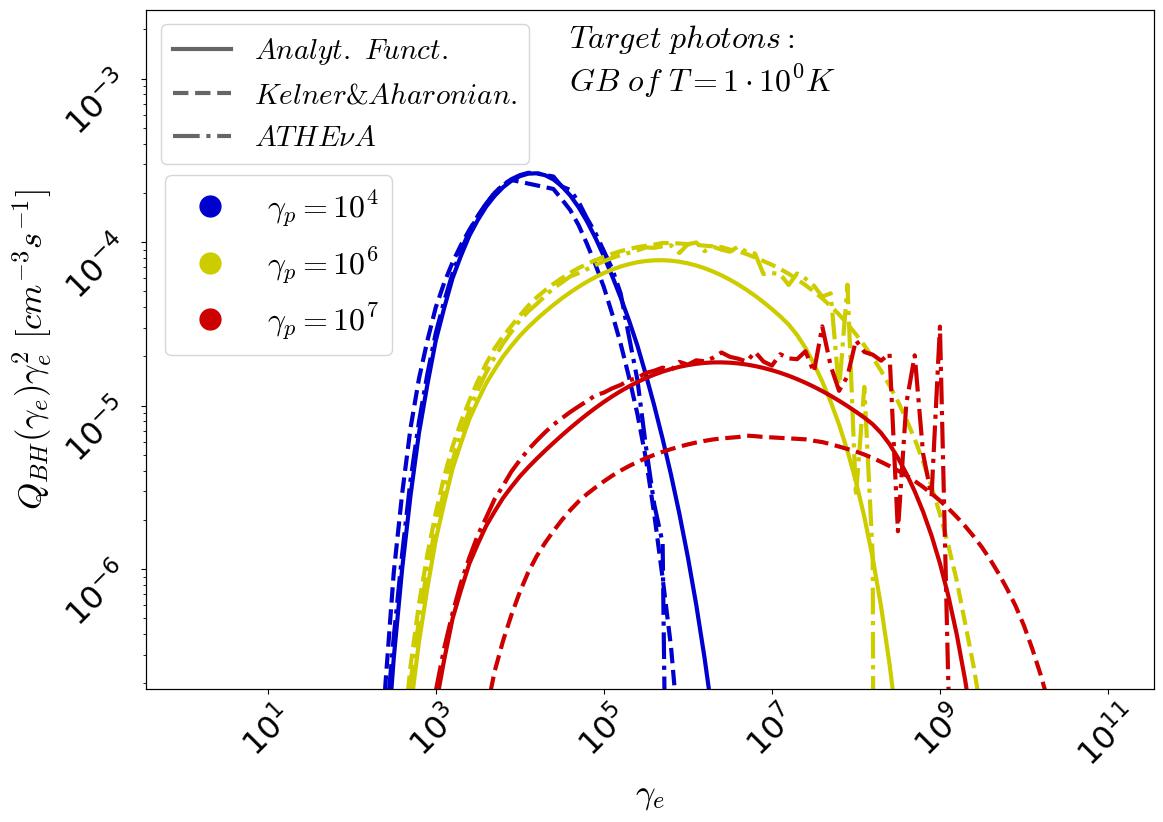}
    \hfill
     \includegraphics[width=0.49\textwidth]{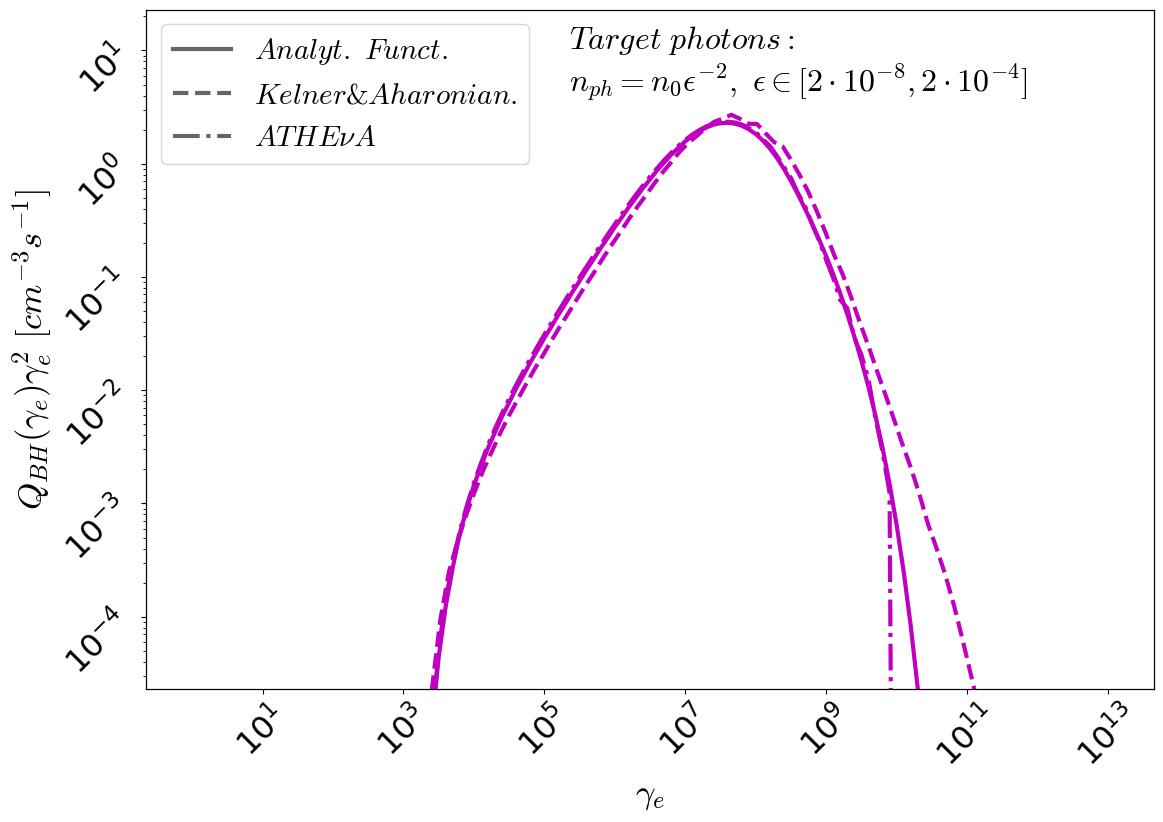}
    \caption{Analytical function approximation, \code \ and Kelner-Aharonian (LeHaMoC) comparison for (i) monoenergetic proton population of various Lorentz factor values interactin with a grey body phoron distribution of $T=10^6$K and (ii) power law protons of $\gamma_p \in [10^4, 10^8]$ interaction with power law photons of $\epsilon \in [10^{-8}, 10^{-4}]$.}
    \label{fig:Ath_LH_analyt_comp}
\end{figure}

Figure \ref{fig:Ath_LH_analyt_comp} shows the comparison of the injection spectrum obtained with the analytical function we have introduced in this work (Eq.~\ref{Q_inj_approx}), the results of the benchmarking code \code, and the results of the KA formula as implemented in {\tt LeHaMoC}. We perform this comparison for two different parameter sets. First, we tested the interaction of a monoenergetic proton population for three different proton Lorentz factor values, $\gamma_p= 10^4,~ 10^6,~10^7$, with a grey body target-photon distribution of temperature $T=10^6$~K (left panel). Second, in the right panel, we show the comparison of the three methods for a power-law proton distribution, $n_p = n_{p,0} \gamma_p^{-2}$ with $\gamma_p \in [10^4, 10^8]$ interacting with a power-law target photon field, $n_{ph}=n_{ph, 0} \epsilon^{-2}$ with $\epsilon \in [10^{-8}, 10^{-4}]$. For all the curves we use a proton compactness of $l_p= 10^{-4}$ and a photon compactness of $l_{ph}=8 \cdot 10^{-6}$ for the grey body photons and $l_{ph}= 10^{-1}$ for the power-law photon distribution. Our analytical approximation matches satisfactorily the \code \ results as expected (see previous sections). More importantly, one notices that the analytical function is in a good agreement with the KA formula in most cases (blue and yellow lines of the left panel and magenta line of the right panel). However, for the red lines of the left panel of figure \ref{fig:Ath_LH_analyt_comp}, where $\gamma_p=10^7$, we see that the KA formula under predicts the production rate of pairs compared to \code, and as a result does not match the results of our analytical function. The latter is due to the fact that the KA formula is not a valid description of the Bethe-Heitler created pair population when the interaction energy, $\gamma_p \epsilon$, reaches values comparable to the mass ratio of protons and electrons, $m_p/m_e$, which is the case for the red lines\footnote{The photon target energy is $\epsilon \approx 3 kT/(m_e c^2) \approx 6\times10^{-4} T_6$ and $\gamma_p \epsilon \simeq 66 \cdot 10^3 \gamma_{p,7} T_6 > m_p/m_e$.} of the left panel in figure \ref{fig:Ath_LH_analyt_comp}. This mismatch does not play in important part in the case of extended proton and/or photon distributions (i.e. right panel of figure \ref{fig:Ath_LH_analyt_comp}) because in such cases the characteristics of the pair distribution are determined of the near-threshold interactions for which the three lines are in good agreement (blue curve of the left panel).

\section{Synchrotron cooling for different B values} \label{App:Bs}

\begin{figure}[h!]
\centering
\includegraphics[width=0.49\textwidth]{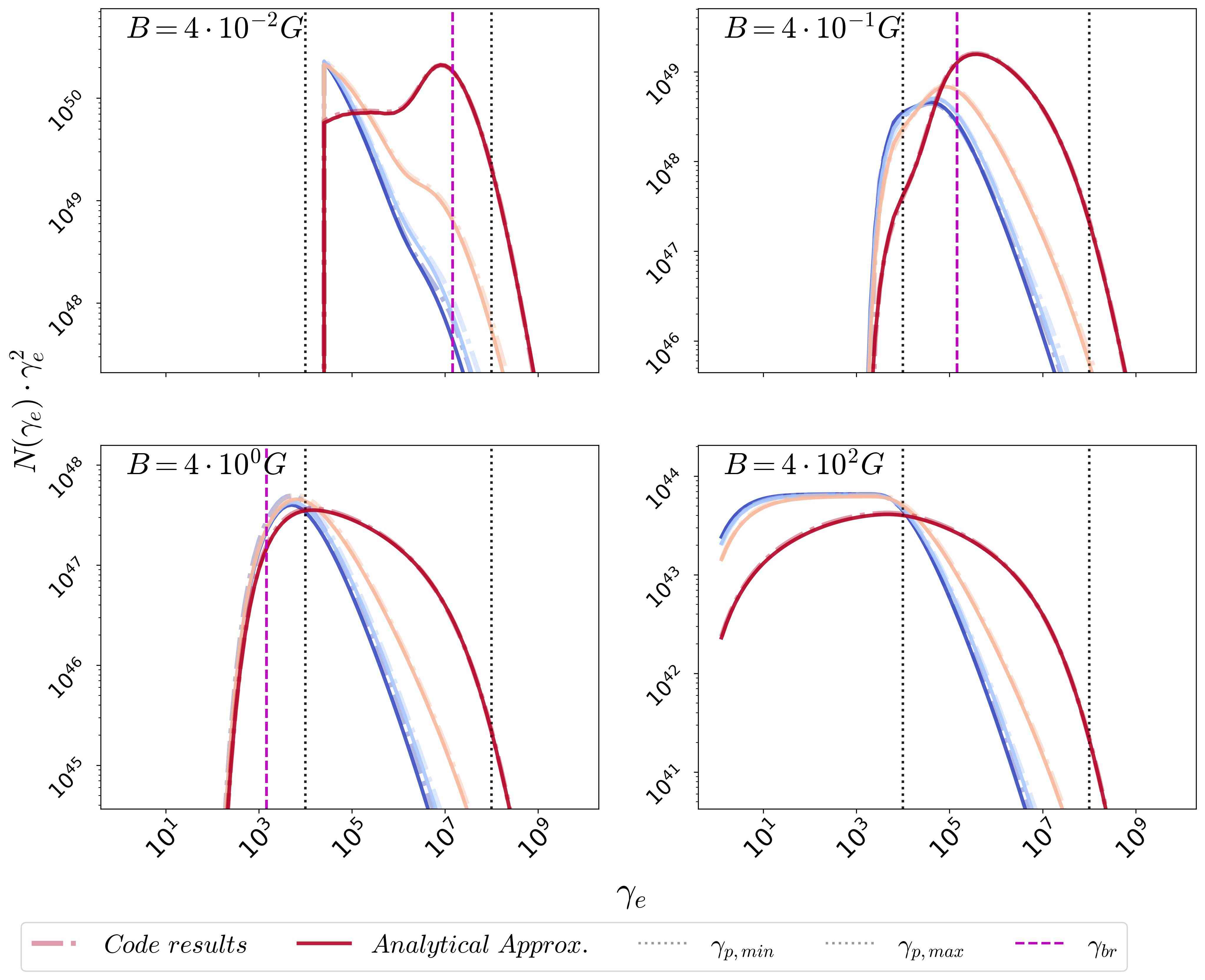}
\includegraphics[width=0.495\textwidth]{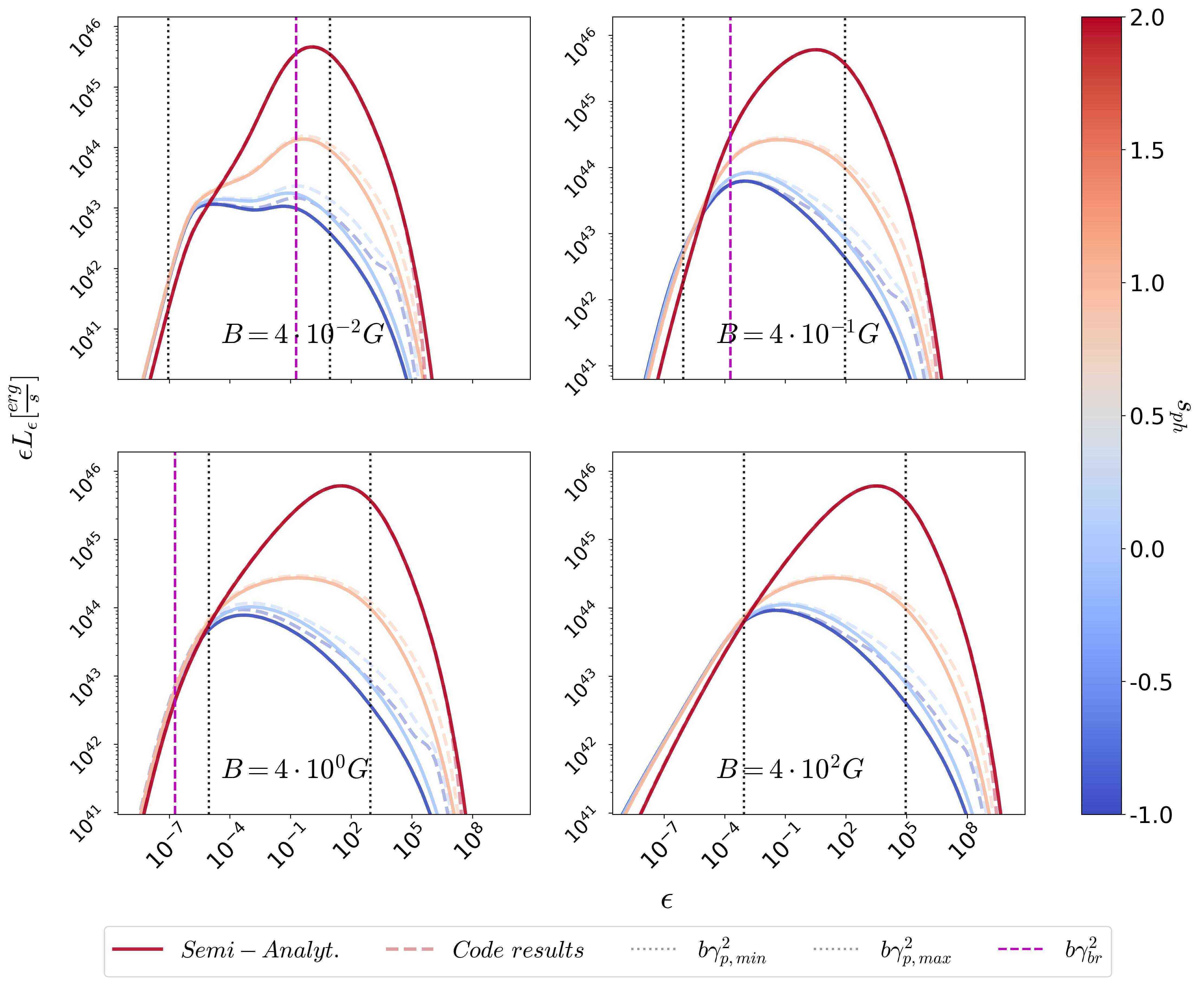}
\caption{Steady-state energy distributions of Bethe-Heitler pairs and their emitted synchrotron spectra for various magnetic field strengths and different power-law slopes of the target photon field  (see color bar). Vertical dotted lines show the synchrotron emission of the pairs with the lowest and highest Lorentz factor values, while vertical magenta lines represent the synchrotron cooling break Lorentz factor, defined in Eq. \ref{syn_cool}. }
\label{fig:I_syn_Bs}
\end{figure}

In section \ref{sec:BH-spec}, we presented steady-state cooled distributions of Bethe-Heitler produced pair populations in a source having a magnetic field strength $B=40$ G (see figure \ref{fig:PL_syn}). In figure \ref{fig:I_syn_Bs} we present steady-state pair distributions with the synchrotron spectra each population emits, for different cooling regimes.

Figure \ref{fig:I_syn_Bs} shows that in sources with high magnetic field values, $B>4$~G, the maximum of the synchrotron spectra does not appear at Lorentz factor value $\gamma_e \approx \gamma_{br}$. On the contrary, both the pair energy distribution and the electromagnetic spectrum peak at either $\gamma_e \approx \gamma_{p,\min}$ or $\gamma_e \approx \gamma_{p,\max}$, depending on the slope of the target photon field (see figure \ref{fig:p_contr}). However, in sources with weaker magnetic fields, $B<4$~G, we observe that $\gamma_{e,\min} \leq \gamma_{br} \leq \gamma_{e, \max}$, so the particles with $\gamma_e < \gamma_{br}$ do not cool significantly in a time frame of $R^\prime_b/c$, where $R^\prime_b=10^{15}$~cm in these examples. As a result the steady-state energy distribution is a combination of the injected energy distribution, as shown in figure \ref{fig:PLs}, with an extra peak originating by the concentration of cooled high-energy pairs around $\gamma_e \approx \gamma_{br}$. The latter feature is more visible for photon power-law slopes $s_{ph} \leq 1$ and it is, also reflected, in the synchrotron spectra of the population.

\section{Additional blazar SEDs} \label{App:SEDs}

\begin{figure}[h!]
\centering
\includegraphics[width=0.95\textwidth]{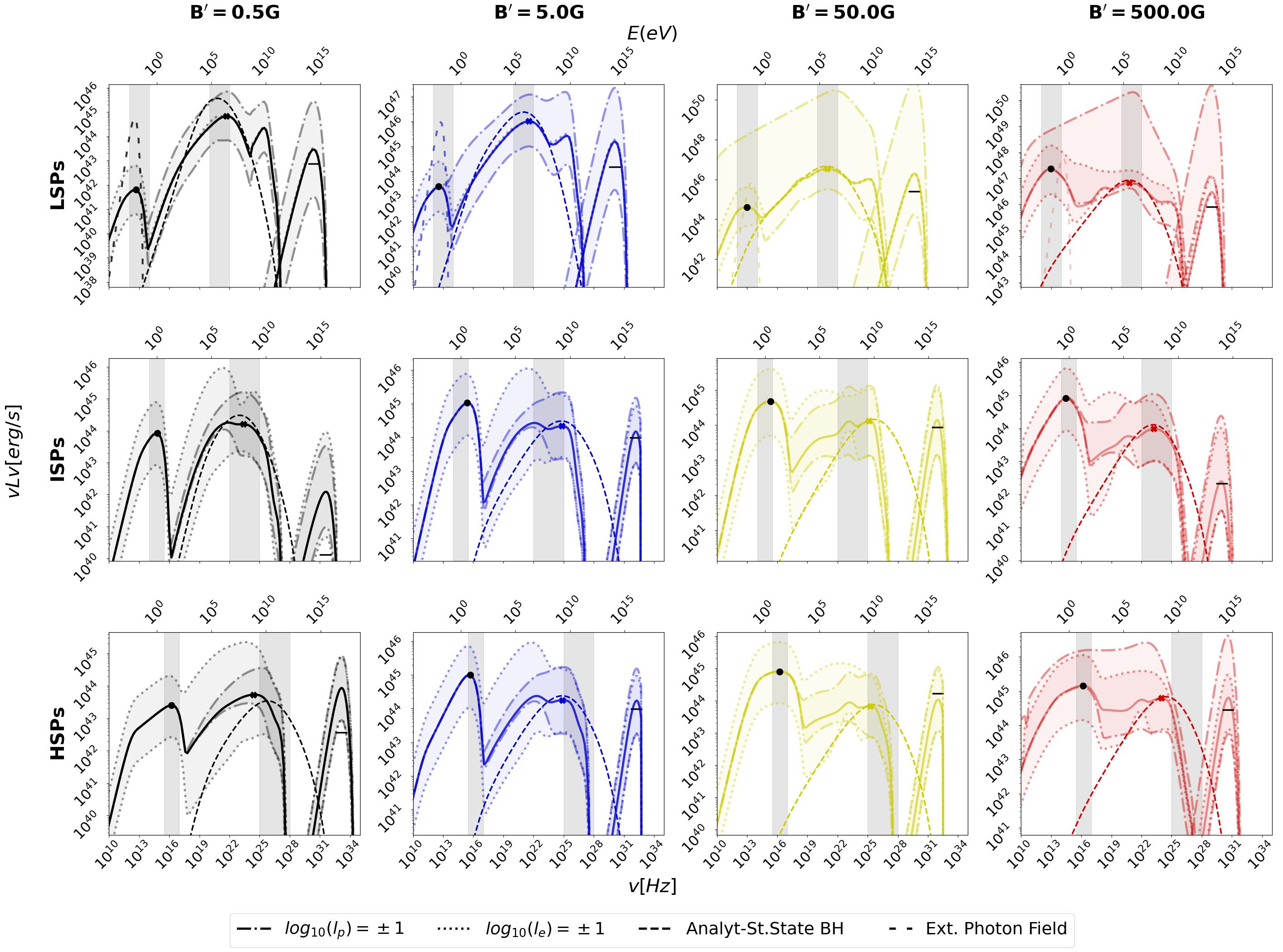}
\caption{LSP-, ISP-, HSP-like cases for a source of radius $R^{\prime}_b=10^{14}$~cm.}
\label{fig:models_var_14}
\end{figure}

\begin{figure}[h!]
\centering
\includegraphics[width=0.95\textwidth]{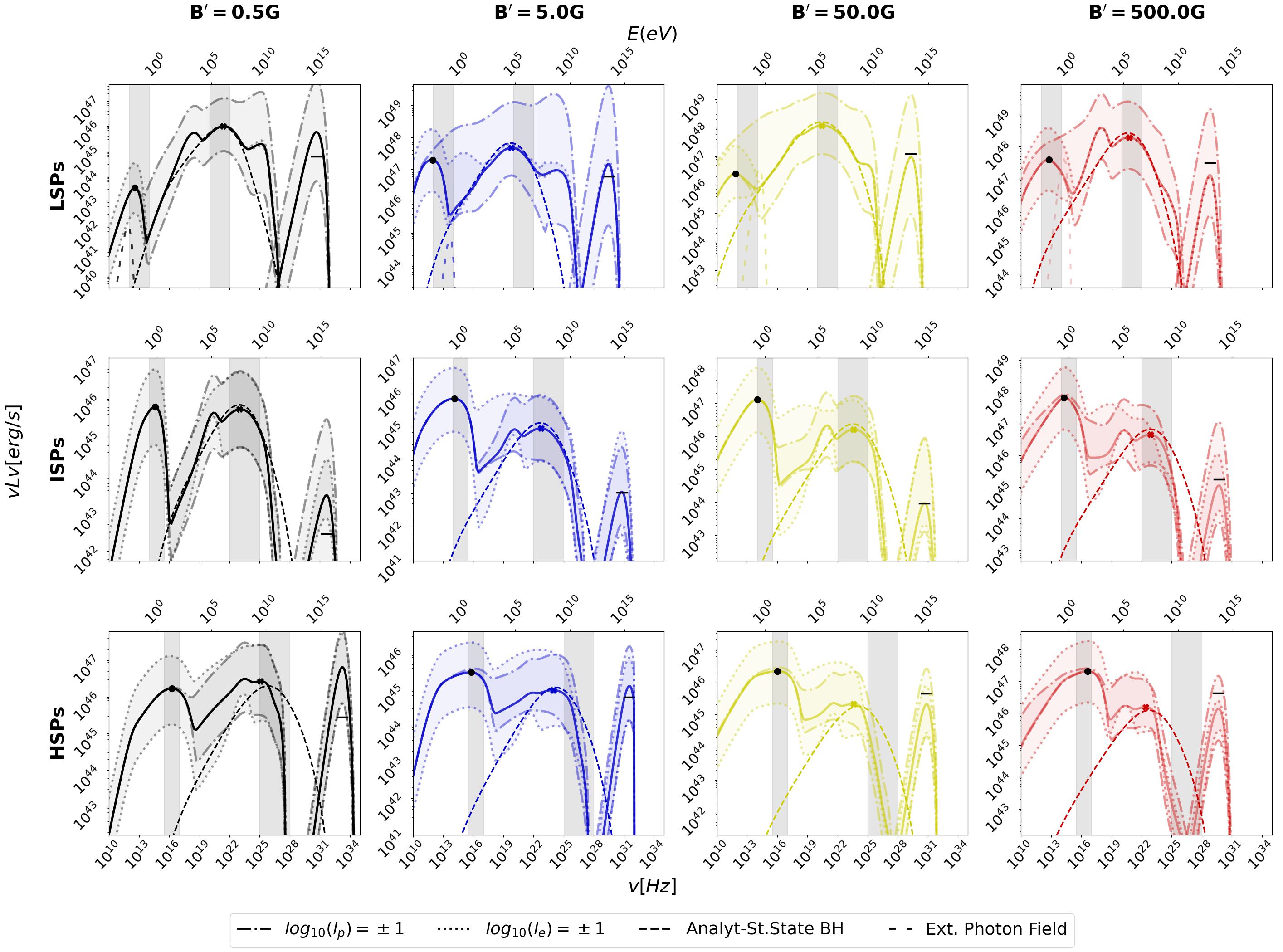}
\caption{LSP-, ISP-, HSP-like cases for a source of radius $R^{\prime}_b=10^{16}$~cm.}
\label{fig:models_var_16}
\end{figure}

In section~\ref{sec:numres} we presented results for a specific source radius. Blazar SEDs computed for smaller and larger emitting regions are presented in figures \ref{fig:models_var_14} and \ref{fig:models_var_16}. For larger emitting volumes ($\propto R_b^{\prime, 3}$) higher photon number densities are required for making the source optically thick to $\gamma \gamma$ pair production. Thus, even by keeping the proton and electron injected luminosities values fixed, we can achieve higher photon luminosity values by increasing the size of the blob. In figures \ref{fig:models_var_14} and \ref{fig:models_var_16} we observe that there are no cases that close to the $\gamma \gamma$ optically thick source limit. When it comes to the variation of $\ell_e$ and $\ell_p$, the behavior of the large sources is similar to the one found for $R_b^{\prime}=10^{15}$~cm.

\clearpage 

\section{SED model parameters} \label{App:Params}
Input parameters to the \code \, code describing the injection populations of electrons, protons and, in some cases, external photons leading to the theoretical models of section~\ref{sec:numres}.

\begin{longtable}{c|c|c|c|c|c|c} \label{tab:model_params}
 & & $B^{\prime} (G)$& 0.5& 5& 50& 500 \\ 
LSPs& & & & & &  \\ \hline 
 & $ R^{\prime}= 1 \cdot 10^{14}$ cm& & & & &  \\ \hline 
 & &$\gamma_e$ &$[10^{1.5}, 10^{2.5}]$ &$[10^{1.0}, 10^{2.0}]$ &$[10^{1.0}, 10^{2.0}]$ &$[10^{1.0}, 10^{2.0}]$ \\ 
 & &$\gamma_p$ &$[10^{6.1}, 10^{6.7}]$ &$[10^{6.0}, 10^{6.5}]$ &$[10^{5.5}, 10^{6.0}]$ &$[10^{5.0}, 10^{5.5}]$ \\ 
 & &$s_e$ &1.8 &1.8 &2.5 &2.5 \\ 
 & &$s_p$ &1.8 &1.8 &2.5 &2.5 \\ 
 & &$l_e$ &$1 \cdot 10^{-5}$ &$3 \cdot 10^{-5}$ &$1 \cdot 10^{-5}$ &$1 \cdot 10^{-2}$ \\ 
 & &$l_p$ &$1 \cdot 10^{-3}$ &$3 \cdot 10^{-3}$ &$1 \cdot 10^{-2}$ &$1 \cdot 10^{-2}$ \\ 
 & &$T^{\prime} (K)$ &$1.8 \cdot 10^{3}$ &$2.0 \cdot 10^{3}$ &$1.2 \cdot 10^{4}$ &$4.0 \cdot 10^{4}$ \\ 
 & &$l_{ext}$ &$1 \cdot 10^{-5}$ &$1 \cdot 10^{-4}$ &$1 \cdot 10^{-4}$ &$3 \cdot 10^{-3}$ \\ 
 & &$l_{ext, max}$ &$1 \cdot 10^{-9}$ &$2 \cdot 10^{-9}$ &$3 \cdot 10^{-6}$ &$9 \cdot 10^{-4}$ \\ 
 & &$\delta$ &80 &60 &65 &40 \\ 
 & $ R^{\prime}= 1 \cdot 10^{15}$ cm& & & & &  \\ \hline 
 & &$\gamma_e$ &$[10^{1.5}, 10^{2.5}]$ &$[10^{1.0}, 10^{2.0}]$ &$[10^{1.0}, 10^{2.0}]$ &$[10^{1.0}, 10^{2.0}]$ \\ 
 & &$\gamma_p$ &$[10^{6.5}, 10^{7.0}]$ &$[10^{5.5}, 10^{6.0}]$ &$[10^{5.5}, 10^{6.0}]$ &$[10^{5.0}, 10^{5.5}]$ \\ 
 & &$s_e$ &1.8 &2.2 &2.5 &2.5 \\ 
 & &$s_p$ &1.8 &2.2 &2.5 &2.5 \\ 
 & &$l_e$ &$1 \cdot 10^{-4}$ &$3 \cdot 10^{-4}$ &$1 \cdot 10^{-5}$ &$3 \cdot 10^{-3}$ \\ 
 & &$l_p$ &$1 \cdot 10^{-2}$ &$3 \cdot 10^{-3}$ &$3 \cdot 10^{-3}$ &$1 \cdot 10^{-3}$ \\ 
 & &$T^{\prime} (K)$ &$3.0 \cdot 10^{2}$ &$1.2 \cdot 10^{4}$ &$3.6 \cdot 10^{4}$ &$1.0 \cdot 10^{5}$ \\ 
 & &$l_{ext}$ &$1 \cdot 10^{-5}$ &$2 \cdot 10^{-5}$ &$3 \cdot 10^{-4}$ &$1 \cdot 10^{-2}$ \\ 
 & &$l_{ext, max}$ &$1 \cdot 10^{-11}$ &$3 \cdot 10^{-5}$ &$3 \cdot 10^{-3}$ &$5 \cdot 10^{-1}$ \\ 
 & &$\delta$ &60 &65 &55 &35 \\ 
 & $ R^{\prime}= 1 \cdot 10^{16}$ cm& & & & &  \\ \hline 
 & &$\gamma_e$ &$[10^{2.0}, 10^{2.5}]$ &$[10^{1.0}, 10^{2.0}]$ &$[10^{1.0}, 10^{2.0}]$ &$[10^{1.0}, 10^{2.0}]$ \\ 
 & &$\gamma_p$ &$[10^{6.5}, 10^{7.0}]$ &$[10^{5.5}, 10^{6.0}]$ &$[10^{5.5}, 10^{6.0}]$ &$[10^{5.0}, 10^{5.5}]$ \\ 
 & &$s_e$ &1.8 &2.2 &2.5 &2.5 \\ 
 & &$s_p$ &1.8 &2.2 &2.5 &2.5 \\ 
 & &$l_e$ &$1 \cdot 10^{-7}$ &$3 \cdot 10^{-4}$ &$3 \cdot 10^{-5}$ &$1 \cdot 10^{-3}$ \\ 
 & &$l_p$ &$1 \cdot 10^{-3}$ &$3 \cdot 10^{-3}$ &$1 \cdot 10^{-3}$ &$1 \cdot 10^{-3}$ \\ 
 & &$T^{\prime} (K)$ &$3.0 \cdot 10^{2}$ &$1.2 \cdot 10^{4}$ &$3.2 \cdot 10^{4}$ &$3.0 \cdot 10^{4}$ \\ 
 & &$l_{ext}$ &$1 \cdot 10^{-6}$ &$3 \cdot 10^{-4}$ &$1 \cdot 10^{-2}$ &$7 \cdot 10^{-2}$ \\ 
 & &$l_{ext, max}$ &$1 \cdot 10^{-10}$ &$7 \cdot 10^{-4}$ &$7 \cdot 10^{-2}$ &$7 \cdot 10^{-2}$ \\ 
 & &$\delta$ &60 &40 &30 &25 \\ 
ISPs& & & & & &  \\ \hline 
 & $ R^{\prime}= 1 \cdot 10^{14}$ cm& & & & &  \\ \hline 
 & &$\gamma_e$ &$[10^{2.5}, 10^{3.5}]$ &$[10^{2.0}, 10^{3.5}]$ &$[10^{2.0}, 10^{3.5}]$ &$[10^{2.0}, 10^{3.0}]$ \\ 
 & &$\gamma_p$ &$[10^{7.5}, 10^{8.0}]$ &$[10^{8.0}, 10^{8.5}]$ &$[10^{8.0}, 10^{8.5}]$ &$[10^{6.5}, 10^{7.0}]$ \\ 
 & &$s_e$ &1.8 &2.01 &1.6 &2.5 \\ 
 & &$s_p$ &1.8 &2.01 &1.6 &2.5 \\ 
 & &$l_e$ &$3 \cdot 10^{-4}$ &$1 \cdot 10^{-4}$ &$1 \cdot 10^{-4}$ &$1 \cdot 10^{-2}$ \\ 
 & &$l_p$ &$3 \cdot 10^{-4}$ &$1 \cdot 10^{-5}$ &$1 \cdot 10^{-5}$ &$1 \cdot 10^{-4}$ \\ 
 & &$\delta$ &60 &60 &30 &10 \\ 
 & $ R^{\prime}= 1 \cdot 10^{15}$ cm& & & & &  \\ \hline 
 & &$\gamma_e$ &$[10^{2.5}, 10^{3.5}]$ &$[10^{1.0}, 10^{3.5}]$ &$[10^{2.0}, 10^{3.0}]$ &$[10^{2.0}, 10^{3.0}]$ \\ 
 & &$\gamma_p$ &$[10^{7.5}, 10^{8.0}]$ &$[10^{7.0}, 10^{7.5}]$ &$[10^{7.0}, 10^{7.5}]$ &$[10^{6.5}, 10^{7.0}]$ \\ 
 & &$s_e$ &1.8 &1.8 &1.6 &2.5 \\ 
 & &$s_p$ &1.8 &1.8 &1.6 &2.5 \\ 
 & &$l_e$ &$1 \cdot 10^{-4}$ &$1 \cdot 10^{-4}$ &$1 \cdot 10^{-3}$ &$1 \cdot 10^{-1}$ \\ 
 & &$l_p$ &$1 \cdot 10^{-4}$ &$1 \cdot 10^{-5}$ &$3 \cdot 10^{-5}$ &$1 \cdot 10^{-4}$ \\ 
 & &$\delta$ &60 &55 &45 &30 \\ 
 & $ R^{\prime}= 1 \cdot 10^{16}$ cm& & & & &  \\ \hline 
 & &$\gamma_e$ &$[10^{2.5}, 10^{3.5}]$ &$[10^{1.0}, 10^{3.5}]$ &$[10^{2.0}, 10^{3.0}]$ &$[10^{2.0}, 10^{3.0}]$ \\ 
 & &$\gamma_p$ &$[10^{7.5}, 10^{8.0}]$ &$[10^{7.0}, 10^{7.5}]$ &$[10^{7.0}, 10^{7.5}]$ &$[10^{6.5}, 10^{7.0}]$ \\ 
 & &$s_e$ &1.8 &1.8 &1.6 &2.5 \\ 
 & &$s_p$ &1.8 &1.8 &1.6 &2.5 \\ 
 & &$l_e$ &$3 \cdot 10^{-6}$ &$1 \cdot 10^{-4}$ &$3 \cdot 10^{-3}$ &$3 \cdot 10^{-1}$ \\ 
 & &$l_p$ &$1 \cdot 10^{-4}$ &$3 \cdot 10^{-5}$ &$3 \cdot 10^{-5}$ &$1 \cdot 10^{-4}$ \\ 
 & &$\delta$ &60 &20 &15 &7 \\ 
HSPs& & & & & &  \\ \hline 
 & $ R^{\prime}= 1 \cdot 10^{14}$ cm& & & & &  \\ \hline 
 & &$\gamma_e$ &$[10^{2.0}, 10^{4.5}]$ &$[10^{1.0}, 10^{3.5}]$ &$[10^{2.0}, 10^{4.0}]$ &$[10^{1.5}, 10^{3.5}]$ \\ 
 & &$\gamma_p$ &$[10^{9.0}, 10^{9.5}]$ &$[10^{8.0}, 10^{8.5}]$ &$[10^{8.0}, 10^{8.5}]$ &$[10^{7.0}, 10^{7.5}]$ \\ 
 & &$s_e$ &2.5 &1.8 &1.8 &1.6 \\ 
 & &$s_p$ &1.8 &1.8 &1.8 &1.6 \\ 
 & &$l_e$ &$3 \cdot 10^{-4}$ &$1 \cdot 10^{-4}$ &$1 \cdot 10^{-4}$ &$1 \cdot 10^{-2}$ \\ 
 & &$l_p$ &$1 \cdot 10^{-5}$ &$1 \cdot 10^{-5}$ &$3 \cdot 10^{-6}$ &$1 \cdot 10^{-4}$ \\ 
 & &$\delta$ &50 &60 &35 &12 \\ 
 & $ R^{\prime}= 1 \cdot 10^{15}$ cm& & & & &  \\ \hline 
 & &$\gamma_e$ &$[10^{2.0}, 10^{4.5}]$ &$[10^{1.5}, 10^{4.0}]$ &$[10^{2.2}, 10^{4.2}]$ &$[10^{1.5}, 10^{3.5}]$ \\ 
 & &$\gamma_p$ &$[10^{9.0}, 10^{9.5}]$ &$[10^{8.0}, 10^{8.5}]$ &$[10^{7.5}, 10^{8.0}]$ &$[10^{6.8}, 10^{7.3}]$ \\ 
 & &$s_e$ &2.5 &1.8 &1.8 &1.6 \\ 
 & &$s_p$ &1.8 &1.8 &1.8 &1.6 \\ 
 & &$l_e$ &$1 \cdot 10^{-5}$ &$3 \cdot 10^{-4}$ &$1 \cdot 10^{-3}$ &$1 \cdot 10^{-1}$ \\ 
 & &$l_p$ &$1 \cdot 10^{-5}$ &$3 \cdot 10^{-5}$ &$1 \cdot 10^{-5}$ &$1 \cdot 10^{-4}$ \\ 
 & &$\delta$ &60 &22 &27 &12 \\ 
 & $ R^{\prime}= 1 \cdot 10^{16}$ cm& & & & &  \\ \hline 
 & &$\gamma_e$ &$[10^{2.0}, 10^{5.0}]$ &$[10^{2.0}, 10^{4.5}]$ &$[10^{2.2}, 10^{4.4}]$ &$[10^{2.0}, 10^{4.0}]$ \\ 
 & &$\gamma_p$ &$[10^{9.0}, 10^{9.5}]$ &$[10^{8.0}, 10^{8.5}]$ &$[10^{7.5}, 10^{8.0}]$ &$[10^{6.5}, 10^{7.0}]$ \\ 
 & &$s_e$ &2.2 &1.8 &1.8 &1.6 \\ 
 & &$s_p$ &1.8 &1.8 &1.8 &1.6 \\ 
 & &$l_e$ &$1 \cdot 10^{-5}$ &$3 \cdot 10^{-4}$ &$1 \cdot 10^{-2}$ &$5 \cdot 10^{-1}$ \\ 
 & &$l_p$ &$1 \cdot 10^{-5}$ &$1 \cdot 10^{-5}$ &$1 \cdot 10^{-5}$ &$1 \cdot 10^{-4}$ \\ 
 & &$\delta$ &60 &12 &8 &5 \\ 

\caption{Parameter values for the injected populations and the source characteristics for the all cases} 
\end{longtable}
\end{document}